\documentclass[reprint,aps,twoside,prd,superscriptaddress,nofootinbib,longbibliography]{revtex4-2}
\pdfoutput=1 %sets output format to pdf. If "0" then output format is DVI.
\usepackage{dcolumn}
\usepackage{physics} %Adds lot of INCREDIBLE commands in math mode for physicists.
\usepackage{makecell} %In tabular env, to format a cell in extremely extensive ways.
\usepackage{enumerate} %Determine Enumeration Style, and nicer back-referening.
\usepackage{amssymb}
\usepackage{amsmath}
\usepackage{amsthm}
\usepackage{tensor,mathtools}
\usepackage{nicefrac}
\usepackage{cancel} %package to draw diagonal lines for cancelling terms
\usepackage[english]{babel}
\usepackage{amsthm}
\newtheorem*{theorem}{Theorem}
\theoremstyle{definition}

\usepackage[version=4]{mhchem}%for chemical symbols
\usepackage[dvipsnames]{xcolor} %to include colours
\usepackage{graphicx}
\usepackage{url}
\usepackage[colorlinks]{hyperref}
\hypersetup{
     breaklinks=true,
    pdfstartview={FitH},    % fits the width of the page to the window
    colorlinks=true,       % false: boxed links; true: colored links
    linkcolor=blue,          % color of internal links
    citecolor=red,        % color of links to bibliography
    filecolor=magenta,      % color of file links
    urlcolor=blue,           % color of external links
    anchorcolor=green,      % Color for anchor text
    linktocpage=true
}

\newcommand{\disd}[1]{\underline{\mathfrak{{#1}}}}
\newcommand{\dis}[1]{\underline{#1}}

\newcommand{\funcal}[2]{\langle \dis{#1},#2 \rangle}
\newcommand{\dfuncal}[2]{\langle \underline{\mathfrak{#1}},#2 \rangle}

\newcommand{\step}{{\theta}}
\newcommand{\dirac}{{\delta}}
\newcommand{\Doublayer}{\Delta}
\newcommand{\stepU}{{\underline{\theta}}}
\newcommand{\diracU}{{\underline{\delta}}}
\newcommand{\DoublayerU}{{\underline{\Delta}}}

\newcommand{\ntime}{{t}}
\newcommand{\nspace}{{s}}

\newcommand{\ifavg}[1]{\left\{#1 \right\}_{\pm}}
\newcommand{\jump}[1]{\left[ #1 \right]_\pm}
\newcommand{\intfc}[1]{\widetilde{#1}}
\newcommand{\dlayer}[1]{\prescript{\mathcal{D}}{}{#1}}

\newcommand{\dotjump}[1]{\left[ #1 \right]^\cdot_\pm}
\newcommand{\hatjump}[1]{\widehat{\left[ #1 \right]}_\pm}

\newcommand{\nolfuncal}[2]{\langle #1,#2 \rangle}
\date{\today}
\begin{document}

\title{A Covariant Distributional Approach for Junctions in Torsional Locally Rotationally Symmetric Class II Spacetimes}

\author{Ujjwal Agarwal}
\email{ujjwal.agarwal@matfyz.cuni.cz}
\affiliation{Institute of Theoretical Physics, Faculty of Mathematics and Physics,
Charles University, Prague, V Hole{\v s}ovi{\v c}k{\' a}ch 2, 180 00 Prague 8, Czech Republic}

\author{Sante Carloni}
\email{sante.carloni@unige.it}
\affiliation{Institute of Theoretical Physics, Faculty of Mathematics and Physics,
Charles University, Prague, V Hole{\v s}ovi{\v c}k{\' a}ch 2, 180 00 Prague 8, Czech Republic}
\affiliation{DIME Sez. Metodi e Modelli Matematici, Universit\`{a} di Genova, Via  All’Opera Pia 15, 16145 - Genoa, (Italy).}
\affiliation{INFN Sezione di Genova, Via Dodecaneso 33, 16146 Genova, Italy}
\affiliation{INAF - Osservatorio Astronomico di Roma, via Frascati 33,
00040 - Monte Porzio Catone (Roma), Italy}
\affiliation{GNFM, Istituto Nazionale di Alta Matematica, P.le Aldo Moro 5,Roma, 00185 Italy}

\begin{abstract}
A rigorous framework for consistent study of junctions in a coordinate independent manner is presented. This is achieved by extending a theory of distributions in curved spacetimes with torsion and combining it with covariant formalism. In this way, one can derive general conditions on the differentiability of the joined manifold. Given a specific theory of gravity, in particular the Einstein-Cartan-Sciama-Kibble one, we evaluate the conditions for obtaining a smooth junction. These conditions can be successfully applied independently of the choice of coordinates used to describe the metric, thereby, evading the drawbacks of coordinate dependent Israel-Darmois framework.
\end{abstract}

\maketitle
\tableofcontents

\clearpage

\section{Introduction}
In many problems in gravitational physics, it is often necessary (and useful) to introduce idealizations that simplify the analysis of spacetime's characteristics and evolution. These idealizations often involve representing a general manifold as a union of several simpler manifolds, each with distinct characteristics. The geometries resulting from joining two different {\it bulk spacetimes} at an interface, which we will call {\it glued spacetimes},  are an example of low-regularity manifolds and find applications in several contexts, from the modeling of relativistic stars to the Oppenheimer-Snyder collapse \citep{OppSnyderColl}. 

The special geometry of glued manifolds implies that standard tensors and their algebra are no longer sufficient. Physical quantities must instead be represented by distributions, a generalization of standard (multilinear) functions that are robust enough to remain well-defined on low-regularity manifolds. Distributions are also central in the description of shock or impulsive gravitational waves, and concentrated sources of matter \citep{Steinbauer2006}.

The use of distributions can substantially simplify the treatment of glued spacetimes. Regardless of the properties of the starting bulk manifolds, they allow not only to find whether the two bulk manifolds can be glued consistently, i.e., the matching conditions, but also to naturally combine the solutions of the field equations in each region and to determine which kind of singular geometry and matter content are induced at the interface. The theory of distributions in the context of general relativity is currently well established \citep{lichnerowicz,Taub,Bruhat,Senovilla2015,SenovillaMars,Marsden}, however, no complete formulation exists in richer spacetimes with a more general connection.

The aim of this paper is to provide such an extension. We will develop the theory of tensor distributions in manifolds with non-vanishing torsion and apply this framework to the study of concentrated matter and singular curvature at junctions between such spacetimes. The bulk geometries considered here are governed by the Einstein--Cartan--Sciama--Kibble (ECSK) theory of gravity \citep{Hehl,HehlSingularity,Kibble,Sciama}. However, several of the results derived below are purely mathematical/geometric, and are therefore independent of the particular gravitational field equations. A central feature of our approach is that it is coordinate-independent: the junction is analyzed in terms of geometric and covariant quantities rather than through a prescribed coordinate representation of the two metrics.

We will begin by developing a pedagogical formulation of tensor distribution theory on manifolds with torsion. The construction is designed to mirror, as closely as possible, the one of Schwartz distributions in Linear Distribution Theory. Schwartz distributions are formally defined as functionals that act on test functions via an integral pairing \citep{schwartzBook,Vladimirov,RobertDistributions}. On a manifold, this naturally leads to the use of integrable densities as the fundamental objects through which distributions are represented. In the given formulation, integrable densities form the basis space for the representation of tensor distributions, and the operations on distributions are derived from their action in this basis space. Particular care is devoted to defining differentiation in the distributional sense, ensuring that the resulting operation remains consistent with the principles of Linear Distribution Theory while correctly incorporating the effects of torsion.

As already mentioned, tensor distribution theory in the absence of torsion has been discussed in Refs. \citep{lichnerowicz,Taub,Bruhat,Senovilla2015,SenovillaMars,Marsden} using also different approaches. For example, the work of Ref. \citep{Marsden} is based on a pointwise convergence topology and generalized exterior algebra. By contrast, our formulation emphasizes representing distributions in the basis space of integrable densities and uses this representation to construct the required operations. In principle, the approach of \citep{Marsden} allows for the introduction of torsion, although this possibility is not developed explicitly.

Once the distributional framework has been established, we use these results to evaluate the geometric singularities and concentrated matter sources at an interface of a glued spacetime. Traditionally, this task is performed using the Darmois--Israel formalism, the most widely used distributional formulation of junctions  \citep{Darmois,Israel,RelToolkitPoisson}. However,  this approach relies on coordinate-dependent quantities, such as the components of the metric tensor. In practice, one must first choose a compatible pair of coordinate systems on the two bulk manifolds so that the junction hypersurface and its normal are described in terms of shared coordinates \citep{Mukohyama}. Since the formalism does not itself provide a systematic way to find such coordinates, this step often becomes a trial-and-error procedure. This can make it difficult to decide whether two bulk geometries are genuinely incompatible or whether any obstruction in joining them is due solely to an unsuitable choice of coordinates. The difficulty becomes particularly acute for junctions involving low-symmetry spacetimes.

To overcome this limitation, in our treatment, we employ the 1+1+2 covariant formalism \citep{EllisElst1998CM,Clarkson_2003}. In this approach, spacetime tensors are decomposed along and orthogonal to preferred congruences. The geometry is then characterized by covariant variables, with direct physical meaning, independent of any coordinate system. This formalism has proved useful in several contexts, including the construction of analytic models for static stellar interiors \citep{CarloniTOVIso1Fluid,CarloniTOVAniso1Fluid,CarloniNaidu}, the study of black-hole perturbations \citep{Clarkson_2003,ClarksonBetschart_2004,Clarkson_2007}, perturbations of static stellar objects \citep{LuzCarloniGaugeInvPert,LuzCarloniComovingPert,LuzCarloniAdiabaticPert}, and cosmological perturbations \citep{Bradley1,Bradley2}. Here we combine the covariant formalism with distributional methods \citep{Israel,RelToolkitPoisson,Senovilla2015,SenovillaMars}, so that the singular curvature and concentrated matter terms at the junction can be encoded directly in coordinate-independent variables.

Although the covariant formalism has previously been used to study junctions in General Relativity \citep{RosaCarloni,Khambule}, the present work focuses on junctions between spacetimes with torsion in the ECSK theory. Junctions in the ECSK framework have been investigated in Refs. \citep{Arkuszewski,Bressange}, but not from a covariant perspective.  

In ECSK theory, the intrinsic spin of matter couples to a non-vanishing torsion field, providing a consistent gravitational framework in which spin matter modifies the differential structure of spacetime \citep{Hehl,HehlSingularity,Kibble,Sciama}. These effects can persist even at macroscopic scales in regimes of extremely high matter density \citep{HehlSingularity}. This is particularly relevant for compact objects, such as neutron stars, whose stability is supported by the neutron degeneracy pressure of a fermionic fluid \citep{PotekhinPhysicsofNS}. Torsion has also been considered in cosmological settings, where it has been used to model late-time accelerated expansion and inflationary behavior in the early universe; see, for instance, Refs. \citep{CaiCapozziello-f(T)gravity,Kopczynski-Singularity} and references therein. Since junctions are central to both astrophysical and cosmological modeling, these applications motivate the extension of junction theory to torsional spacetimes. In this paper, the spin content of matter is modeled by a Weyssenhoff fluid, a semi-classical spin fluid composed of a collection of spinors \citep{WeyssenhoffRaabe,Korotky,Halbwachs}.

A further simplifying assumption in this work is local rotational symmetry, which implies the existence of a local axis of rotation at each point, represented by a spacelike vector field. Many spacetimes of astrophysical and cosmological interest admit this symmetry. In General Relativity, such locally rotationally symmetric (LRS) spacetimes have been studied and classified in Refs. \citep{Ellis1967,EllisStewart1968}, and many physically relevant exact solutions have been organized within the corresponding LRS taxonomy \citep{EllisElst}. Here, we exploit this symmetry together with the $1+1+2$ covariant decomposition, which allows the geometry to be described in terms of a set of scalar variables. Torsional locally rotationally symmetric, or TLRS, spacetimes governed by ECSK theory with Weyssenhoff matter were introduced in Ref. \citep{UjjwalSante}, where their governing equations, classification, and representative examples were discussed. Static stellar models in ECSK theory sourced by Weyssenhoff fluids and belonging to the TLRS class geometries have also been studied in Ref. \citep{CarloniLuz2019}.

The junction conditions derived in this paper are organized into three classes. The first class consists of \textit{fundamental junction conditions}. These are the minimal requirements that must hold for the glued spacetime to be well defined within tensor distribution theory. As in the torsion-free analyses of Refs. \citep{Clarke_1987,Israel,RelToolkitPoisson}, their derivation relies on arguments independent of the gravitational dynamics. In the torsional case, we obtain an additional fundamental condition that must be imposed on the torsion tensor. 
The second class consists of \textit{type-I covariant junction conditions}, which express the consequences of the fundamental conditions for the variables used in the covariant formalism. The third class consists of \textit{type-II covariant junction conditions}, which yield further relations involving discontinuities of the covariant variables and determine the singular contributions at the interface. These conditions also lead to an algorithmic procedure for identifying coordinate systems suitable for describing the junction between two metric solutions. We illustrate the procedure through a representative example in which two spacetimes, each initially given in coordinates, are joined using the covariant junction conditions.

The fundamental junction conditions are required, in part, to exclude products of singular distributions, in particular, quadratic and higher-order powers of Dirac delta distribution. Such products lie outside the ordinary theory of distributions and require a broader framework, such as the Colombeau algebra of generalized functions \citep{Colombeau,Gsponer}. The development of such algebras on manifolds is naturally motivated by the non-linearity of gravitational field equations and has been discussed in Refs. \citep{Steinbauer2001,Steinbauer2006}. Nevertheless, ordinary distribution theory is sufficient for a broad class of problems, including junctions between spacetimes, gravitational shock waves, and impulsive gravitational waves. For the latter, continuous metric representations have been constructed in Refs. \citep{Penrose1972,PodolskyBook,PodolskyPaper1,PodolskyPaper2,PodolskyPaper3}, ensuring that their analysis remains consistent within Linear Distribution Theory. General conditions for the consistent application of distribution theory in gravitational settings have been discussed in Refs. \citep{Geroch,SteinbauerComparison}.

Finally, we note that several approaches to junctions avoid the explicit use of distributions, including the boundary-term approach \citep{Ramirez}, the geometric deformation approach \citep{Huber}, and variational-principle methods \citep{Mukohyama}. These methods are partly motivated by the limitations of ordinary distribution theory when products of Dirac delta distributions arise, a problem that becomes especially important in extensions of General Relativity whose field equations contain non-linear curvature terms. Even in such approaches, however, sufficient regularity of the glued manifold, in particular the continuity of the metric assumed in this work, remains essential for the consistent evaluation of integrals and the use of the divergence theorem. The coordinate-independent distributional and covariant framework developed here is therefore intended to provide a systematic and physically transparent treatment of junctions in torsional gravitational theories.

%Structure of the Paper
The structure of the paper is as follows:
In Section \ref{Sec:LinearDistTheory}, a review of the Linear Distribution Theory is presented.
In Section \ref{sec:IntegrableDensities}, we discuss integration on the manifold, the space of integrable densities, and the divergence theorem.
The information of the last two sections is then combined to describe the Tensor Distribution Theory in Section \ref{Sec:TensorDistTheory}.
In Section \ref{Sec:JunctionsAndDistributionsFull}, we describe the fundamental junction conditions and the distributions necessary to characterize the glued spacetime, formed by joining two torsional spacetimes.
In Section \ref{Sec:TLRS-spacetimes}, a short review of ECSK theory of gravity, covariant formalism, Weyssenhoff fluid and the TLRS spacetimes is provided.
In Section \ref{Sec:JuncStarVacFunda}, we discuss the implications of fundamental junction conditions in the context of covariant formalism.
In Section \ref{Sec:TypeIIJC}, utilizing covariant formalism, we derive the properties of the concentrated matter shell and singular curvature at the interface between two torsional spacetimes with local rotational symmetry, governed by ECSK theory.
In Section \ref{Sec:Examples}, a simple example has been discussed to further elaborate on how the coordinate independent approach to study junctions can be translated to the usual problem of joining two spacetimes described by a metric in specific coordinates.
In Section \ref{Sec:Conclusion}, we discuss the results and conclusions of this paper.
The following information is presented in appendices: 
description of space of tensor densities and formal introduction of explicitly tensorial distributions (App. \ref{App:TensorDenDis});
the details on mappings of integrable densities to $d-$forms and functions (App. \ref{App:DensityComp}); 
the general relation between Lie and Covariant derivatives (App. \ref{App:LieCovDerivRel});
pedagogical derivation of fundamental distributions based on the continuity properties of the manifold (App. \ref{App:DistributionsDerivations}); 
and certain elaborate evaluations and equations, related to Sec. \ref{SubSec:DivPartofWeylEqn} (App. \ref{App:DivPartofWeylEqn}).

%Conventions
In this paper, we shall utilise the natural units $8\pi G=1=c$ and metric signature ${(- + + +)}$. In Sections \ref{sec:IntegrableDensities} and \ref{Sec:TensorDistTheory}, we represent tensors without the indices, in particular, since we deal mostly with $d-$forms and densities (unless specified otherwise). From Section \ref{Sec:JunctionsAndDistributionsFull}, we shall represent the tensors fields utilising abstract indices (unless specified otherwise). Abstract indices help in describing the rank and symmetries of the tensor field and facilitate operations on tensorial objects, but do not denote components in a specific basis \citep{WaldBook}.

\section{Linear Distribution Theory} \label{Sec:LinearDistTheory}

We begin with a brief review of linear distribution theory aimed at introducing the concepts and notation needed for the tensor distribution theory developed in Section \ref{Sec:TensorDistTheory}. More detailed treatments can be found in Refs. \cite{Vladimirov,RobertDistributions,Gsponer,Colombeau,Steinbauer2006}. For further discussion of non-linear distribution theory, in particular Colombeau generalized functions and their applications, see Refs. \citep{Gsponer, Colombeau,Steinbauer2006}.

Let $O\subseteq\mathbb{R}^n$ be an open set. We denote by $\mathcal{C}^m(O)$ the space of $m$-times continuously differentiable functions on $O$, with $m=\infty$ corresponding to smooth functions and $m=0$ to continuous functions. The space of test functions on $O$ is denoted by $\mathcal{D}(O)$. Its elements are smooth functions with compact support in $O$, and hence vanish at the boundary of their support. A {\it Schwartz distribution} on $O$ is a linear and continuous functional on $\mathcal{D}(O)$. Thus, an element $\dis{f}\in\mathcal{D}'(O)$ acts as
\begin{align}
    \mathcal{D}'(O) \ni \dis{f}: \varphi \mapsto \funcal{f}{\varphi} \in \mathbb{R}  \ , \label{defoffunctional}
\end{align}
where $\varphi\in\mathcal{D}(O)$\footnote{By convention, all distributional quantities will be underlined throughout this work.}. Here $\funcal{f}{\varphi}$ denotes the action of the functional $\dis{f}$ on the test function $\varphi$. Linearity and continuity mean, respectively, that
\begin{gather}
    \funcal{f}{a\varphi+b\psi}
    =
    a\funcal{f}{\varphi}+b\funcal{f}{\psi}
    \ , \\
    \lim_{k\to\infty}\funcal{f}{\varphi_k}
    =
    \funcal{f}{\varphi}
    \ ,
\end{gather}
for all $\varphi,\psi,\varphi_k\in\mathcal{D}(O)$, with $a,b\in\mathbb{R}$ and with $\{\varphi_k\}_{k\in\mathbb{N}}$ a convergent sequence of test functions:
\begin{equation}
    \lim_{k\to\infty}\varphi_k=\varphi\,.
\end{equation}

A Schwartz distribution may also be represented as the weak limit of a sequence of basis functions. More precisely, defining $\tilde{\mathcal{D}}(O)$ as the space of the basis functions $f_\epsilon$, one writes
\begin{align}
    \mathcal{D}'(O) \ni \dis{f} = \lim_{\epsilon \to 0} f_\epsilon \ , \quad \text{where } f_\epsilon \in \tilde{\mathcal{D}}(O) \ ,\label{RepresentationOfDistributionViaBasis}
\end{align}
provided the weak convergence condition
\begin{align}
 \lim_{\epsilon \to 0} \int_O f_\epsilon (x) \varphi(x) dx \in \mathbb{R} \ , \quad \forall \varphi \in \mathcal{D}(O) \ , \quad x\in O \ ,
\end{align}
is satisfied \citep{Gsponer,Vladimirov}. In this representation, the action of $\dis{f}$ is given by
 \begin{align}
    \funcal{f}{\varphi} = \lim_{\epsilon \to 0} \int_O f_\epsilon (x) \varphi(x) dx \in \mathbb{R} \ . \label{Distributionoperation}
\end{align}
A given Schwartz distribution need not have a unique sequence representation. Rather, the admissible sequences form an equivalence class of representations of the same distribution. In spite of this fact, the space $\mathcal{D}'(O)$ can be characterized abstractly, independently of any particular representative sequence.

The Dirac delta distribution provides a standard example. For simplicity, let $O=\mathbb{R}$. The distribution $\diracU$ is defined by \citep{Gsponer}
\begin{align}
    \mathcal{D}'(O) \ni \dirac: \varphi \mapsto 
    \funcal{\dirac}{\varphi}
    =
    \varphi(0)
    \in \mathbb{R} \ ,
    \label{lindiracdefinition}
\end{align}
and can be generated by either of the following two sequences:
\begin{equation} \begin{split}
    {}^1\dirac_\epsilon(x)
    &=
    \frac{1}{\pi\epsilon}
    \frac{\epsilon^2}{x^2+\epsilon^2}
    \ ,
    \\
    {}^2\dirac_\epsilon(x)
    &=
    \frac{1}{\sqrt{\pi}\epsilon}
    \exp\left(-\frac{x^2}{\epsilon^2}\right)
    \ .
\end{split} \end{equation}
This example also illustrates a general feature of Schwartz distributions: both the test functions and the basis functions may be identified with smooth functions that vanish at the boundary of the open set under consideration. A similar identification applies to the derivatives of the Dirac delta distribution, which also belong to $\mathcal{D}'(O)$ \citep{Gsponer,RobertDistributions}. In what follows, we shall therefore denote both the space of test functions and the space of basis functions by $\mathcal{D}(O)$.

For our purposes, two elementary properties of Schwartz distributions will be important. The first is that Schwartz distributions are uniquely determined by their action on all test functions. If
\[
\funcal{f}{\varphi}=\funcal{g}{\varphi}
\quad \forall \varphi \in \mathcal{D}(O) \ ,
\]
then $\dis{f}$ and $\dis{g}$ are equal on $O$, that is, $\dis{f}=\dis{g}$. 

Second, $\mathcal{D}'(O)$ is ``weakly complete''. Given a sequence of Schwartz distributions $\{\dis{f_k}\}_{k\in\mathbb{N}}$, if the numerical sequence
\begin{align}
    \lim_{k\to\infty}\funcal{f_k}{\varphi}
\end{align}
converges for every $\varphi\in\mathcal{D}(O)$, then one can define
\begin{align}
    \dis{f}
    =
    \lim_{k\to\infty}\dis{f_k}
    \in
    \mathcal{D}'(O)
    \ .
\end{align}

A {\it null distribution} can also be defined through its action on test functions. If  $\tilde{O}\subset O$ is open and
\begin{equation}
\funcal{f}{\varphi}=0, \quad\forall \varphi \in \mathcal{D}(\tilde{O}) \ ,
\end{equation}
then we say that $\dis{f}=0$ in $\mathcal{D}'(\tilde{O})$. This leads to the definition of the {\it support of a distribution}, i.e. the closed set
\begin{align}
Supp(\dis{f}) = O\setminus O_f \ ,
\end{align}
where $O_f$ is the zero set of $\dis{f}$, defined as the union of all open subsets $\tilde{O}_i\subset O$, with $i\in I\subset\mathbb{N}$, on which $\dis{f}=0$.

An algebra on $\mathcal{D}'(O)$ can be constructed by defining operations through the action of distributions on test functions. The addition is the closed binary operation
\begin{align}
    \dis{h}
    =
    \dis{f}+\dis{g}
    \ ,
    \qquad
    \dis{h}\in\mathcal{D}'(O)
    \ ,
\end{align}
such that
\begin{align}
    \funcal{h}{\varphi}
    =
    \funcal{f}{\varphi}
    +
    \funcal{g}{\varphi}
    \ .
    \label{BinaryOperationDistVecSpace}
\end{align}
The zero distribution, which has null support, acts as the additive identity and it is always possible to define an additive inverse. Multiplication of a distribution by a smooth function $\Phi\in\mathcal{C}^{\infty}(O)$ is defined by
\begin{align}
    \funcal{\Phi f}{\varphi}
    =
    \funcal{f}{\Phi\varphi}
    \ ,
    \label{ScalarmultDistVecSpace}
\end{align}
where $\dis{\Phi f}\equiv\Phi\dis{f}\in\mathcal{D}'(O)$. If $\dis{f}$ is represented by the sequence $f_\epsilon$, then $\dis{\Phi f}$ is represented by the sequence $\Phi f_\epsilon$. With the addition law \eqref{BinaryOperationDistVecSpace} and the scalar multiplication \eqref{ScalarmultDistVecSpace}, the space of Schwartz distributions forms a vector space over $\mathcal{C}^{\infty}(O)$\footnote{Strictly speaking, the space of smooth functions is a ring, rather than a field, since a smooth function does not necessarily admit a well-defined multiplicative inverse. Accordingly, the space of distributions forms a \textit{module} over the ring of smooth functions.}.

A differential algebra on   $\mathcal{D}'(O)$ can be introduced by defining the {\it derivative in the distributional sense}. Denoting by the usual partial derivative $\partial$, one sets
\begin{align}
    \funcal{\partial f}{\varphi}
    =
    -
    \funcal{f}{\partial\varphi}
    \ ,
\end{align}
where $\partial$ acts on $\varphi\in\mathcal{D}(O)\subset\mathcal{C}^{\infty}(O)$ as the ordinary partial derivative. Higher-order derivatives are defined recursively:
\begin{align}
    \funcal{\partial^n f}{\varphi}
    =
    (-1)^n
    \funcal{f}{\partial^n\varphi}
    \ .
    \label{DerivativeofLinearDistribution}
\end{align}
The derivative of a distribution obeys the following properties \citep{Vladimirov}:
\begin{enumerate}
    \item The operation $\partial$ is linear and continuos on distributions and maps $\mathcal{D}'(O) \rightarrow \mathcal{D}'(O)$.
    \item The Leibniz formula holds true for the derivative of the product $\dis{\Phi f}$
    \begin{align}
        \partial (\dis{\Phi f}) = \dis{f}\partial\Phi + \Phi \dis{\partial f} \ . \label{SchDistLeibniz}
    \end{align}
    \item In general, $Supp(\partial f) \subset Supp(f)$
    \item Distributions can be indefinitely differentiated in the distributional sense.
\end{enumerate}

The representation of Schwartz distributions by sequences of smooth functions allows many operations on smooth functions to be extended to distributions. The guiding principle is that an operation on a distribution is performed by applying the corresponding operation to a representative sequence of basis functions. Equations \eqref{ScalarmultDistVecSpace} and \eqref{DerivativeofLinearDistribution}, for example, follow from this prescription together with \eqref{Distributionoperation} and suitable boundary conditions\footnote{The differential algebra described above can also be formulated for broader classes of generalized functions that do not necessarily admit a sequence representation in terms of smooth functions. In that setting, the space of basis functions is described as the topological dual of the space of generalized functions, with \eqref{defoffunctional} playing the role of the inner product \citep{Vladimirov, Gsponer}. This observation underlies the construction of Colombeau generalized functions \citep{Colombeau}.}. However, not every operation on representative sequences yields a new sequence satisfying the weak convergence condition. When this happens, the corresponding operation is ill-defined within Linear Distribution Theory. The most important example is the product of two Schwartz distributions. In general, the product of two representative sequences does not define a new weakly convergent representative sequence \citep{Gsponer,Colombeau} and, therefore, the product of two distributions is not itself a distribution.

Products of distributions nevertheless arise frequently in physical applications. They are usually handled through additional, ad hoc prescriptions. For instance, the step distribution $\stepU$ is defined by
\begin{align}
    \mathcal{D}'(\mathbb{R}) \ni \stepU:
    \varphi
    \mapsto
    \funcal{\step}{\varphi}
    =
    \int_0^\infty \varphi\,dx
    \ .
    \label{linstep}
\end{align}
One commonly imposed prescription is
\begin{align}
    \stepU^2
    =
    \stepU
    \ .
    \label{ThetaThetaRule}
\end{align}
The tension between such prescriptions and the differential structure becomes apparent when one differentiates products of distributions. To see this explicitly, let us consider the derivative of the step distribution. Since test functions vanish at the boundary, its derivative  is the Dirac delta distribution:
\begin{equation}
\begin{aligned}
    \funcal{\partial\step}{\varphi}
    &=
    -
    \funcal{\step}{\partial\varphi}
    =
    -
    \int_0^\infty
    \frac{\partial\varphi(x)}{\partial x}\,dx
    =
    \varphi(0)
    =
    \funcal{\dirac}{\varphi}
    \ .
\end{aligned}
\label{lindirac}
\end{equation}
Now consider the derivative of $\stepU^3$. If one applies the Leibniz rule before using the prescription \eqref{ThetaThetaRule}, one obtains
\begin{equation}
\begin{aligned}
    \partial(\stepU^3)
    &=
    3\stepU^2\diracU
    =
    3\stepU\diracU
    \ ,
\end{aligned}
\end{equation}
while applying the prescription \eqref{ThetaThetaRule} before the Leibniz rule, leads to
\begin{equation}
\begin{aligned}
    \partial(\stepU^3)
    &=
    \partial(\stepU^2)
    =
    2\stepU\diracU
    \ .
\end{aligned}
\end{equation}
These two expressions would imply
\begin{align}\label{inconsistencyDistLeibniz}
    3\stepU\diracU
    =
    2\stepU\diracU
    \implies
    \stepU\diracU
    =
    0
    \implies
    \diracU
    =
    0
    \ ! \ .
\end{align}
This inconsistency is a manifestation of the Schwartz Impossibility Result, discussed in Refs. \citep{schwartz1954, Colombeau, Steinbauer2006}, which shows that the product of distributions cannot be incorporated into the standard distributional framework while retaining all the usual algebraic and differential properties.

Thus, in Linear Distribution Theory, the product of two Schwartz distributions is not defined in general and must be supplemented by additional prescriptions when it appears in applications. The most commonly used rules are \cite{Steinbauer2006}
\begin{align}
    \stepU^2
    &=
    \stepU
    \ ,
    &
    \stepU\diracU
    &=
    \frac{1}{2}\diracU
    \ .
    \label{LinAdHocRules}
\end{align}
By contrast, $\diracU^2$ and higher powers of the Dirac delta distribution are usually removed by imposing suitable constraints\footnote{For further details, see Ref. \citep{Gsponer}. In that work, the consistency of the ad hoc prescriptions in \eqref{LinAdHocRules} is established within the Colombeau algebra. In contrast, terms such as $\diracU^2$ do not define Schwartz distributions, although they may be described as generalized functions.}. Because of the Schwartz Impossibility Result, the Leibniz rule cannot be maintained for the product of two arbitrary distributions. Hence, for $f,g\in\mathcal{D}'(O)$, one has in general
\begin{equation}
\begin{aligned}
    \partial(\dis{fg})
    \neq
    \dis{g}\,\partial\dis{f}
    +
    \dis{f}\,\partial\dis{g}
    \ .
\end{aligned}
\end{equation}
In practical calculations, one first resolves the product by an appropriate prescription and only then differentiates the resulting distribution.

A key link between finite differentiability and distribution theory is provided by the following result \citep{Gsponer,Colombeau,schwartzBook}.

\begin{theorem}[Schwartz Local Structure (SLS) Theorem]
    Any Schwartz distribution is locally a partial derivative of a continuous function in $\mathcal{C}^0(O)$.
\end{theorem}

The SLS theorem shows that Schwartz distributions naturally extend the algebra of continuous and finitely differentiable functions. It also clarifies the relation between the formal definition of distributions as functionals and their familiar pointwise representations. Consider, for example, the continuous function
\begin{align}
\xi(x)
=
    \begin{cases}
    & 0 \ , \quad \forall x \in (-\infty,0)
    \\
    & x \ , \quad \forall x \in [0,\infty)
    \end{cases}
    \ ,
\end{align}
whose pointwise derivative is
\begin{align}
    \partial\xi
    =
    \frac{\partial\xi(x)}{\partial x}
    =
    \vartheta(x)
    =
    \begin{cases}
        & 0 \ , \quad \forall x \in (-\infty,0)
        \\
        & 1 \ , \quad \forall x \in (0,\infty)
        \\
        & \text{ill-defined} \ , \text{ for }x=0
    \end{cases}
    \quad .
    \label{linsteppointwise}
\end{align}
Irrespective of the value assigned at $x=0$, the integral properties of $\vartheta(x)$ coincide with those of the step distribution \eqref{linstep}\footnote{Any finite value may be assigned to $\vartheta(0)$, since the value of a function at a single point does not affect the integral.}. In this sense, $\vartheta$ is a pointwise representative of $\stepU$. This representation is useful, but it should not obscure the importance of the functional definition. As a distribution, the derivative of $\stepU$ is the well-defined Dirac delta distribution. By contrast, differentiating the pointwise representative $\vartheta(x)$ gives the formal object
\begin{align}
\partial\vartheta(x)
=
\dirac_p(x)
=
    \begin{cases}
        \infty \ , \text{ for } x=0
        \\
        0 \ , \text{ otherwise}
    \end{cases}
    \ ,
\end{align}
which is not a well-defined function. Consequently, standard expressions such as $x\diracU(x)$ become indeterminate if interpreted purely pointwise.

We close this review with two comments that will be relevant later. First, the pointwise representative $\vartheta(x)$ might appear to justify some of the ad hoc rules in \eqref{LinAdHocRules}. Indeed, if one chooses $\vartheta(0)=1$, then
\begin{align}
    \vartheta^2(x)
    =
    \vartheta(x)
    \ .
\end{align}
However, this argument only reproduces the first rule in \eqref{LinAdHocRules}. More importantly, the inconsistency displayed in \eqref{inconsistencyDistLeibniz} reappears as soon as one differentiates $\vartheta^3$ using the Leibniz rule. This result illustrates a broader point: the Schwartz Impossibility Result does not merely state that the Leibniz rule fails for products of distributions. Rather, it reflects the deeper incompatibility between differentiation and multiplication for finitely continuous functions \citep{Colombeau}. Through the SLS theorem, this incompatibility is inherited by distributions and is responsible for the general ill-definedness of their products. Nevertheless, the pointwise representation of the step distribution remains useful in applications of Linear Distribution Theory and, in the present work, of Tensor Distribution Theory; see Section \ref{Sec:JunctionsAndDistributionsFull}.

Second, although in Linear Distribution Theory the space of basis functions is usually identified with the space of test functions, namely $\mathcal{D}(O)$, these two spaces are conceptually distinct. This distinction will become essential in the formulation of Tensor Distribution Theory in Section \ref{Sec:TensorDistTheory}.

\section{Integrable Densities}\label{sec:IntegrableDensities}

In the next section, we aim to provide a description of distributions on a curved manifold. Since, as discussed in the previous section, linear distributions are defined via functionals acting on test functions as integrals, this task requires the introduction of some important mathematical objects associated with integration on the manifold, i.e., integrable densities \citep{NomizuInterscience,JohnLeeBook,Nicolaescu,SchutzBook}. In addition, we will need to generalize these quantities and the corresponding theorems to the case in which torsion is present.  With this objective in mind, in the following, we will review integration on a manifold with non-vanishing torsion, using integrable densities (see \citep{NomizuInterscience} for additional information). 

The key element in the integration operation on a manifold is the {\it volume form}. A volume form of a $d-$dimensional orientable manifold $\mathcal{M}$ is typically described by a $d-$form with a vanishing covariant derivative \citep{NomizuInterscience}. 

One of the crucial reasons why $d-$forms are used to describe volume forms and integrals can be understood by looking at the transformation law for components of $d-$forms. Given a $d-$form $\alpha \in \Omega^d_\mathcal{M}$
\begin{equation}\label{genericdform}\begin{aligned}
     \alpha 
     =& \frac{1}{d!}\alpha_{i_0\, i_1 ...i_{(d-1)}} (x) dx^{i_0}\wedge dx^{i_1}\wedge ... \wedge dx^{i_{d-1}} 
     \\
     =& \alpha_{0\,1 ...(d-1)} (x) dx^{0}\wedge dx^{1}\wedge ... \wedge dx^{d-1} \ ,
\end{aligned}\end{equation}
where $\wedge$ is the exterior (wedge) product, the components of a $d-$form transform as
\begin{align}
    \bar{\alpha}_{0 ... (d-1)} (\bar{x}) = J {\alpha}_{0 ... (d-1)} ({x}) \ , \label{TransDForm}
\end{align}
where $J$ is the determinant of transformation matrix $\partial x^i / \partial \bar{x}^j$ for the coordinate systems $x$ and $\bar{x}$. Since the quantity $dx^{i_0}\wedge dx^{i_1}\wedge ... \wedge dx^{i_{d-1}} $ transforms like \eqref{TransDForm}, but with the factor $J^{-1}$, the above definition implies that any $d-$form, as well as its integral, is coordinate independent\footnote{This statement is of course true if the orientation of the manifold is fixed.}.

To study integration on the manifold, it is prudent to map the $d-$forms to a space of {\it integrable densities} on the manifold. An integrable density $\mathfrak{a}$ can be expressed as 
\begin{align}
    \mathfrak{a} = \mathfrak{a}(x) d^dx \ , 
\end{align}
where $\mathfrak{a}(x) \in \mathbb{R}$\footnote{Strictly speaking, one describes the operation $\mathfrak{a}:e_i \mapsto \mathfrak{a}(e_i) \in \mathbb{R}$, where $e_i$ describes a holonomic frame, given by $e_i=\partial / \partial x^i$ for a coordinate system $x$.}, and possess the following transformation law:
\begin{align}
    \bar{\mathfrak{a}}(\bar{x}) = \left| J \right| \mathfrak{a}(x) \ . \label{TransDensity}
\end{align}
The space of integrable densities $\mathbb{D}_\mathcal{M}$ forms a vector space\footnote{One typically defines the basis for the vector space via a so-called coordinate density $\mathfrak{e} = d^dx$, for a given coordinate basis $x$, with the property that $\mathfrak{e}(x) = 1$.} over the space of functions on the manifold, which we will denote, from now on, as $\mathcal{F}_\mathcal{M}$.

An integrable density $\mathfrak{a}$ is formally associated with $d-$form $\alpha$ as
\begin{align}
    \alpha=\varepsilon[\mathfrak{a}] \ , \label{DensityDformRel}
\end{align}
where $\varepsilon[\cdot]$, from now on referred to as \textit{orientation tensor} for convenience, operates on a basis as\footnote{To be precise, the map $\varepsilon$ describes an isomorphism from $\mathbb{D}_\mathcal{M}$ to the space of $d$-forms $\Omega^d_\mathcal{M}$ up to continuous transformations for orientable manifolds (or orientable sub-regions of a non-orientable manifold), i.e., for which globally non-vanishing $d$-forms can be defined. More details can be found in Appendix \ref{App:DensityComp}.}
\begin{align}
    \varepsilon[d^dx] = dx^{0}\wedge ... \wedge dx^{d-1} \ .
\end{align}
In terms of components, and choosing some coordinate $x$,  the above mapping  reads
\begin{align}\label{DensityDformRelComp}
    \alpha_{i_0 i_1 ...i_{(d-1)}} (x) = \mathfrak{a}(x) \varepsilon_{i_0 i_1 ...i_{(d-1)}} \ ,
\end{align}
where $\varepsilon_{i_0 i_1 ...i_{(d-1)}}$ is given by the Levi-Civita symbol, defined in any coordinate basis as \citep{SchutzBook}:
\begin{align}
    \varepsilon_{i_0 i_1 ...i_{(d-1)}} = 
    \begin{cases}
        +1 \text{, for } i_0 i_1 ...i_{(d-1)} =               
             \\  \qquad\text{even permutation}
             \\ \qquad \text{of }0,1...(d-1); 
        \\
        -1 \text{, for } i_0 i_1 ...i_{(d-1)} = 
                                                \\ \qquad \text{odd permutation}   
                                                \\ \qquad\text{of }0,1...(d-1); 
        \\
        0 \text{, otherwise.}
    \end{cases}
\end{align}
Note that the Levi-Civita symbol is not a $ d$-form nor does it follow its transformation law. In addition,  both the covariant and the Lie derivative $\mathcal{L}_X$ of Levi-Civita symbol vanish 
\begin{equation}
\mathcal{L}_X\varepsilon = \nabla_X\varepsilon=0 \ .
\end{equation}  
 Because of this property $\varepsilon$  is commonly referred to as a {\it structure preserving isomorphism}.

We can easily verify that the integral of a generic integrable density $\mathfrak{a}$ is independent of the choice of coordinates. In fact:
\begin{equation}\begin{aligned}
     \int_\mathcal{M} \mathfrak{a} &= \int_\mathcal{M} \mathfrak{a} (x) d^dx \ ,
     \\
      \int_\mathcal{M} \bar{\mathfrak{a}}(\bar{x}) d^d\bar{x} 
   &= \int_\mathcal{M} \mathfrak{a}(x) \left|J\right| \left|J^{-1}\right| d^dx =\int_\mathcal{M} \mathfrak{a} 
    \ .
\end{aligned}\end{equation}
i.e., like in the case of the d-forms, we obtain the same result irrespective of which coordinate system is utilized to evaluate the integral. Indeed, this result is the very reason why the expression $\int_\mathcal{M} \mathfrak{a}$ can be written without referring to coordinates.

Using integrable densities, and given a function $f\in\mathcal{F}_\mathcal{M}$ on a manifold, the integral 
\begin{align}
    \int_\mathcal{M} f \mathfrak{a} = \int_\mathcal{M} f\mathfrak{a} (x) d^dx \ , \text{ where } \mathfrak{a}\in \mathbb{D}_\mathcal{M} \ ,
\end{align}
is also well defined since, as we have seen, $f \mathfrak{a}\in \mathbb{D}_\mathcal{M}$.

While the choice of a volume form in a manifold is completely arbitrary, for the metric-compatible connections ($\nabla_c g_{ab}=0$), a  specific volume form, called the {\it Levi-Civita $d-$form}, is normally employed. This $d-$form is given by
\begin{align}
    \eta = \sqrt{|\mathfrak{g}| (x)}  dx^0 \wedge dx^1 \wedge ... \wedge dx^{d-1} \ , \label{LeviCivitaDForm}
\end{align}
where $\mathfrak{g}(x)$ is the determinant of the metric in a given coordinate system. 
The key property of the Levi-Civita tensor is that it is the only $d-$form which has a vanishing covariant derivative $\nabla_a \eta = 0$. This can be seen  immediately by noticing that any generic $d-$form $\alpha$ can always be written as
\begin{align}
    \alpha = k \eta \ ,
\end{align}
where $k \in \mathcal{F}_\mathcal{M}$. Hence, by imposing the condition $\nabla \alpha = 0$, we have
\begin{align}
\nabla k = 0 \implies k=\text{constant} \ .
\end{align}
Therefore, any other $d-$form with vanishing covariant derivative differs from $\eta$ only trivially by a constant factor, and $\eta$ describes a unique volume form for any given torsion spacetime.

The integrable density 
\begin{align}
   d\mathcal{V} = \mathfrak{G} = \sqrt{|\mathfrak{g}|(x)} d^dx\ ,
\end{align}
associated to $\eta$ as
\begin{align}
    \eta = \varepsilon \left[\mathfrak{G}\right]
\end{align}
is called {\it Levi-Civita volume element} or just {\it volume element}. The two separate notations are beneficial for intuitive understanding, $d\mathcal{V}$ for writing integrals and $\mathfrak{G}$ for differential operations. From the properties of the Levi-Civita symbol, we have that the property  $\nabla_a \eta = 0$ implies directly $\nabla_a \mathfrak{G} = 0$. In addition, it can be shown,  using the transformation law of the metric tensor, that $\mathfrak{G}$ transforms like a density as one would expect (see Appendix \ref{App:TensorDenDis}).

Using the volume element, the integral of a function $f\in\mathcal{F}_\mathcal{M}$ on the manifold now takes the well-known form, given as 
\begin{align} \label{IntegalFuncDef}
  \int_\mathcal{M} f d\mathcal{V} 
  = \int_\mathcal{M} f\mathfrak{G} = \int_\mathcal{M} f \sqrt{|\mathfrak{g}|(x)} d^dx \ .
\end{align} 
In addition, it can be proven that a function can be naturally mapped to an integrable density via $\mathfrak{G}$ as
\begin{align}
    a \mathfrak{G} = \mathfrak{a} \ . \label{DensityFunctionRel}
\end{align}
This result is formally derived in Appendix \ref{App:DensityComp}.

It should then be clear that, strictly speaking, the integral \eqref{IntegalFuncDef} of a function $f$ is directly related to the integral of its associated density  $\mathfrak{f}$, obtained by applying the formal mapping \eqref{DensityFunctionRel}. The existence of a formal mapping between functions and integrable densities plays a crucial role in the development of Tensor Distribution Theory, as it provides a direct relationship between distributions (which must be integrable) and the description of non-continuous functions on the manifold.

Finally, we look at the divergence theorem on the manifold. The divergence of a vector field $X^k$, denoted by $\div{X}$, is a function on the manifold defined via the Lie derivative with respect to the vector field $X^k$ (denoted by $\mathcal{L}_X$) of the volume form $\eta$:
\begin{align}\label{definitiondivergence}
    (\div{X}) \eta = \mathcal{L}_X \eta \ ,
\end{align}
where the Lie derivative of $\eta$ can be evaluated in terms of components or via Cartan's Lemma \citep{WaldBook}. However, for the purpose of further evaluating the divergence ${\div{X}}$ in component form, it is easier to utilize the relation between the covariant derivative and the Lie derivative operating on a $d-$form, which is given as (see Appendix \ref{App:LieCovDerivRel} for further details):
\begin{equation}\label{LieCovDerivRelationDform}\begin{aligned}
    \mathcal{L}_X \alpha - \nabla_X \alpha &= \mathbb{A}_X \alpha \ ,
    \\
    (\mathbb{A}_X \alpha)_{a_0 ... a_{d-1}} &= (\nabla_n X^n + T^k{}_{nk}X^n) \alpha_{a_0 ... a_{d-1}} \ ,
\end{aligned}\end{equation}
where $\alpha \in \Omega^d_\mathcal{M}$, the second equation is written in abstract indices and $T^a{}_{bc}$ is the torsion tensor defined in Section \ref{SubSec:ECTheory} [cf. \eqref{DefofTorsion}] \citep{NomizuInterscience}. Applying relation \eqref{LieCovDerivRelationDform} to $\eta$, keeping in mind the vanishing of the covariant derivative of the volume form, and using Equation \eqref{definitiondivergence}, we obtain (in abstract indices):
\begin{align}
\div{X} = \nabla_n X^n + T^{k}{}_{nk}X^n \ . \label{DivVctrFldDef}
\end{align}

Using the relation \eqref{DensityDformRel}, \eqref{LieCovDerivRelationDform} and the property that isomorphisms (like $\varepsilon$) are structure preserving, the Lie derivative of integrable densities can be evaluated as
\begin{equation}\begin{aligned}
    \mathcal{L}_X\mathfrak{a} 
    &= \nabla_X \mathfrak{a} + \mathfrak{a} (\nabla_k X^k + T^a{}_{ka}X^k) 
    \\ &= \nabla_k (X^k \mathfrak{a}) + T^a{}_{ka} \mathfrak{a} X^k 
    \\ &= \div (\mathfrak{a}X) \ ,
\end{aligned} \label{LieDerivDensity} \end{equation}
where $\nabla_X (\cdot)=X^k\nabla_k(\cdot)$ and we introduced the shorthand notation 
\begin{align}
    \div (\mathfrak{a}X) = \nabla_k (X^k \mathfrak{a}) + T^a{}_{ka} \mathfrak{a} X^k \ .
\end{align}

Using the above definitions, the divergence theorem for manifold $\mathcal{M}$ can be presented as \citep{NomizuInterscience}
\begin{equation} \label{DivergenceTheorem}
\begin{aligned}
    \int_\mathcal{M} \div(fX) d\mathcal{V}
        &= \int_\mathcal{M} \left[\div(fX \mathfrak{G}) - f\nabla_X \mathfrak{G}) \right]
        \\
        &=\int_\mathcal{M} \div(\mathfrak{f}X)= \int_\mathcal{M} \mathcal{L}_X \mathfrak{f}
        \\
        &= \int_\mathcal{M} \mathcal{L}_X (fd\mathcal{V})= \int_{\partial \mathcal{M}} \epsilon fX^k n_k d\sigma \ ,
\end{aligned} \end{equation}
where $f\in\mathcal{F}_\mathcal{M}$, $n_a$ is the (local) normal to the boundary $\partial \mathcal{M}$ of the manifold $\mathcal{M}$ and $\epsilon$ is defined as
\begin{align}
    \epsilon = 
    \begin{cases}
        +1 \ , \text{ for spacelike } n_a
        \\
        -1 \ , \text{ for timelike } n_a
    \end{cases}
    \ .
\end{align}
The boundary $\partial \mathcal{M}$ is described (locally) by the induced metric
\begin{align*}
    P_{ab} = g_{ab} - \epsilon n_a n_b \ ,
\end{align*} 
where, similarly to the case of the volume element $d\mathcal{V}\equiv\mathfrak{G}$, we have defined the volume element $d\sigma \equiv \mathfrak{P}$ as the square root of the modulus of the determinant of $P_{ab}$ in a given coordinate system.

\section{Tensor Distribution Theory }\label{Sec:TensorDistTheory}

We are now ready to develop a consistent description of tensorial objects on non-smooth manifolds. Such a \textit{Theory of Tensor Distributions} has been discussed in previous works in absence of torsion \citep{lichnerowicz,Taub,Bruhat,Senovilla2015,SenovillaMars,Marsden}. Here, we provide a brief yet rigorous treatment that reveals several salient points, which will be used later to formulate the covariant junction conditions.

Following the strategy given in Section \ref{Sec:LinearDistTheory} for Schwartz distributions, we shall begin our description of tensor distributions fundamentally as an extension of integrable densities. In this way, we ensure that distributions,  as functionals, are integrable objects. Furthermore, we aim to develop a differential algebra for distributions on the manifold, analogous to that in Linear Distribution Theory. This will allow us to extend the SLS theorem to distributions on the manifold.

The \textit{space of density distributions} $\mathbb{D}'_\mathcal{M}$ is defined as an extension of the space of integrable densities $\mathbb{D}_\mathcal{M}$, serving as the basis space. The \textit{density distribution} $\disd{a}\in\mathbb{D}'_M$ acts on the test functions $\varphi\in\mathcal{F}_\mathcal{M}$ as:
\begin{align}
    \mathbb{D}'_\mathcal{M} \ni \disd{a}: \varphi \mapsto  \dfuncal{a}{\varphi} \in \mathbb{R}  \ ,
\end{align}
where $\dfuncal{a}{\varphi}$ denotes the action of density distribution $\disd{a}$ on the test function $\varphi$. For the purpose of applications in most areas in Physics, we can limit ourselves to considering density distributions $\disd{a}$ for which there exists a representation given by a sequence $\mathfrak{a}_\epsilon\in\mathbb{D}_\mathcal{M}$ in the basis space of integrable densities\footnote{Distributions without a basis representation are useful primarily for developing broader algebras, such as Colombeau algebra.}, i.e., such that
\begin{align}
    \disd{a} &= \lim_{\epsilon \to 0} \mathfrak{a}_\epsilon \ ,
\end{align}
the action of the density distribution $\disd{a}$ on the test function $\varphi$ is evaluated as:
\begin{align}
    \dfuncal{a}{\varphi} &= \lim_{\epsilon \to 0} \int_\mathcal{M} \varphi(x) \mathfrak{a}_\epsilon(x) d^dx = \lim_{\epsilon \to 0} \int_\mathcal{M} \varphi \mathfrak{a}_\epsilon  \in \mathbb{R} \ . \label{DensityDistFundOp}
\end{align}
The integral \eqref{DensityDistFundOp} must satisfy the weak convergence condition [cf. \eqref{Distributionoperation}]. In practice, this means that the members of the test space and the basis space must also be smooth and satisfy the same boundary conditions as in Linear Distribution Theory, i.e., they belong to a subset of $\mathcal{F}_\mathcal{M}$ and $\mathbb{D}_\mathcal{M}$. In the following, for brevity, we will simply indicate the test space and basis space as  $\mathcal{F}_\mathcal{M}$ and $\mathbb{D}_\mathcal{M}$ respectively. Also notice that, differently from the case of Linear Distribution Theory, now the space of test functions and the space of basis functions are very different.

As in Linear Distribution Theory, we can now construct an algebra for density distributions. In particular, the addition of a density distribution and its multiplication with  a smooth function can be given as  ($\varphi, \Phi\in\mathcal{F}_\mathcal{M}$):
\begin{equation}\label{DensityDistVctrSpace}\begin{gathered}
    \mathbb{D}'_\mathcal{M} \ni \disd{c} = \disd{a} + \disd{b}: \varphi \mapsto \dfuncal{c}{\varphi} = \dfuncal{a}{\varphi} + \dfuncal{b}{\varphi}
    \ , \\
    \mathbb{D}'_\mathcal{M} \ni \disd{c}=\Phi\disd{a}: \funcal{\mathfrak{c}}{g} = 
    \langle \Phi\disd{a},g\rangle = \dfuncal{a}{\Phi g}
    \ ,
\end{gathered}\end{equation}
and are essentially defined by their counterpart on the sequence representation of the distribution in the basis space. Using these operations, we can construct a vector space of density distributions over the space of smooth functions $\mathcal{F}_\mathcal{M}$. 

Next, we can develop the differential algebra of the space of density distributions. However, this analysis is slightly more complicated for density distributions: on one hand, the partial derivative is coordinate-dependent and therefore ill-suited in a covariant setting; on the other hand, we have a few options for coordinate-independent (tensor) derivatives, such as the  Lie derivative and the covariant derivative. 

In solving this problem, our guiding principle will be to obtain a derivative operator that mimics the derivative in Linear Distribution Theory \eqref{DerivativeofLinearDistribution}. Such an operator is the Lie derivative. The reason is that, from Section \ref{Sec:LinearDistTheory}, we know that the derivative of a distribution ultimately requires the application of the divergence theorem, and we know, from Section \ref{sec:IntegrableDensities}, that the generalization of this theorem on a manifold is directly related to the Lie derivative. 

In fact, one can prove explicitly that the Lie derivative yields a derivative rule for density distributions, similar to that for Schwartz distributions [cf. \eqref{DerivativeofLinearDistribution}]. Given $\varphi\in\mathcal{F}_\mathcal{M}$ we have, using the definition of Lie derivative of a density,
\begin{align}
\funcal{\mathcal{L}_X \mathfrak{a}}{\varphi} 
    &= \lim_{\epsilon \to 0} \int_\mathcal{M} \varphi \mathcal{L}_X \mathfrak{a}_\epsilon
    \nonumber \\
    &= \lim_{\epsilon \to 0} \int_\mathcal{M} \left\{ \div(\varphi \mathfrak{a}_\epsilon X) - \mathfrak{a}_\epsilon X^a\nabla_a\varphi \right\}
    \nonumber \\
    &= - \lim_{\epsilon \to 0} \int_\mathcal{M} \mathfrak{a}_\epsilon \nabla_X \varphi
    \nonumber \ .
  \end{align}
Now, using $\nabla_X \varphi = \mathcal{L}_X \varphi$ for $\varphi\in\mathcal{F}_\mathcal{M}$, we have\footnote{The integral over boundary $\partial\mathcal{M}$ vanishes since the test functions satisfy similar boundary conditions as $\mathcal{D}(O)$ in Linear Distribution Theory.} 
 \begin{align}
\funcal{\mathcal{L}_X \mathfrak{a}}{\varphi}
    &= -\dfuncal{a}{\mathcal{L}_X\varphi} \ .
\label{DensityDistDeriv}
\end{align}
For this reason, we have
\begin{align}
    \mathcal{L}_X: \mathbb{D}'_\mathcal{M} \mapsto \mathbb{D}'_\mathcal{M}\ , \label{DensityDerivativeProp2} 
\end{align}
 and in what follows, the Lie derivative will be utilized as the \textit{derivative in the distributional sense} for the density distributions on the manifold.
 
It is also easy to verify that the Lie derivative satisfies the Leibniz rule for product with a test function $\varphi\in\mathcal{F}_\mathcal{M}$:
\begin{gather}
    \mathcal{L}_X (f \disd{a}) = (\mathcal{L}_X f) \disd{a} + f \mathcal{L}_X \disd{a} \ ,
    \label{DensityDerivativeProp1}
\end{gather}
and thus it matches the properties of the distributional derivative in Linear Distribution Theory \eqref{SchDistLeibniz}. Hence, using the Lie derivative, the addition $+$ and the tensor product $\otimes$, a differential algebra $\mathcal{A} = (+,\otimes,\mathcal{L}_X)$ for the space of density distributions $\mathbb{D}'_\mathcal{M}$ can be constructed, which is analogous to the one described in Section \ref{Sec:LinearDistTheory}.

Finally, using the above results and employing \eqref{LieDerivDensity}, \eqref{DensityDistVctrSpace} and \eqref{DensityDistDeriv}, we can derive the action of the covariant derivative on the density distribution as:
\begin{equation}\begin{aligned}
    \funcal{\nabla_X \mathfrak{a}}{f} 
    &= \langle \dis{\mathcal{L}_X \mathfrak{a}} - \disd{a}\div X,f\rangle
    \\
    &= -\dfuncal{a}{\mathcal{L}_X f + f\div X}
    \\
    &= -\dfuncal{a}{\div(fX)}
    \ .
\end{aligned}\end{equation}

The existence of a consistent differential algebra for density distributions allows us to extend the SLS theorem also to this type of distribution. The SLS theorem can be used to derive a pointwise representation of a distribution via the derivative of continuous $\mathcal{C}^0$ objects, and, as we will see in Section \ref{Sec:JunctionsAndDistributionsFull}, it will be extremely important for our purposes. 

As in the case of Schwartz distributions, and for the same reasons, the product of tensor distributions is, in general, not defined. Furthermore, the Schwartz Impossibility Result also holds in this context. As a consequence, in performing derivatives of tensor distributions, the Leibniz rule does not apply \citep{Steinbauer2006,schwartz1954,Colombeau}. In fact, the product of two density distributions introduces an additional problem: the resulting object cannot be integrated in a coordinate-independent manner, as it describes a density of weight 2. See Appendix \ref{App:TensorDenDis} for additional details.

In typical applications of distribution theory on a manifold, it is often preferable to describe density distributions in terms of their function counterparts given by the mapping \eqref{DensityFunctionRel}. The \textit{function distribution} $\dis{f} \in \mathcal{F}'_\mathcal{M}$ associated to a density distribution $\disd{f}$, is given as
\begin{align}
\dis{f} &= \frac{\disd{f}}{\mathfrak{G}} \ , \label{DensityFunctionRelDist}
\end{align}
and operates on the functions on the manifold as
\begin{align} \label{FunctionDistFundOp}
\underline{f}:& \varphi \mapsto \funcal{f}{\varphi} = \lim_{\epsilon \to 0} \int_\mathcal{M} f_\epsilon \varphi d\mathcal{V} \in \mathbb{R} \ ,
\end{align}
where $\varphi \in \mathcal{F}_M$ and $f_\epsilon$ is the sequence function associated with sequence density $\mathfrak{f}_\epsilon$ [see \eqref{DensityDistFundOp}].

Let us now evaluate the derivative of a function distribution. To this end, we can employ the basis function representation and the integral action in \eqref{FunctionDistFundOp}:
\begin{equation}\begin{aligned}\label{PreDerFunDistr}
    \langle \mathcal{L}_X \dis{f}, \varphi \rangle &=
    \lim_{\epsilon \to 0} \int_\mathcal{M} (\mathcal{L}_X f_\epsilon) \varphi d\mathcal{V}
    \\ &=
    \lim_{\epsilon \to 0} \int_\mathcal{M} \left\{ \mathcal{L}_X (f_\epsilon \varphi d\mathcal{V}) - f_\epsilon \mathcal{L}_X (\varphi d\mathcal{V}) \right\} \ ,
\end{aligned}\end{equation}
where the first term can be eliminated using the divergence theorem \eqref{DivergenceTheorem} and the second term is resolved using \eqref{LieDerivDensity} as:
\begin{align}
    \mathcal{L}_X (\varphi d\mathcal{V}) 
    &= \nabla_X (\varphi d\mathcal{V}) + \varphi d\mathcal{V} \div{X}=
  \nonumber \\
    &=d\mathcal{V} \,\div{(\varphi X)}   \ ,
\end{align}
which, substituted in \eqref{PreDerFunDistr}, leads to
\begin{align}
    \langle \mathcal{L}_X \dis{f}, \varphi \rangle &= - \funcal{f}{\div{(\varphi X)}} \ .
\end{align}
It can be directly verified that the covariant derivative of a function distribution leads to the same result as its Lie derivative:
\begin{align}
   \funcal{\nabla_X f}{\varphi} = - \funcal{f}{\div{(\varphi X)}} =  \funcal{\mathcal{L}_X f}{\varphi} \ , \label{nablaXoffunctiondis}
\end{align}
where $\dis{\mathcal{L}_X f},\dis{\nabla_X f} \in \mathcal{F}'_\mathcal{M}$ because of  \eqref{DensityDerivativeProp2}.

Since the Lie derivative and covariant derivative of a function distribution both evaluate to the divergence of the vector field $X^k$, it is clear that without introducing densities distributions it would not have been obvious to define a proper derivative in the distributional sense, as in Eq. \eqref{DensityDistDeriv}, nor construct a differential algebra for function distributions. In this sense, this result justifies the necessity to introduce density distributions even if we will mainly deal with function distributions.
 
Seen their connection with density distributions, it is not difficult to prove that the product of two function distributions is ill-defined in general, and the Schwartz Impossibility Result prevents the applicability of the Leibniz rule in derivation.  In addition, as in the case of Schwartz distributions,  the derivative of a function distribution is itself a distribution.

For later applications, it is useful to introduce the co-vector distribution induced by the covariant derivative $\dis{\nabla_X f}$. We start with the covariant derivative of the function distributions  $\dis{f}$ along $X^k$, and we have 
\begin{align}
    \langle \nabla_X\dis{f},\varphi \rangle
    = \langle X^a \nabla_a \dis{f},\varphi\rangle 
    = \langle \nabla_a \dis{f},\varphi X^a\rangle \ ,
\end{align}
and then using \eqref{nablaXoffunctiondis}, we obtain
\begin{align}
     \langle\nabla_a\dis{f},\varphi X^a\rangle = - \langle\dis{f},\div(\varphi X)\rangle \ . \label{CovDevDist}
\end{align}
Remembering property \eqref{DensityDerivativeProp2}, we can conclude that in the above operation, the covariant derivative $\nabla_a$ maps from a function distribution $\dis{f}$ to a co-vector distribution. Hence, one can naturally define an induced co-vector distribution 
\begin{equation}
\nabla_a \dis{f} \equiv \dis{\nabla_a f} \in \mathcal{T}'{}^0_1(\mathcal{M}) \ ,
\end{equation}
where $\mathcal{T}'{}^0_1(\mathcal{M})$ is the space of co-vector distributions and the  following property 
\begin{align}
     \funcal{\nabla_af}{\varphi X^a} = - \funcal{f}{\div(\varphi X)} \label{CovDevDistCoVec}
\end{align}
holds. 

The definition of $ \dis{\nabla_a f} $ indicates that explicit tensorial distributions can and should be formally introduced. This extension simply requires using the concept of tensor densities, which naturally allows for the description of tensor distributions. More details on this topic can be found in Appendix \ref{App:TensorDenDis}. For our purposes, we only need to describe function distributions, and therefore we will focus exclusively on these objects.

An important final detail that will be fundamental for our purposes is that a consistent description of the function distribution requires continuity of the metric tensor. From the relation \eqref{DensityFunctionRelDist}, it is evident that if the metric tensor is not at least continuous $\mathcal{C}^0$ on the compact support of the density distribution, then the function distribution is ill-defined and requires ad hoc operations for its manipulation. In the next section, we will present several specific fundamental distributions whose compact support will be determined by the underlying geometry of the manifold.

\section{Glued Spacetimes, Interfaces and Tensor Distributions}\label{Sec:JunctionsAndDistributionsFull}

As already mentioned, in many instances, the analytical descriptions of physically relevant spacetimes are more conveniently obtained by solving the field equations of gravity for two different \textit{bulk spacetimes} $(\mathcal{M}^\pm,g^\pm)$ (with different metrics,  matter content, symmetries, etc.), and later joining them at a hypersurface called an \textit{interface}. In the \textit{glued spacetime} $(\mathcal{M},g)$ obtained in this way, some quantities may differ across the interface $\mathcal{I}$. Therefore, generically, an object on the glued spacetime $\mathcal{M}$ would be discontinuous and must be described as a distribution. 

It is then clear why, when treating glued spacetimes, one must employ Tensor Distribution Theory. In doing so, one must impose certain geometrical and differentiability conditions on bulk spacetimes and the glued manifold to ensure a consistent mathematical treatment. Collectively, we call these conditions \textit{Fundamental Junction Conditions}.

A distributional object $\dis{O}$ on the glued spacetime $\mathcal{M}$ must correspond to its respective value $O^\pm$ within each part $\mathcal{M}^\pm$. For example, the metric $\dis{g_{ab}}$ of the glued spacetime $\mathcal{M}$ must correspond to the metric $g_{ab}^\pm$ in each of the parts $\mathcal{M}^\pm$. The local representation of the distributions encountered in the previous sections is crucial to ensure that this requirement is met. 

Some distributions may also present a \textit{singular part}. This occurs most commonly when considering derivatives of discontinuous tensor distributions. As we will see, these singular parts are associated with a generalization of the Dirac delta distribution (centered at the interface).  In general, the fact that tensor distributions on the glued spacetime $\mathcal{M}$ present singular quantities at the interface is referred to as the fact that the glued spacetime presents a \textit{thin shell} (see e.g. \citep{RelToolkitPoisson}). While the term thin shell is motivated by the interpretation of the singular quantities as the existence of a thin matter distribution at the interface, in general, other quantities might also behave in the same way. For example, as we shall see, the Weyl tensor might possess non-vanishing singularities both in General Relativity and Torsion spacetimes. In what follows, we will evaluate how discontinuities give rise to singular quantities on the interface, and we show that working with tensor distributions allows us to correctly incorporate the effects of non-vanishing torsion in the formation of the thin shell. 

We will also derive certain conditions which, if satisfied, ensure that there are no singular quantities at the interface and the thin shell disappears\footnote{Note that, for some cases, the conditions for smooth matching cannot be satisfied and the thin shell cannot be eliminated. See Section VI A in Reference \citep{RosaCarloni}.}. We call these conditions \textit{smooth matching conditions}. In the case that smooth matching conditions are satisfied, we say that two spacetimes $\mathcal{M}^\pm$ are \textit{smoothly matched}.

Also note that in literature, one typically refers to the fundamental junction conditions and smooth matching conditions collectively as {\it junction conditions} \citep{Huber}. We differentiate between fundamental junction conditions and smooth matching conditions as follows: given two spacetimes to be joined, firstly, the fundamental junction conditions must be satisfied to form the interface and the glued spacetime. Then after evaluating the singular quantities at the interface, smooth matching conditions may be imposed in order to eliminate the thin shell.

We will also find it useful to introduce the following notations to facilitate the study of the interface and the glued spacetime, composed of two separate spacetimes joined together (indices are omitted for brevity):
\begin{align}
    \ifavg{A} 
    &= \frac{1}{2} 
        \left( 
        \lim_{x^+ \xrightarrow[\mathcal{M}^+]{} \mathcal{I}} 
        +
         \lim_{x^- \xrightarrow[\mathcal{M}^-]{} \mathcal{I}}
        \right) \underline{A}
    \nonumber \\
    &= \frac{1}{2} \left( A^+ \middle|_\mathcal{I} + A^- \middle|_\mathcal{I} \right)
    \ ,
    \label{InterfaceAverage} \\
    \jump{A} 
    &= 
    \left(
    \lim_{x^+ \xrightarrow[\mathcal{M}^+]{} \mathcal{I}}
    -
    \lim_{x^- \xrightarrow[\mathcal{M}^-]{} \mathcal{I}}
    \right) \underline{A}
    = \left( A^+ \middle|_\mathcal{I} - A^- \middle|_\mathcal{I} \right)
    \ , 
\end{align}
where $\dis{A}\in \mathcal{T}'^r_s(\mathcal{M})$ is an order $(r,s)$ tensor distribution, $A^\pm \in \mathcal{T}^r_s(\mathcal{M^\pm})$ are the order $(r,s)$ tensors associated to the local representation of $\dis{A}$, and  $x^\pm \in \mathcal{M}^\pm$ is a point on the manifold. We also introduced the shorthand notation
\begin{align}
    \lim_{x^\pm \xrightarrow[\mathcal{M}^\pm]{} \mathcal{I}} \dis{A} = A^\pm \big|_\mathcal{I}  \ ,
\end{align}
i.e., $A^\pm \big|_\mathcal{I}$ corresponds to the value of tensor $A^\pm$ in the vicinity of the interface $\mathcal{I}$. The structure of the manifold provided later will clarify this operation further. We then call $\jump{A}$ a {\it jump} in the tensor distribution $\underline{A}$ across the interface. A useful relation for the jump of the product of two tensor distributions $\dis{A}$, $\dis{B}$ is
\begin{equation}\begin{aligned}
    \jump{AB} 
        &= (A^+ B^+) |_\mathcal{I} - (A^- B^-) |_\mathcal{I}
    \\
        &= \ifavg{A}\jump{B} + \ifavg{B}\jump{A}
    \ . \label{jumpofproductdist}
\end{aligned}\end{equation}

In the following, we shall assume that both $\mathcal{M}^{-}$ and $\mathcal{M}^{+}$ are $4-$dimensional, possess non-vanishing torsion, and are metric compatible ($\nabla g=0$). We shall focus on describing the glued spacetime manifold $\mathcal{M}$, obtained by joining two spacetimes $\mathcal{M}^-$ and $\mathcal{M}^+$ along the interface  $\mathcal{I}$. We then formally define certain fundamental distributions on the glued spacetime and introduce some additional notations used throughout this paper. Finally, we derive the covariant derivative of a generic tensor distribution, a key result for practical applications and the study of glued spacetimes and interfaces.

From this section onward, we use abstract indices to denote tensorial quantities, unless specified otherwise.

\subsection{Fundamental Junction Conditions}\label{SubSec:JunctionsFundamentals}

Let us now give the fundamental junction conditions which must be satisfied to ensure that two given spacetimes can be successfully joined at a well-described $3-$dimensional interface\footnote{A discussion on the $3-$dimensional interface is sufficient for most applications, including junctions at the surface of black holes and stars. A lower dimensional surface (for a light-like normal) requires a different geometrical set up which is not reported here.}. We shall also discuss the physical and mathematical implications of these conditions.

The \textbf{first fundamental junction condition} requires that for each spacetime $\mathcal{M}^\pm$, at the would-be-interface and in its neighborhood \citep{Clarke_1987}, one can describe a normal form $n_a^\pm$ orthogonal to which a $3-$hypersurface $\mathcal{I}^\pm$ can be described. Such a decomposition requires that the following so-called \textit{hypersurface orthogonality condition} is satisfied at each hypersurface $\mathcal{I}^\pm$ for each normal $n_a^\pm$\footnote{This condition is better given in terms of one-forms as ${n} \wedge D {n} = 0$ where $D()$ is the exterior derivative.} \citep{WaldBook,LuzHypersurface,UjjwalSante}
\begin{align}
    \bigg(2n_{[a} \nabla_b n_{c]} + n_{[a} T^{k}_{\ bc]} n_k\bigg)\bigg|_\mathcal{I} = 0
    \ ,
    \label{HypersurfaceOrthogonalityCondition}
\end{align}
where we have omitted $\pm$ for clarity. For each spacetime $\mathcal{M}^\pm$, we also define the projection tensor $P_{ab}^\pm$ as
\begin{align}
    P^\pm_{ab} &= g^\pm_{ab} - \epsilon n^\pm_a n^\pm_b \ ,
\end{align}
where $\epsilon$ is given as (we omit $\pm$ for brevity):
\begin{align}
    \epsilon &= n^a n_a =
                    \begin{cases}
                        1, \text{ for spacelike normal} \\
                        -1, \text{ for timelike normal}
                    \end{cases}
    \ , \label{defofepsilon}
\end{align}
and describes the normalization of the normal form. Just like the normal form $n^\pm_a$, the projection tensor $P^\pm_{ab}$ is also described on the hypersurface $\mathcal{I}^\pm$ and in its neighborhood. These hypersurfaces $\mathcal{I}^\pm$ are to be identified to form an interface $\mathcal{I}$ in $\mathcal{M}$. 

The \textbf{second fundamental junction condition} requires that the identification of the hypersurfaces $\mathcal{I}^+$ and $\mathcal{I}^-$ to form the interface $\mathcal{I}$ must be given by a diffeomorphic mapping $\mathcal{C}^1$ between $\mathcal{I}^\pm$. The structure of the glued manifold is given as \cite{Clarke_1987}:
\begin{align}
    \mathcal{M} = (\mathcal{M}^- - \mathcal{I}^- ) \cup \mathcal{I} \cup (\mathcal{M}^+ - \mathcal{I}^+ ) \label{StructureManifold}
    \ .
\end{align}
The second fundamental junction condition constrains the differentiability properties of tensorial objects on the manifold. It implies that any vector field, which can be described by a generating curve/flow on the manifold, must be at least continuous ($\mathcal{C}^0$) across the interface. As a consequence:
\begin{itemize}
    \item the normal $n_a$ is $\mathcal{C}^0$ across the interface. That is, $n_a^{+} |_q = n_a^{-} |_q \ , \ \forall q \in \mathcal{I}$. \label{jump-n-zero}
    \item the projection tensor $P_{ab}$, which describes the geometry of $3-$hypersurface $\mathcal{I}$ (spanned by three vector fields), is $\mathcal{C}^0$ across the interface. Therefore, $P_{ab}^{+} |_q = P_{ab}^{-} |_q \ , \ \forall q \in \mathcal{I}$. \label{jump-P-zero} 
\end{itemize}
From the previous two results, it is evident that the metric tensor $g_{ab}$ is $\mathcal{C}^0$ across the interface. Therefore, $g_{ab}^{+} |_q = g_{ab}^{-} |_q \ , \ \forall q \in \mathcal{I}$.

Furthermore, the continuity of the projection tensor $P_{ab}$ ensures that integrals over the interface are uniquely defined.  This is an important result, as the divergence theorem relies on these integrals and plays a key role in evaluating the derivatives of distributions. Hence, a unique integral on the interface allows us to consistently define glued spacetimes with distributions.

In General Relativity, the Levi-Civita connections are discontinuous across the interface, since they involve (partial) derivatives of the metric, but not singular. Therefore, the second fundamental junction condition ensures that while the Riemann curvature tensor (distribution) may be singular (due to the derivative of the connections), the quadratic terms in the Dirac delta distribution do not appear in the Riemann curvature tensor (distribution).

However, for torsion spacetimes, the Christoffel symbols $\Gamma_a{}^b{}_c$ will be the sum of the Levi-Civita connection related to the metric and the contorsion tensor related to torsion. The fundamental junction conditions described above do not constraint the singular part of torsion, which, if non-zero, would introduce quadratic Dirac delta terms in Riemann curvature tensor distribution on the glued spacetime through the quadratic terms in Christoffel symbols. Consequently, in torsional spacetimes, to apply the theory of tensor distributions, one needs to impose a \textbf{third fundamental junction condition} that the torsion tensor has no singular part at the interface.

The three fundamental junction conditions are, therefore, necessary to prevent the appearance of ill-defined objects in tensor distribution theory. If they are satisfied, we can safely treat the glued spacetime within tensor distribution theory. Should these conditions be relaxed, one should employ more complex formalisms such as the Colombeau algebra (and its extension to curved manifolds).

Finally, if we assume that there are no junctions within the bulk manifolds $\mathcal{M}^\pm$, the fundamental junction conditions imply that the bulk manifolds are at least $\mathcal{C}^3$ differentiable, the metric $g_{ab}^\pm$ is $\mathcal{C}^2$ differentiable, as described in Ref. \citep{Clarke_1987} for torsion-free spacetimes, and additionally for torsion spacetimes, the torsion tensor must be at least $\mathcal{C}^1$. Such a regularity property is sufficient to omit the need for distribution theory, since the Riemann curvature tensor is continuous, and no singular terms appear in the Bianchi identities, which describe the local conservation laws.

We can now rewrite the second and third fundamental junction conditions in a different form, which will be more useful when we give the junction conditions in the covariant formalism. More specifically, the  second fundamental junction condition can be written as:
\begin{equation}
\jump{n_a} = \jump{P_{ab}} = \jump{g_{ab}} = 0 \ , \label{jump-npg-zero}
\end{equation}
 while the third reads
\begin{equation}
\intfc{T}_{abc}=0 \ , \label{Singular-Tor-Zero}
\end{equation}
where $\intfc{T}_{abc}$ denotes the singular part of torsion.
Note that, due to \eqref{jump-npg-zero}, we also obtain
\begin{align}
    \ifavg{n_a} = n^+_a |_\mathcal{I} =  n^-_a |_\mathcal{I} \ . \label{normalabuseofnotation}
\end{align}
As a consequence, for brevity, we simply represent normal as $n_a$ and do not distinguish between its values in different regions of the manifold \eqref{StructureManifold} and also write ${\ifavg{n_a} = n_a}$. A similar choice will be made for the normal vector $n^a = g^{ab}n_b$  with the additional assumption that it points from $\mathcal{M}^-$ towards $\mathcal{M}^+$ at the interface $\mathcal{I}$.

Some remarks are now in order, which will be useful later. The condition $\intfc{T}_{abc} = 0$ acquires further physical meaning once a theory of gravity is chosen, as only then does torsion couple to the matter content of the spacetime. The condition $\intfc{T}_{abc} = 0$ demands a certain differentiability property from the torsion tensor of the manifold $\mathcal{M}$, that is, torsion can be at most discontinous but non-singular\footnote{In contrast, this condition does not restrict the $3-$torsion (even at the interface), which is indeed why the $3-$divergence described in equation \eqref{3-divergenceformula} includes a $3-$torsion term. Indeed, this is true in general,  i.e., given a quantity, its projected $3-$quantity and its singular part at the interface $\mathcal{I}$ are not related.}.

\subsection{Fundamental Distributions, Tensor Distributions and Their Covariant Derivative}

As discussed, the bulk manifolds $\mathcal{M}^\pm$ are considered to be smooth (specifically, they must be at least $\mathcal{C}^3$). Therefore, a generic quantity $A^\pm$ on the bulk manifolds belongs to the space of smooth  $(r,s)$ tensors. Upon creating the glued manifold $\mathcal{M}$, the global description of the quantity $A$ requires the introduction of the tensor distribution $\dis{A}\in \mathcal{T}'^r_s(\mathcal{M})$ because the interface $\mathcal{I}$ is only $\mathcal{C}^1$ differentiable.

On the glued spacetimes, tensor distributions can be written in a specific and advantageous way that highlights the structure  \eqref{StructureManifold} of the glued spacetime and makes it evident that, on the bulk spacetimes, the distribution reduces to its local representation. This is achieved by using the properties of the functionals to express tensor distributions on $\mathcal{M}$ as combinations of the so-called fundamental distributions. These fundamental distributions carry information about the manifold's structure and its low-regularity region, since their compact support is defined in accordance with it.

For brevity, here we limit ourselves to a brief description of these fundamental distributions. A rigorous, step-by-step process for developing such distributions is provided in Appendix \ref{App:DistributionsDerivations}.

The first fundamental distribution is given by the step distribution $\stepU$ (associated with a step density distribution), which is defined as 
\begin{align}
   \mathcal{F}'_\mathcal{M} \ni \stepU : \varphi \mapsto \funcal{\step}{\varphi} = \int_\mathcal{M} \varphi\stepU d\mathcal{V} = \int_{\mathcal{M}^+} \varphi d\mathcal{V} \ . \label{StepDist}
\end{align}
However, the above definition of the  step distributions is not enough. As mentioned in the beginning of the section, we must ensure that the tensor distributions $\dis{A}\in\mathcal{T}'^r_s(\mathcal{M})$ must correspond to its respective value $A^\pm\in\mathcal{T}^r_s(\mathcal{M^\pm})$ in each part $\mathcal{M}^\pm$. To ensure this, we require a local representation of the step distribution $\stepU$. As we have seen in Section \ref{Sec:TensorDistTheory}, such a representation can be determined by extending the SLS theorem to tensor distribution theory. In this way, the analogous of \eqref{linsteppointwise} can be given as\footnote{Note that we use the same notation $\stepU$ for the function distribution in \eqref{StepDist} and its local representation in \eqref{StepDistPointwise}.}:
\begin{align}
\stepU = \begin{cases}
    0 \ , \quad \text{for } x\in\mathcal{M}^--\mathcal{I}
    \\
    \frac{1}{2} \ ,  \quad \text{for } x\in\mathcal{I}
    \\
    1 \ ,  \quad \text{for } x\in\mathcal{M}^+-\mathcal{I}
\end{cases}  
\ , \label{StepDistPointwise}
\end{align}
where we have chosen the value at the interface $\mathcal{I}$ to complete the definition. 

We then define Dirac delta distribution $\diracU$ with its compact support on the interface $\mathcal{I}$ as
\begin{align}
\mathcal{F}'_\mathcal{M} \ni \diracU : \varphi \mapsto \funcal{\dirac}{\varphi} = \int_\mathcal{I} \varphi d\sigma \ . \label{DiracDist}
\end{align}
The step and Dirac delta distributions are related via a covariant derivative as
\begin{align}
    \underline{\nabla_a\step} = \epsilon n_a \diracU \ . \label{CovDerofStep}
\end{align}

With the fundamental distributions ${\{ \stepU,\diracU \}}$ above and the local representation of the step distribution \eqref{StepDistPointwise} at hand, we can now represent most (but not all) tensor distributions on $\mathcal{M}$  as a combination of the two local representations and its singular part. For example, in the simple case of a $\dis{A}\in\mathcal{T}'^r_s(\mathcal{M})$, we could have (indices are omitted): 
\begin{align}
    \dis{A} &= A^+ \stepU + A^- (1 - \stepU) + \intfc{A} \diracU \ , \label{GeneralTensorDist}
\end{align}
where we have indicated the singular part of $\dis{A} $ as $\intfc{A}$. Using this representation, operations on distributions can be performed using standard algebra. Naturally, in the case of product of distributions, one will have to use the equivalent of the ad-hoc operations already defined in \eqref{LinAdHocRules}:
\begin{align}
    \stepU \ \stepU &= \stepU^2 = \stepU \ , \label{stepsquareadhoc}
    \\ 
    \stepU \ \diracU &= \frac{1}{2}\diracU \ . \label{stepdiracadhoc}
\end{align}
Notice that the chosen value of local representation of step distribution $\stepU$ at the interface $\mathcal{I}$ in Equation \eqref{StepDistPointwise} brings ``regularity'' of both the left-hand and right-hand derivative of $\stepU$ being the same ($\lim_{\epsilon\rightarrow 0} \frac{1/2}{\epsilon}$). Also, this chosen value makes one of the ad-hoc operations \eqref{stepdiracadhoc} \textit{appear} to be ``natural'' as 
\begin{align}
    \stepU \ \diracU &= \stepU|_\mathcal{I} \diracU = \frac{1}{2}\diracU \ . \label{adhocNatural}
\end{align}
This result simplifies its application by making it more intuitive.

On $\mathcal{M}$, derivatives of distributions may also play an important role, and if these distributions have a singular part, we will also need to describe the distribution associated with the covariant derivative of the Dirac delta distribution $\diracU$.  For example, the Riemann curvature tensor can have a singular term in general. Since the Bianchi identity involves the covariant derivative of the Riemann tensor, we also need to describe the distribution associated with the covariant derivative of this tensor. 

The covariant derivative of $\diracU$ is given by (see Appendix \ref{App:DistributionsDerivations} for details)
\begin{align}
    \nabla_a \diracU = \epsilon n_a \DoublayerU + \epsilon n^l \bigg( \ifavg{\nabla_l n_a} + n^k \ifavg{T_{kla}} \bigg) \diracU \ , \label{CovDerofDirac}
\end{align}
where we defined the third fundamental distribution, the so-called \textit{double layer distribution} $\DoublayerU$, as:
\begin{align}
    \mathcal{F}'_\mathcal{M} \ni \DoublayerU : \varphi \mapsto \funcal{\Doublayer}{\varphi} = - \int_\mathcal{I} d\sigma \div(\varphi n) \ , \label{defdoublelayer}
\end{align}
where $\varphi \in \mathcal{F}_M$. Notice that, in the above formula, we have:
\begin{align}
\div(\varphi n) = \nabla_a(\varphi n^a) + \varphi n^a  T^k{}_{ak} \ ,
\end{align}
which does not represent the $3-$divergence term. We can further improve the readability of the above equation by noticing that the ad-hoc operation \eqref{stepdiracadhoc} allows us to write, as detailed in Appendix \ref{App:DistributionsDerivations},
\begin{align} \label{DoubleLayer}
    \nabla_a \diracU = \epsilon n_a \DoublayerU + \epsilon n^l ( \dis{\nabla_l n_a} + n^k \dis{T_{kla}} ) \diracU \ . 
\end{align}
The above definition of the double-layer distribution $\DoublayerU$ differs from that given in \citep{Senovilla2015}, in which these distributions were first employed in the context of gluing spacetimes. In particular, the double-layer distribution in  \citep{Senovilla2015} is a co-vector distribution, whereas in our case, we have a function distribution, and in \citep{Senovilla2015}, torsion was set to zero. However, it is straightforward to verify that our definition of the double-layer distribution $\DoublayerU$ is equivalent to that in \citep{Senovilla2015}, as shown in Appendix \ref{App:DistributionsDerivations}.

Including also the double-layer terms, we can write an extension of equation \eqref{GeneralTensorDist}. For a generic tensor distribution $\dis{A}\in\mathcal{T}'^r_s(\mathcal{M})$, we have:
\begin{align}\label{GeneralTensorDist2}
    \dis{A} &= A^+ \stepU + A^- (1 - \stepU) + \intfc{A} \diracU + \dlayer{A}\DoublayerU \ ,
\end{align}
where $\dlayer{A}$ notation is used to indicate the double layer part of $A$.

The representations \eqref{GeneralTensorDist} (and \eqref{GeneralTensorDist2}) are particularly useful as one can use them to treat distribution in a fashion that is very similar to the standard differential algebra for tensors, because each term is a product of a sufficiently differentiable part and a fundamental distribution. To see this explicitly, we can evaluate the covariant derivative of a generic tensor distribution $\dis{A}$ given in Equation \eqref{GeneralTensorDist}. The first term on the RHS is evaluated as 
\begin{align}
    \nabla(A^+\stepU) = \stepU\nabla(A^+) + A^+\nabla\stepU \ ,
\end{align} 
since the Leibniz rule is applicable for a product of distribution $\stepU$ and a smooth tensor field $A^+$ [cf. \eqref{DensityDerivativeProp1}]. We also highlight the obvious equivalence of $\nabla(A^+)=(\nabla A)^+$ since the covariant derivative of $A^+$ would be taken in $\mathcal{M}^+$. We shall similarly evaluate the covariant derivative of the each term in Equation \eqref{GeneralTensorDist}. Further, equations \eqref{CovDerofStep} and \eqref{DoubleLayer} are employed to resolve derivative of fundamental distributions.

As an example, with explicit abstract indices, let us look at a tensor distribution $\dis{A}^a{}_b$ of rank-$(1,1)$ given by
\begin{align}
    \dis{A}^a{}_b &= (A^+)^a{}_b \stepU + (A^-)^a{}_b (1 - \stepU) + \intfc{A}^a{}_b \diracU \ .
\end{align}
Its covariant derivative $\nabla_a \dis{A}^b{}_c$ is obtained using Equations \eqref{CovDerofStep} and \eqref{DoubleLayer}, as follows:
\begin{equation}
\begin{aligned}
    \dis{\nabla_a A^b{}_c}  &= 
        (\nabla_a A^b{}_c)^+ \stepU + (\nabla_a A^b{}_c)^- (1-\stepU) + \epsilon n_a \jump{A^b{}_c} \diracU
        \\ & \qquad
        + (\nabla_a \intfc{A}^b{}_c ) \diracU + \epsilon n^l \intfc{A}^b{}_c \left( \dis{\nabla_l n_a} + n^k \dis{T_{kla}} \right) \diracU
        \\ & \qquad
        + \epsilon n_a \intfc{A}^b{}_c \DoublayerU \ . 
\end{aligned}
\label{covariantderivativeofdistribution}
\end{equation}
Notice that the only contribution to the singular part of $\dis{\nabla_a A^b{}_c}$, under the assumption $\intfc{A}^b{}_c=0$, comes from the jump $\jump{A^b{}_c}$.

\section{Torsion LRS Spacetimes with Weyssenhoff Fluid}\label{Sec:TLRS-spacetimes}

In the previous sections, we have described the theory of tensor distributions. Such a formalism is necessary to study non-smooth manifolds, such as the glued spacetime. In the remainder of this work, we apply these tools to explore junction conditions in the context of the Einstein-Cartan-Sciama-Kibble (ECSK) Theory of Gravity \citep{Hehl} sourced by a Weyssenhoff Fluid \citep{WeyssenhoffRaabe,Korotky}.

\subsection{Torsion Spacetimes and Einstein-Cartan-Sciama-Kibble Theory of Gravity}\label{SubSec:ECTheory}

Given a $4-$dimensional pseudo-Riemannian manifold $(g_{ab}, \mathcal{M})$ equipped with a covariant derivative ($\nabla$), which is metric compatible ($\nabla g = 0$), the torsion tensor $T^k{}_{ab}$ and the Riemann tensor $R_{abcd}$ are described as
\begin{gather}
    \nabla_a\nabla_b \psi - \nabla_b \nabla_a \psi = -T^k{}_{ab} \nabla_k \psi \ , \label{DefofTorsion}
    \\
    \nabla_a \nabla_b X_c - \nabla_b \nabla_a X_c + T^k{}_{ab} \nabla_k X_c = -R_{ab}{}^d{}_c X_d \ , \label{DefofReimann}
\end{gather}
where $\psi$ is a generic function on the manifold and $X^a$ is a generic vector field on the manifold \citep{UjjwalSante,Hehl,HawkingEllisBook,WaldBook}. We can further decompose the Riemann tensor into the Ricci tensor $R_{ab}$ and the Weyl tensor $C_{abcd}$, which are defined as
\begin{equation}
\begin{aligned}
    R_{ab} &= g^{mn}R_{manb} \ , \qquad R = g^{ab} R_{ab} \ , 
    \\
    C_{abcd} 
    &=  R_{abcd} - \frac{1}{2}\left( R_{ac}g_{bd} + R_{bd}g_{ac} - R_{ad}g_{bc} - R_{bc}g_{ad} \right) 
    \\ & \quad
    + \frac{R}{6}\left( g_{ac}g_{bd} - g_{ad}g_{bc} \right)
     \ , \label{DefWeylTensor}
\end{aligned}
\end{equation}
where we also defined the Ricci Scalar $R$.

The Riemann tensor possesses following symmetries:
\begin{align}
    R_{abcd} &= -R_{bacd} = -R_{abdc} \ .
\end{align}
The Weyl tensor has the same symmetries as the Riemann tensor and, additionally, it is completely traceless by definition.

The \textit{algebraic} and the \textit{differential} Bianchi identities are given as, respectively,
\begin{align}
    R_{[ab}{}^{n}{}_{c]} &= \nabla_{[a}T^n{}_{bc]} - T^k{}_{[ab} T^n{}_{c]k}
    \ , \label{BianchiIdType1}
    \\
    \nabla_{[a}R_{bc]}{}^{k}{}_{l} &= T^n{}_{[ab}R_{c]n}{}^{k}{}_{l}
    \ . \label{BianchiIdType2}
\end{align}
The differential Bianchi identity \eqref{BianchiIdType2} can be re-written in terms of Weyl tensor using \eqref{DefWeylTensor} and further, upon contracting the indices $a$ and $l$, we obtain the generalization of Trumper-Kundt formula \citep{TrumperKundt} 
\begin{align}
\nabla^d C_{abcd} 
    &= - \nabla_{[a} R_{b]c} - \frac{1}{6} g_{c[a} \nabla_{b]} R - \frac{1}{2} T_{nmk} R_{[a}{}^{nmk} g_{b]c}
    \nonumber \\ & \quad
    - T_{nm[a} \left( g_{b]c} R^{nm} + 2 R_{b]}{}^{nm}{}_{c} \right) - T^n{}_{ab} R_{nc}
    \label{oncecontractedBianchiId} \ ,
\end{align}
which will be referred to as the \textit{Weyl equation} in this paper.

The ECSK theory aims to incorporate spin effects into General Relativity and postulates that this quantum property of matter must be coupled to a non-vanishing torsion tensor on the manifold. In this theory, the Ricci and torsion tensors are related to the matter content of spacetime, as follows
\begin{align}
    R_{ab} - \frac{1}{2}g_{ab} R &= S_{ab}
    \ , \label{EinsteinEqn} \\
    T^c{}_{ab} + 2\delta^c_{[a}T^d{}_{b]d} &= 2\Delta^c{}_{ab}
    \ , \label{HypermomentumEq}
\end{align}
where $S_{ab}$ is the canonical energy-momentum tensor and $\Delta^c{}_{ab}$ is the hypermomentum tensor \citep{Sciama,Kibble,Hehl,Capozziello,CarloniLuz2019}\footnote{Note that these references use different index conventions compared to this work. Reference \citep{UjjwalSante} provides more details with the same conventions.}. The hypermomentum tensor follows the anti-symmetric property 
\begin{align}
    \Delta^c{}_{ab}=-\Delta^c{}_{ba} \ ,
\end{align}
while $S_{ab}$ has no symmetries. Further, the matter conservation equation can be obtained by the unique non-vanishing contraction of \eqref{oncecontractedBianchiId}:
\begin{align}
    \nabla_b S^{ab} &= \frac{1}{2} T^{n}{}_{n}{}^{a} g^{km}S_{km} - T^{nma} S_{nm} - \frac{1}{2} T_{nmk} R^{anmk}
    \label{ConservationEqn} \ ,
\end{align}
where we used \eqref{EinsteinEqn} to replace Ricci tensor $R_{ab}$ and Ricci Scalar $R$ with canonical energy-momentum tensor $S_{ab}$.

\subsection{$1+1+2$ Covariant Approach and Local Rotational Symmetry}\label{SubSec:CovariantApproachLRS}

Traditionally, in relativistic gravitation, a coordinate system is used from the beginning to solve the field equations and to describe operations (such as time derivatives). However, this procedure inherits the features (and pathologies) of these coordinates and may therefore complicate the understanding of the physics of a given gravitational system. Alternative approaches use coordinate systems only at the very last steps of calculations and can therefore be more effective in complex/more general settings. One such approach, called the \textit{1+1+2 Covariant Approach}, involves defining a set of orthogonal \textit{timelike and spacelike congruences} ${(u_a,e_a)}$, and a spacelike $2-$surface (locally) orthogonal to them \citep{CarloniLuz2019,Clarkson_2003,ClarksonBetschart_2004,Clarkson_2007}.  The geometry of the manifold and the behavior of the matter fluid are then characterized by objects, the \textit{covariant variables}, defined along the congruences and on the orthogonal $2-$surface which, as we will see, possess a distinct physically meaning and a mathematically rigorous definition.

These variables satisfy a series of equations, the so-called \textit{covariant equations} obtained from purely geometrical relations, like the Bianchi identities (\ref{BianchiIdType1}-\ref{BianchiIdType2}), and are weakly related to the gravitational field equations in the sense that these last equations are only used to connect the components of the Riemann and torsion tensors to the energy-momentum and hypermomentum tensors of the source, respectively. As a consequence, the covariant equations can be easily generalized to any theory of gravity.

The covariant formalism allows one to freely choose the congruences ${(u_a,e_a)}$. As such, one can choose a pair of congruences which describe a \textit{specific frame} identified by a unique property or proceed with a generic pair of congruences which describe a \textit{generic frame}. One typically fixes the frame by assigning a physical description to the timelike congruence $u_a$. For example, the timelike congruence $u_a$ can be freely identified with the fluid source $4-$velocity, the $4-$velocity of any other observer, or a timelike Killing field if one exists. Therefore, in the covariant approach, the choice of the reference frame and its physical implications play a central role. In this section, we present a brief review of the covariant decomposition in a generic frame. Instead, we shall choose a specific frame in Section \ref{SubSec:FluidOfTheSpacetime}.

We start by defining a pair of orthogonal normalized timelike vector field $u_a$ and the normalized spacelike vector field $e_a$, 
\begin{align}
    u^a u_a &= -1 \ , & e^a e_a &= 1 \ , & u^a e_a &= 0 \ ,
\end{align}
along with the projection operators $h_{ab}$ and $N_{ab}$, defined as
\begin{align}
    h_{ab} &= g_{ab} + u_a u_b 
    \ , &
    u^a h_{ab} &= 0
    \ , \\
    N_{ab} &= h_{ab} - e_a e_b
    \ , &
    u^a N_{ab} &= 0
    \ , &
    e^a N_{ab} &= 0
    \ .
\end{align} 
Locally, $h_{ab}$ describes the metric of the $ 3$-hypersurface orthogonal to $u_a$ and $N_{ab}$ describes the metric of the $ 2$-surface orthogonal to $u_a$ and $e_a$. Similarly, one can define the local volume form $\eta_{ab}$ for the $2-$surface $N_{ab}$ by projecting the Levi-Civita tensor for $4-$dimensional spacetime $\eta_{abcd}$, defined in equation \eqref{LeviCivitaDForm}. We define $\eta_{ab}$ as follows,
\begin{equation} \label{LeviCivita2Form} \begin{aligned}
    \eta_{abc} &= \eta_{[abc]} = \eta_{dabc} u^d
    \ , \qquad
    \eta_{ab} = \eta_{[ab]} = \eta_{abc} e^c
    \ , \\
    \eta^{ab}\eta_{pq} &= N^a_p N^b_q - N^a_q N^b_p
    \ , \qquad
    \eta^{ab}\eta_{pb} = N^a_{p}
    \ ,
    \\
    \eta_{abc} &= e_a \eta_{bc} - e_b \eta_{ac} + e_c \eta_{ab} \ ,
\end{aligned} \end{equation}
where we have also provided the well-known relation for the tensor product of the Levi-Civita tensor $\eta_{ab}$ (of the $2-$surface) with itself.

The tensorial objects on the manifold can now be projected along $u_a$, $e_a$, and on the $2-$surface $N_{ab}$, thereby decomposing them into scalars and $2-$vectors/$2-$tensors on the $2-$surface $N_{ab}$. This decomposition is presented in detail in Reference \citep{CarloniLuz2019} without pre-imposing any symmetries on the spacetime. However, for the brief review presented in this work, we shall discuss the $1+1+2$ covariant decomposition for spacetimes that possess local rotational symmetry. \textit{Local rotational symmetry} implies that, at each point of the $3-$hypersurface $h_{ab}$, there exists a local axis of rotational symmetry described by a spacelike vector field \citep{Ellis1967,EllisStewart1968,EllisElst}.  
The congruence $e_a$ is then chosen to be parallel to this preferred direction. The local rotational symmetry and the above choice of congruence $e_a$ imply that the aforementioned $2-$vectors and $2-$tensors on the $2-$surface $N_{ab}$, defined in the $1+1+2$ covariant decomposition, must vanish. Lastly, the spacetimes possessing local rotational symmetry and non-vanishing torsion tensor are called \textit{Torsional Locally Rotationally Symmetric (TLRS) Spacetimes}. This terminology was introduced in Reference \citep{UjjwalSante}, which discusses these spacetimes and their properties in more detail.

The first set of variables of the 1+1+2 covariant approach can be obtained from the decomposition of covariant derivatives of the congruence vector fields and of the Weyl tensor for TLRS spacetimes, as follows:
\begin{equation} \label{DecompositionSpaceTimeTensors} \begin{aligned}
    \nabla_a u_b &= - \mathcal{A} u_a e_b + \left( \frac{\Theta}{3} + \Sigma \right) e_a e_b + \left( \frac{\Theta}{3} - \frac{\Sigma}{2} \right) N_{ab}
    \\ & \quad
    + \Omega \eta_{ab}
    \ , \\
    \nabla_a e_b &= - \mathcal{A} u_a u_b + \left( \frac{\Theta}{3} + \Sigma \right) e_a u_b + \frac{\phi}{2} N_{ab} + \xi \eta_{ab}
    \ , \\
    C_{abcd} &= -2 u_a E_{b[c} u_{d]} + 2 u_b E_{a[c} u_{d]} - 2 \eta_{ab}{}^{e} H_{e[c} u_{d]} 
    \\ & \quad
    - 2 \eta_{cd}{}^{e} \overline{H}_{e[a} u_{b]} - \eta_{abp} \eta_{cdq} E^{qp}
    \ , \\
    E_{ab} &= u^c u^d C_{acbd} = \mathcal{E} \left( e_a e_b - \frac{N_{ab}}{2} \right) + \mathbb{E} \eta_{ab} 
    \ , \\
    H_{ab} &= \frac{1}{2}\eta_a{}^{pq} C_{pqbr} u^r = \mathcal{H}_r e_a e_b + \frac{1}{2} \mathcal{H}_t N_{ab}
    \ , \\
    \overline{H}_{ab} &= \frac{1}{2} \eta_a{}^{rs} C_{bqrs} u^q = \mathbb{H}_r e_a e_b + \frac{1}{2} \mathbb{H}_t N_{ab} 
    \ .
\end{aligned} \end{equation}
The quantities $\{ \mathcal{A}, \Theta, \Sigma, \Omega, \phi, \xi \}$ and $\{ \mathcal{E}, \mathbb{E}, \mathcal{H}_r, \mathcal{H}_t, \mathbb{H}_r, \mathbb{H}_t \}$  are called {\it kinematic variables} and {\it Weyl variables} respectively. For a given congruence of worldlines $u^a$, $\mathcal{A}$ is the acceleration felt by an observer following the worldline, $\Theta$ is the rate of isotropic expansion of the congruence, $\Sigma$ is its rate of shear along the local axis of symmetry and $\Omega$ is the vorticity, which describes the rotation of the congruence relative to a non-rotating frame \citep{EllisElst1998CM}. Similarly, on the $3-$hypersurface $h_{ab}$, $\phi$ is the expansion and $\xi$ is the rotation of the congruence $e^a$ \citep{Clarkson_2003}. Tensors $E_{ab}$ and $\{H_{ab}, \overline{H}_{ab}\}$ are the electric and magnetic part of the Weyl tensor, an interpretation which is based on the analogy with decomposition of Maxwell field strength tensor \citep{EllisElst1998CM}.

The general decomposition of the canonical energy-momentum tensor for TLRS spacetimes is given as \citep{UjjwalSante,CarloniLuz2019}
\begin{align}
    S_{ab} 
    &= \mu u_a u_b + p (e_a e_b + N_{ab}) + \Pi \left( e_a e_b - \frac{N_{ab}}{2} \right) 
    \nonumber \\ & \quad
    + 2 Q e_{(a} u_{b)} + 2\mathbb{Q} e_{[a} u_{b]} + M \eta_{ab}
    \ ,
    \label{EnergyMomentum-LRS}
\end{align}
where $\mu$ is the energy density, $p$ is isotropic pressure, $\Pi$ is the anisotropic pressure, and $Q$ is the energy flux. The quantities $\mathbb{Q}$ and $M$ appear only for torsion spacetimes due to the non-symmetric nature of the canonical energy-momentum tensor. We shall call the variables obtained in decomposition \eqref{EnergyMomentum-LRS}, that is $\{ \mu, p, \Pi, Q, \mathbb{Q}, M \}$, {\it matter variables}. The collection of kinematic, Weyl, and matter variables is sufficient to determine all the properties of a TLRS spacetime and constitutes the covariant variables mentioned above. The covariant variables, describing the TLRS spacetimes, are scalars in the sense that they represent full contractions between tensorial objects. Notice, however, that these quantities depend on the choice of frame $(u_a,e_a)$ used for covariant decomposition 
of the spacetime. 

It is convenient to utilize a shorthand notation for the covariant derivative of a generic scalar $\psi$ along the timelike congruence $u^a$ (evolution) and along the spacelike congruence $e_a$ (propagation), given as
\begin{align}
    u^a \nabla_a \psi &= \dot{\psi} 
    \ , &
    e^a \nabla_a \psi &= \hat{\psi} \ ,
\end{align}
referred to as \textit{dot-derivative} and \textit{hat-derivative} respectively\footnote{The description for hat-derivative here is well-suited for $1+1+2$ covariant approach only with local rotational symmetry. See reference \citep{UjjwalSante,CarloniLuz2019} for the general $1+1+2$ decomposition of the covariant derivative.}. These will be important for a compact formulation of the covariant equations.

Finally, one can also describe covariantly the generalized Gaussian curvature of the $2-$surface described by $N_{ab}$\footnote{Note that we utilize the notion of \textit{generalized} Gaussian curvature here, which can be described irrespective of whether or not the $2-$surface $N_{ab}$ forms an integral submanifold. See Section \ref{SubSec:TLRSClassII}.}. For a TLRS spacetime, with general torsion and energy-momentum tensor \eqref{EnergyMomentum-LRS}, the generalized Gaussian curvature can be evaluated to be \citep{EllisBruniHwang,ClarksonBetschart_2004}
\begin{align}
    K = \frac{\phi^2}{4} + \xi^2 - \Omega^2 + \frac{\mu}{3} - \mathcal{E} -\frac{\Pi}{2} - \left( \frac{\Sigma}{2} - \frac{\Theta}{3} \right)^2 \ . \label{gaussiancurvatureLRS}
\end{align}
We shall discuss the physical meaning of generalized Gaussian curvature in Section \ref{SubSec:TLRSClassII}.

\subsection{Matter Content of the Spacetime - Weyssenhoff Fluid} \label{SubSec:FluidOfTheSpacetime}

For our system, we choose the matter content to be the (uncharged) \textit{Weyssenhoff fluid} \citep{WeyssenhoffRaabe,Korotky,Halbwachs}. The Weyssenhoff fluid is a semi-classical fluid representation of matter that carries spin. The spin is introduced by means of an anti-symmetric spin density tensor $L_{ab}$ orthogonal to its fluid $4-$velocity in accordance with the Frenkel condition \citep{Mathisson1937,Frenkel,Paparetrou}. This orthogonality property can be used to simplify the covariant decomposition of the spin density tensor $L_{ab}$ (and, as we will see soon, those of $S_{ab}$ and $\Delta_{abc}$) by choosing a specific frame. 

While the congruence $u_a$ in the previous section \ref{SubSec:CovariantApproachLRS} describes a generic frame, we will now choose the congruence $u_a$ to be described by the Weyssenhoff fluid $4-$velocity. As such, the following covariant decomposition and analysis is done in a specific frame, termed as \textit{comoving frame} or \textit{fluid rest frame}. With this choice of frame, the properties of the spin density tensor $L_{ab}$ are given as
\begin{align}
    L_{ab} &= - L_{ba}
    \ , &
    u^a L_{ab} &= 0
    \ . \label{PropertiesofSpinDensity}
\end{align}
Furthermore, the hypermomentum tensor and the energy-momentum tensor of the Weyssenhoff fluid are postulated to be given as \citep{Korotky,WeyssenhoffRaabe}
\begin{align}
    \Delta^c{}_{ab} &= u^c L_{ab} \ ,
    &
    S_{ab} &= -u_a P_b + p (g_{ab} + u_a u_b) \ , \label{WeyssenhoffPostulateEnergyMom}
\end{align}
where $P_a$ is the $4-$vector density of energy-momentum\footnote{Historically, the 4-vector density of energy-momentum was introduced to generalize the energy momentum tensor of a pressureless perfect fluid, so as to obtain an explicitly non-symmetric energy-momentum tensor described as a bi-vector. Indeed, in Equation \eqref{WeyssenhoffPostulateEnergyMom}, as in Equation \eqref{PropertiesofSpinDensity}, fundamentally, $u_a$ is the fluid 4-velocity.}. Here, $p$ is the isotropic pressure, and ${\mu = u^a u^b S_{ab} = u^a P_a}$ is the energy density in the rest frame of the fluid.

The properties of Weyssenhoff fluid greatly simplify the field equation \eqref{HypermomentumEq}. Taking the $g^b_c$ trace of equation \eqref{HypermomentumEq} and using \eqref{PropertiesofSpinDensity} and \eqref{WeyssenhoffPostulateEnergyMom}, reveals that the torsion tensor compatible with Weyssenhoff fluid satisfies
\begin{align}
    T^k{}_{ak} = 0 \ .
\end{align}
Consequently, equation \eqref{HypermomentumEq} simplifies to
\begin{align}
    T^c{}_{ab} = 2\Delta^c{}_{ab} = 2 u^c L_{ab} \ . \label{HypermomentumEqSimplified}
\end{align} 
Finally, the relation between the $4-$vector density of energy-momentum $P_a$ and the spin density tensor $L_{ab}$ can be evaluated by taking a contraction of the algebraic Bianchi identity \eqref{BianchiIdType1}, which allows one to relate the anti-symmetric part of the canonical energy-momentum tensor $S_{ab}$ to the hypermomentum using the field equations (\ref{EinsteinEqn}-\ref{HypermomentumEq}) (see Reference \citep{UjjwalSante} for details). We therefore obtain
\begin{align}
    P_a &= -\mu u_a - 2 u^m L_{na} \nabla_m u^n \ ,
\end{align}
which leads to
\begin{align}
    S_{ab} = \mu u_a u_b + p (g_{ab} + u_a u_b) - 2 u_a L_{bn} u^m \nabla_m u^n
    \ . \label{WeysennFluidEMTnsr}
\end{align}
For a TLRS spacetime, the covariant decomposition of the spin density tensor in the comoving frame is given as
\begin{align}
    L_{ab} = \tau \eta_{ab} \ , \label{SpinAngMom-GF-LRS}
\end{align}
where we utilised equation \eqref{PropertiesofSpinDensity}. Consequently, using the above decomposition along with equations \eqref{HypermomentumEqSimplified} and \eqref{WeysennFluidEMTnsr}, we obtain the covariant decomposition of the energy-momentum tensor and hypermomentum of the Weyssenhoff fluid in the comoving frame as
\begin{align}
    S_{ab} &= \mu u_a u_b + p (e_a e_b + N_{ab}) \ , \label{EnergyMomentum-WF-LRS}
    \\
    T^a{}_{bc} &= 2\Delta^a{}_{bc} = 2\tau u^a \eta_{bc} \ . \label{SimplifiedTorsionofTheory}
\end{align}
Notice that $\tau$ determines both torsion and hypermomentum; therefore, it simultaneously describes a property of the spacetime and a property of the matter. In the following, we will consider it essentially as a matter variable. In the Weyssenhoff fluid case, therefore, the only non-zero matter variables in the comoving frame will be\footnote{In the context of gluing spacetimes, this statement will remain true for the bulk manifolds $\mathcal{M}^\pm$ described as TLRS spacetimes filled with Weyssenhoff fluid. However, for the glued spacetime $\mathcal{M}$, the energy-momentum tensor and the matter variables must be raised to distributions as previously discussed. For these distributions, the behavior of the singular part of the matter variables must be carefully considered. This is further discussed in Section \ref{Sec:JuncStarVacFunda}.} $\{ \mu, p, \tau \}$.

\subsection{TLRS Class II Spacetimes}\label{SubSec:TLRSClassII}
Like their counterparts in General Relativity \citep{EllisElst}, TLRS spacetimes can be classified globally by the values of certain covariant variables. A complete classification of TLRS spacetimes is given in Reference \citep{UjjwalSante}. In the following, we summarize the main ideas of this classification; then we focus on a specific TLRS class, on which we construct the covariant junction conditions. The generalization of these conditions to other TLRS classes will be presented elsewhere.

A manifold admits a foliation into an integral submanifold and a one-form $X_a$ if the one-form $X_a$ satisfies the following hypersurface orthogonality condition\footnote{This condition is similar to Equation \eqref{HypersurfaceOrthogonalityCondition}, except that \eqref{HypersurfaceOrthogonalityCondition} was only evaluated locally at the interface since, in that instance, we did not require global foliation.} \citep{UjjwalSante,LuzHypersurface,WaldBook}
\begin{equation} \label{HyOrthoCondG}\begin{aligned}
    2X_{[a} \nabla_b X_{c]} + X_{[a} T^{k}_{\ bc]} X_k = 0 \ .
\end{aligned} \end{equation}
If the one-form $X_a$ satisfies the above equation, then we say $X_a$ (and equivalently $X^a=g^{ab} X_b$) is \textit{hypersurface orthogonal} to the family of integral submanifolds. Now we apply the above condition \eqref{HyOrthoCondG} to congruences $u_a$ and $e_a$, and use Equations \eqref{DecompositionSpaceTimeTensors} and \eqref{SimplifiedTorsionofTheory} to re-write the condition in terms of covariant variables. We obtain that congruences $u_a$ and $e_a$ are hypersurface orthogonal if they satisfy, respectively,
\begin{equation} \label{congruencesorthognality} \begin{aligned}
    2u_{[a} \nabla_b u_{c]} + u_{[a} T^{k}_{\ bc]} u_k = 0 \ &\implies \ \Omega-\tau =0 \ ,
    \\
    2e_{[a} \nabla_b e_{c]} + e_{[a} T^{k}_{\ bc]} e_k = 0 \ &\implies \ \xi = 0 \ .
\end{aligned} \end{equation}

In Reference \citep{UjjwalSante}, it was found that the TLRS spacetimes, when described in the comoving frame, can be classified into four classes. Each class admits a characteristic foliation such that both, only one or neither of the congruences ${(u_a,e_a)}$ are hypersurface orthogonal. In particular, the so-called TLRS class II spacetimes are defined by the property that both the congruences $(u_a,e_a)$ are hypersurface orthogonal, that is,
\begin{align}
    \Omega - \tau = 0 =\xi \ .  \label{DefinitionClassII-TLRS}
\end{align}
Therefore, these spacetimes admit a foliation into a timelike vector field, a spacelike vector field and a $2-$dimensional Riemannian submanifold. Many spacetimes of cosmological and astrophysical significance, including most of the interior solutions of stellar objects, are members of TLRS class II \citep{EllisElst,CarloniTOVIso1Fluid,CarloniTOVAniso1Fluid,CarloniLuz2019}. Compared to the other classes, TLRS class II also has two special properties. Firstly, the matter fluid of a TLRS class II spacetime does not need to satisfy any additional conditions \citep{UjjwalSante}. Secondly, since both congruences ${(u_a,e_a)}$ are hypersurface orthogonal, any vector field defined as a combination of these congruences would also be hypersurface orthogonal.

For TLRS class II, since the $2-$surface described by $N_{ab}$ forms an integral submanifold, the Gaussian curvature $K$, defined in Equation \eqref{gaussiancurvatureLRS}, can be related to the curvature of the $2-$surface as
\begin{align}
    \prescript{2}{}{R}_{ab} &= K N_{ab}
    \ , & 
    \prescript{2}{}{R}&=2K
    \ ,
\end{align}
where $\prescript{2}{}{R}_{ab}$ and $\prescript{2}{}{R}$ are the Ricci tensor and Ricci scalar of the $2-$surface $N_{ab}$ respectively \citep{ClarksonBetschart_2004}. In addition, if we assume that matter is represented by Weyssenhoff fluid and that we are in a frame comoving with such fluid, the Gaussian curvature in Equation \eqref{gaussiancurvatureLRS} simplifies to 
\begin{align}
    K = \frac{\phi^2}{4} - \tau^2 + \frac{\mu}{3} - \mathcal{E} - \left( \frac{\Sigma}{2} - \frac{\Theta}{3} \right)^2 \ ,
\end{align}
where we used Equations \eqref{EnergyMomentum-WF-LRS} and \eqref{DefinitionClassII-TLRS}. Using the covariant equations in Reference \citep{UjjwalSante}, the dot-derivative and hat-derivative of Gaussian curvature $K$ for the TLRS class II spacetime filled with the Weyssenhoff fluid can be evaluated to be
\begin{align}
    \dot{K} &= K\left( \Sigma - \frac{2}{3}\Theta \right)
    \ , &
    \hat{K} &= -\phi K
    \ . \label{gaussiancurvatureLRSDerivatives}
\end{align}

In what follows, we will consider non-vacuum spacetimes as TLRS class II, but we will also deal with vacuum TLRS spacetimes. As the classification of TLRS spacetimes depends on the matter variables \citep{UjjwalSante}, these geometries cannot be ascribed to any specific class. In this sense, we will consider vacuum spacetimes as a {\it generic} TLRS spacetime and we will assume that they can have 
\begin{align}
    (\Omega - \tau), \xi \neq 0\ .  \label{DefinitionVacuum-TLRS}
\end{align}
We will find this property useful to determine some general constraint on geometries that can be glued to each other.

\section{Type-I Covariant Junction Conditions}\label{Sec:JuncStarVacFunda}

We are now ready to formulate the junction between two spacetimes using the $1+1+2$ covariant formalism. In pursuit of this goal, we first specialize the fundamental junction conditions, described in Section \ref{Sec:JunctionsAndDistributionsFull}, to study the properties of the congruences and covariant variables at the interface. 

Since the glued spacetime $\mathcal{M}$ must be $\mathcal{C}^1$ differentiable at the interface $\mathcal{I}$, any vector field, which can be described by a generating curve/flow on the manifold, would be at least $\mathcal{C}^0$ across the interface. We can then define the distributional congruences ${(\dis{u_a}, \dis{e_a})}$ on the glued spacetime $\mathcal{M}$ which is aligned with the congruences $(u_a^\pm,e_a^\pm)$ in the respective manifolds $\mathcal{M}^\pm$. Since ${(\dis{u_a}, \dis{e_a})}$ must be continuous across the interface, we have
\begin{align}
\jump{u_a} = \jump{e_a} = \jump{N_{ab}} = \jump{\eta_{ab}} &= 0 \ , \label{ZeroJumpCovariantTensor}
\\
\intfc{u}_a = \intfc{e}_a = \intfc{N}_{ab} = \intfc{\eta}_{ab} &= 0 \ ,\label{ZeroDivergenceCovariantTensor}
\end{align}
 which imply
 \begin{equation}
 \jump{g_{ab}}=0 = \intfc{g}_{ab} \ .
 \end{equation}
Hence, consistently with the notation adopted for the normal $n_a$ in Equation \eqref{normalabuseofnotation}, in the following we do not utilize different notations for $u_a$ in the bulk manifold $\mathcal{M}^\pm$:
\begin{align}
    u_a &= u^+_a = u^-_a \ ,
    &
    \ifavg{u_a} &= u_a \bigg|_\mathcal{I} \ ,
\end{align}
and (ab)use similar brevity in notations for $e_a$, $N_{ab}$ and $\eta_{ab}$. 

From the fundamental junction conditions in Section \ref{Sec:JunctionsAndDistributionsFull}, we also know that the singular part of the torsion tensor must vanish as given in Equation \eqref{Singular-Tor-Zero}. This condition can be easily rewritten in terms of covariant variables associated with the torsion tensor \citep{CarloniLuz2019}. In the case of TLRS spacetimes with a Weyssenhoff fluid source,  taking into account \eqref{SimplifiedTorsionofTheory}, we obtain
\begin{align}
    \intfc{T}_{abc} = 0 \ \implies \intfc{\tau} = 0 \ . \label{zerodivergencetorsion}
\end{align}
The relations \eqref{ZeroJumpCovariantTensor}, \eqref{ZeroDivergenceCovariantTensor}, and \eqref{zerodivergencetorsion}, along with the hypersurface orthogonality condition \eqref{HypersurfaceOrthogonalityCondition} on the normal, collectively characterize the so-called \textit{Type-I Covariant Junction Conditions (CJC-I)}. While these conditions are direct consequences of fundamental junction conditions, we treat them separately to explicitly highlight their form in the covariant formalism.

Next, we must discuss the normal $n_a$ since it explicitly characterizes the interface as evident by Equation \eqref{covariantderivativeofdistribution}. As we are considering an interface between two TLRS spacetimes,  the normal must be orthogonal to the two-surface $N_{ab}$ and can be written parametrically as 
\begin{align}
    n_a = \ntime u_a + \nspace e_a \ . \label{NormalParametric}
\end{align}
Then we have two cases:
\begin{enumerate}
    \item Timelike normal $(\ntime,\nspace)=(1,0)$: $n_a = u_a$, $\epsilon=-1$.
    \item Spacelike normal $(\ntime,\nspace)=(0,1)$: $n_a = e_a$, $\epsilon=1$.
\end{enumerate}
where, as already stated, $\epsilon = n^a n_a$. For the parametric form \eqref{NormalParametric}, the condition \eqref{jump-npg-zero} is satisfied due to \eqref{ZeroJumpCovariantTensor}. However, as we shall see, we will have to impose condition \eqref{HypersurfaceOrthogonalityCondition} at a later stage. The advantage of the decomposition \eqref{NormalParametric} is that it allows us to study the properties of the interface in terms of covariant variables, for both spacelike and timelike normals.

Using these tools and combining them with the property of the ECSK theory of gravity and local rotational symmetry, we will be able to determine some properties for the distributions associated with the covariant variables. We will dedicate the following subsections to this task.

\subsection{The Singular Part of Kinematic Variables}\label{SubSec:No2quantities}

We can deduce the singularity structure of the 1+1+2  kinematical variables in a relatively easy way using the operations that we have defined in the previous sections, in particular, the derivative and the decomposition \eqref{DecompositionSpaceTimeTensors}. In fact, because of the properties \eqref{ZeroJumpCovariantTensor}, we have 
\begin{equation}
 \begin{aligned}
  \underline{\nabla_a w_b} 
    &=  (\nabla_a w_b)^+ \stepU + (\nabla_a w_b)^- (1-\stepU)  \ ,
\end{aligned}
\end{equation}
where $w_a=u_b ,e_b$. This property immediately implies that the singular part of the kinematic variables is zero. In fact
\begin{align}
    \intfc{(\nabla_a u_b)} = \epsilon n_a \jump{u_b} = 0 \,,
    \quad
    \intfc{(\nabla_a e_b)} = \epsilon n_a \jump{e_b} = 0
    \,, \label{ZeroDivKinematic}
\end{align}
where we used \eqref{covariantderivativeofdistribution}. Therefore, for the covariant decomposition of $\underline{\nabla_a u_b}$ and $\underline{\nabla_a e_b}$ as given in \eqref{DecompositionSpaceTimeTensors}, we can conclude
\begin{align}
    \intfc{\mathcal{A}} = \intfc{\Theta} = \intfc{\Sigma} = \intfc{\Omega} = \intfc{\phi} = \intfc{\xi} = 0 \ . \label{ZeroDivKinematicCov}
\end{align}

As we have seen, if we consider both manifolds $\mathcal{M}^\pm$ to be TLRS spacetimes, $2-$vectors and $2-$tensors on the surface $N_{ab}$ vanish for both $\mathcal{M}^\pm$. Therefore, the only vectorial or tensorial quantities that can appear on  $\mathcal{M}$ are the derivatives of the scalar covariant variables.  Since we have seen that the derivative of the step distribution is the Dirac delta, we can use the rules defined in the previous section to derive the distributional form of these objects and calculate their singular part.

Considering a generic $1+1+2$ variables $\psi$, its associated distribution is
\begin{align}
    \underline{\psi} = \psi^+ \stepU + \psi^- (1-\stepU) \ .
\end{align}
Upon taking the covariant derivative, we obtain
\begin{equation} \label{GlueManLRSInterM}
 \begin{aligned}
  \underline{\nabla_a \psi} 
    &=  (\nabla_a \psi)^+ \stepU + (\nabla_a \psi)^- (1-\stepU) + \epsilon n_a [\psi] \diracU \ .
\end{aligned}
\end{equation}
From the structure of the above expression, we can see that only some of the components of the vector distribution $\underline{\nabla_a \psi}$ can have a singular part.  

In particular, the component on the $2-$surface $N_{ab}$,  has no singular part as $N^a_b n_a = 0$ due to \eqref{NormalParametric}. In addition, since the bulk manifold has local rotational symmetry, $N^a_b (\nabla_a \psi)^\pm=0$, we can conclude that
\begin{align}
    N^a_b \underline{\nabla_a \psi} = 0 \ .
\end{align}
This result is important as it implies that the emergence of a preferred direction on $N_{ab}$ is avoided, thereby extending the local rotational symmetry to the glued spacetime $\mathcal{M}$. Instead, the components of $\underline{\nabla_a \psi} $ along $u^a$ or $e^a$ can have, depending on the nature of the normal, a singular part. 

Relation \eqref{GlueManLRSInterM} also has the consequence, via \eqref{DefofReimann}, that the Riemann tensor can contain singular terms: they derive from the second derivative of the congruence distributions. The next question, which we will address in the following subsection, is whether the Riemann tensor and other geometrical objects can also have double layers.

\subsection{Geometrical Double Layers}\label{SubSec:NoDoubleLayer}

As discussed, it is possible to show that the Riemann tensor can contain singular terms. In addition, as  $(T_{abc}, \nabla_a u_b, \nabla_a e_b)$ do not have singular terms, one can conclude, using \eqref{DefofReimann}, that no double layers are present in the Riemann tensor. This result, however, could be deduced in general from the  $\mathcal{C}^1$ differentiability of the glued manifold at the interface already discussed in  Section \ref{Sec:JunctionsAndDistributionsFull}. The same reasoning also allows us to conclude that no double layers can appear in the Riemann tensor, due to the absence of singular terms in the Levi-Civita connection and torsion.

As we will see, this is not sufficient to completely exclude the double layers from our treatment of junction conditions. In fact, equations \eqref{oncecontractedBianchiId} that will lead to the full junction conditions will contain the derivative of the Riemann tensor,  which generates double-layer terms. 

It should be remarked that the presence of double layers depends on the underlying theory of gravity. For example, a non-trivial double-layer structure can emerge in other theories of gravity, like, e.g., quadratic theories of gravity \citep{Senovilla2015}. Indeed, as we will see in the next section, the structure of the field equations of a given theory can have a subtle influence on the junction conditions, but, most of all, on the compatibility of matter and geometry, a topic that is often taken for granted in relativistic gravitation.

\subsection{Irregularities in the Matter Variables} \label{SubSec:SingularMatterAtInterface}

While the continuity properties of the manifold are sufficient to determine the regularity of the geometric quantities, the same is not true for the regularity of the matter fields. The reason is that the singularities of matter fields are ultimately derived from their field-theory description, an aspect often taken for granted in relativistic gravitation.

Considering the matter singularities offers an interesting interpretation of the (distributional) field equations on $\mathcal{M}$: they can be seen as a compatibility constraint between the singular structure of the geometry and the matter fields. For example, from the field equation \eqref{HypermomentumEq} of ECSK theory and Equation \eqref{zerodivergencetorsion}, we can also conclude that 
\begin{align}
    \intfc{\Delta}_{abc} = 0 \ . \label{zerointerfacehypermomentum}
\end{align}

Additionally, the field equations (\ref{EinsteinEqn}-\ref{HypermomentumEq}), and the absence of double layer term in the Riemann Curvature tensor and torsion tensor implies 
\begin{align}
 \dlayer{S_{ab}}=0=\dlayer{\Delta_{abc}}\ ,
\end{align} 
that is, the double layer terms in tensor distributions describing matter ($\dis{S_{ab}}$, $\dis{\Delta_{abc}}$) also vanish.

Probably one of the most intriguing outcomes in the distributional description of the matter energy-momentum tensor  $\underline{S_{ab}}$ appears when one evaluates its singular part $\intfc{S}_{ab}$. In fact, one can verify that the properties of the matter fluid in the bulk manifolds $\mathcal{M}^\pm$ and the resulting simplifications in the energy-momentum tensor $S_{ab}^\pm$ do not always carry over to the singular quantities describing the shell at the interface.

To see this, we start with the contraction $g_n^c$ of the algebraic Bianchi identity \eqref{BianchiIdType1}:
\begin{align}
    R_{[ab]} = \frac{1}{2}\nabla_k T^k{}_{ab} + \nabla_{[a}T^k{}_{b]k} - \frac{1}{2} T^m{}_{ab}T^n{}_{nm} \ .
\end{align}
Using the distribution $\underline{T_{abc}}$ and $\underline{R_{ab}}$, given as 
\begin{align}
    \underline{R_{ab}} &= R_{ab}^+ \stepU + R_{ab}^- (1-\stepU) + \intfc{R}_{ab}\diracU \ ,
    \\
    \underline{T_{abc}} &= T_{abc}^+ \stepU + T_{abc}^- (1 - \stepU) \ ,
\end{align}
we can relate the singular part of the energy-momentum tensor to the jump of torsion tensor as
\begin{align}
    \intfc{R}_{[ab]} = \intfc{S}_{[ab]} = \frac{ \epsilon}{2}n_k \jump{T^k{}_{ab}} \diracU  +  \epsilon n_{[a} \jump{T^k{}_{b]k}} \diracU \ ,
\end{align}
where we used \eqref{covariantderivativeofdistribution} and \eqref{EinsteinEqn}. Using the torsion tensor \eqref{SimplifiedTorsionofTheory}, the above relation simplifies to
\begin{align}
    \intfc{S}_{[ab]} = \epsilon \diracU n_k u^k \eta_{ab} \jump{\tau} \ .
\end{align}
Therefore, we arrive at an interesting result. Although the energy-momentum tensor is symmetric for the bulk manifold $\mathcal{M}^\pm$, energy-momentum tensor distribution $\underline{S_{ab}}$ may possess a non-vanishing anti-symmetric part due to the singular part at the interface $\mathcal{I}$. 

Let us understand this result in terms of the covariant matter variables in Equation \eqref{EnergyMomentum-LRS} via a concrete  example made within our setting. The matter variable $\underline{M}$, raised to a distribution, is defined via $\eta^{ab}$ projection of distribution $\underline{S_{ab}}$. By using the above equation and \eqref{NormalParametric}, we obtain
\begin{align}
    \underline{M} = \intfc{M} \diracU \ ,  \qquad \intfc{M} = -\epsilon \ntime \jump{\tau} \ .
\end{align}
While the distribution $\underline{M}$ vanishes in the bulk manifold $\mathcal{M}^\pm$ ($M^\pm=0$), in certain cases it may possess a non-vanishing singular part $\intfc{M}$ at the interface. Consequently, \textit{a priori}, we must assume that the singular part of all matter variables is non-vanishing.

In summary, the energy-momentum tensor distribution is given as
\begin{align}
    \underline{S_{ab}} = S_{ab}^+ \stepU + S_{ab}^- (1-\stepU) + \intfc{S}_{ab}\diracU \ ,
\end{align}
where the singular part $\intfc{S}_{ab}$ decomposes as
\begin{equation}\begin{aligned}
    \intfc{S}_{ab} &=   \intfc{\mu} u_a u_b + \intfc{p} (e_a e_b + N_{ab}) + \intfc{\Pi} \left( e_a e_b - \frac{N_{ab}}{2} \right) 
    \\ & \quad
                        + 2 \intfc{Q} e_{(a} u_{b)} + 2\intfc{\mathbb{Q}} e_{[a} u_{b]} + \intfc{M} \eta_{ab}
    \ .
\end{aligned} \label{EnergyMomentum-singular} \end{equation}

\section{Type-II Covariant Junction Conditions} \label{Sec:TypeIIJC}

While the type-I junction conditions are necessary for the geometric compatibility of the two bulk spacetimes, the glued spacetime can satisfy the gravitational field equations only if additional conditions are met. In the traditional Israel--Darmois approach, these conditions are obtained by identifying the singular parts of the geometric quantities appearing in the field equations—terms related to the covariant derivative of the normal form—and requiring them to match the singular part of the matter source tensor. When these conditions are trivially satisfied, that is, when they reduce to a $0=0$ identity, we refer to them as smooth matching conditions.
In the covariant formalism, an equivalent result is obtained by using the covariant equations to relate the jump of the kinematic variables to the jump or the singular part of the matter and Weyl variables. These relations will be called henceforth \textit{Type-II Covariant Junction Conditions (CJC-II)}\footnote{ In fact, CJC-II are really just relations, and the moniker ``conditions'' only applies when one requires that the bulk spacetimes $\mathcal{M}^\pm$ are smoothly matched.} and this section is dedicated to its derivation.

We will obtain the CJC-II by taking projections of  the singular part of equations \eqref{BianchiIdType1}, \eqref{oncecontractedBianchiId}, \eqref{ConservationEqn}, and the so-called \textit{Ricci identities} given as
\begin{align}
    R_{abcd}u^d &= \nabla_a \nabla_b u_c - \nabla_b \nabla_a u_c + T^{k}{}_{ab} \nabla_k u_c
    \label{RicciIdentityU}
     \ , \\
    R_{abcd}e^d &= \nabla_a \nabla_b e_c - \nabla_b \nabla_a e_c + T^{k}{}_{ab} \nabla_k e_c
    \label{RicciIdentityE}
     \ .
\end{align}
The procedure is similar to the one followed in Reference \citep{UjjwalSante} to evaluate the covariant equations; i.e., taking into account the parametric form \eqref{NormalParametric} for the normal, we project the above equations, select their singular part, and employ the CJC-I to simplify the final result\footnote{One could also start directly from the covariant equations to evaluate the CJC-II. However, the covariant equations in \citep{UjjwalSante} are not well-suited for such a procedure because the projection of the energy-momentum tensor \eqref{EnergyMomentum-WF-LRS} was used in their derivation, and this choice makes it impossible to include the effect of the singular part of $\dis{S_{ab}}$. Therefore, one should first re-derive the covariant equations in full generality and then select their singular part. The approach we are proposing, while equivalent, has the advantage of being faster.}.

In the following, we shall derive the CJC-II for interfaces between a TLRS class II spacetime permeated by a Weyssenhoff fluid (the {\it interior spacetime}) and a generic vacuum TLRS spacetime (the {\it exterior spacetime}). We choose to refer to the interior spacetime as $\mathcal{M}^-$ and the exterior spacetime as $\mathcal{M}^+$. Both $\mathcal{M}^\pm$ are governed by field equations of ECSK theory (\ref{EinsteinEqn}-\ref{HypermomentumEq}).

For manifold $\mathcal{M}^-$, the congruences ${(u^-_a, e^-_a)}$ describe the comoving frame and, by definition \eqref{DefinitionClassII-TLRS}, ${(u^-_a, e^-_a)}$ are hypersurface orthogonal. Therefore, TLRS class II spacetime $\mathcal{M}^-$ is characterised by
\begin{align}
\Omega^- - \tau^- &= 0 \ ,
&
\xi^- &= 0 \ . \label{InnerClassIIConditon}
\end{align}

The exterior spacetime $\mathcal{M}^+$ is considered to be a {(true-)}vacuum spacetime possessing local rotational symmetry which, as discussed, has no specific class. The vacuum condition implies a vanishing $S^+_{ab}$ and $\Delta_{abc}^+$ \citep{UjjwalSante} and, combined with the field equation \eqref{HypermomentumEq}, also leads to vanishing torsion tensor $T_{abc}^+=0$. In terms of covariant matter variables, the vacuum condition in $\mathcal{M}^+$ implies
\begin{align}
    \mu^+=p^+=\Pi^+=Q^+=\mathbb{Q}^+=M^+=\tau^+=0 \ . \label{OuterVacuumCondition}
\end{align}
Lastly,  we consider the congruences $(u^+_a, e^+_a)$  as a generic frame.

In this setting, the singular part of all the Weyl covariant variables and the matter covariant variables \textit{a priori} are considered to be non-vanishing.  Additionally, we  have
\begin{equation}
\jump{\Pi} = \jump{Q}=\jump{\mathbb{Q}} = \jump{M} = 0
\end{equation}
due to the energy-momentum tensor $S_{ab}^-$ given by \eqref{EnergyMomentum-WF-LRS} and $S^+_{ab}=0$. Also, from the discussion in Section \ref{Sec:JuncStarVacFunda}, none of the covariant variables possesses a double layer term.

The parametric form \eqref{NormalParametric} is applicable only if a pair of compatible frames is chosen to describe the bulk manifolds $\mathcal{M}^\pm$. For the given problem, a suitable frame must be chosen in $\mathcal{M}^+$ such that it matches the congruences of the comoving frame in $\mathcal{M}^-$ at $\mathcal{I}$. However, since the congruences $(u_a^+,e_a^+)$ describe a generic frame in $\mathcal{M}^+$, we can evaluate the CJC-II for the generic frame and use the CJC-II  to find the condition for which these frames are compatible. We shall come back to this point of choosing a suitable frame when we discuss the application of covariant junction conditions in Section \ref{Sec:Examples} and provide further insights into how it leads to a major advantage for the covariant approach to study junctions in comparison to the traditional (coordinate-dependent) approach.

Finally within the covariant formalism, and using \eqref{NormalParametric}, the condition \eqref{HypersurfaceOrthogonalityCondition}  takes the form
\begin{align}
    \ntime (\Omega-\tau)^\pm \bigg|_\mathcal{I} + \nspace \xi^\pm \bigg|_\mathcal{I} = 0 \ , \label{NormalParametricOrthogonality}
\end{align}
where we used \eqref{DecompositionSpaceTimeTensors} and \eqref{SimplifiedTorsionofTheory}.
The interesting aspect of this relation is that once the normal has been chosen, it demands that only one of the variables ($\Omega-\tau$ or $\xi$) vanishes at the interface $\mathcal{I}$. In turn, this implies that if the normal is timelike (spacelike), the spacetime can have $\xi\neq0$ ($\Omega-\tau\neq0$). Such a property is important in the study of junctions in more complex TLRS spacetimes. We will pursue such a study elsewhere.

The evaluation of the CJC-II is organized into various subsections below for clarity.

\subsection{Singular Part of Ricci Identities}\label{SubSec:DivPartofRicciId}

The singular part for the equation \eqref{RicciIdentityU} is given as
\begin{align}
    \intfc{R}_{abcd}u^d &= \epsilon (n_a \jump{\nabla_b u_c} - n_b\jump{\nabla_a u_c}) \ , 
\end{align}
where we used \eqref{covariantderivativeofdistribution}, \eqref{zerodivergencetorsion} and \eqref{ZeroDivKinematic}. The following projections
\begin{align}
    \{ u^a h^{bc}, u^a N^{bc}, u^a \eta^{bc}, e^a N^{bc}, \eta^{abc}, \eta^{ab} e^c \} \ ,
\end{align}
lead to independent equations at the interface $\mathcal{I}$. For each projection above, we obtain, respectively, 
\begin{equation} \label{singRicci-U-Eqns} \begin{aligned}
    \epsilon \ntime \jump{\Theta} +\epsilon \nspace \jump{\mathcal{A}} = \frac{1}{2}\intfc{\mu} + \frac{3}{2}\intfc{p}
    \ , \\
    \epsilon \ntime \jump{\Sigma - \frac{2}{3}\Theta} = \intfc{\mathcal{E}} - \frac{1}{3}\intfc{\mu} - \intfc{p} - \frac{1}{2}\intfc{\Pi}
    \ , \\
    \epsilon \ntime \jump{\Omega} = \intfc{\mathbb{E}} - \frac{1}{2}\intfc{M}
    \ , \\
    \epsilon \nspace \jump{\Sigma - \frac{2}{3}\Theta} = -\intfc{Q} -\intfc{\mathbb{Q}}
    \ , \\
    \epsilon \nspace \jump{\Omega} = \frac{1}{2}\intfc{\mathcal{H}_r} + \frac{1}{2}\intfc{\mathcal{H}_t}
    \ , \\
    \intfc{\mathcal{H}_r} = 0
    \ .
\end{aligned}\end{equation}

Similarly, for the singular part of the equation \eqref{RicciIdentityE} 
\begin{align}
    \intfc{R}_{abcd}e^d &= \epsilon (n_a \jump{\nabla_b e_c} - n_b\jump{\nabla_a e_c}) \ ,
\end{align}
obtained using \eqref{covariantderivativeofdistribution}, \eqref{zerodivergencetorsion} and \eqref{ZeroDivKinematic}, we take the projections
\begin{align}
    \{ u^a N^{bc}, e^a N^{bc}, u^a \eta^{bc}, e^a \eta^{bc} \}
\end{align}
leading to, respectively,
\begin{equation}\label{singRicci-E-Eqns}\begin{aligned}
    \epsilon \ntime \jump{\phi} = \intfc{\mathbb{Q}} - \intfc{Q}
    \ , \\
    \epsilon \nspace \jump{\phi} = -\intfc{\mathcal{E}} - \frac{2}{3} \intfc{\mu} - \frac{1}{2}\intfc{\Pi}
    \ , \\
    2 \epsilon \ntime \jump{\xi} = \intfc{\mathbb{H}_t}
    \ , \\
    \epsilon \nspace \jump{\xi} = - \intfc{\mathbb{E}} - \frac{1}{2}\intfc{M}
    \ .
\end{aligned}\end{equation}

\subsection{Singular Part of Algebraic Bianchi Identity}\label{SubSec:DivPartofBianchiI}

The singular part of the algebraic Bianchi identity \eqref{BianchiIdType1} is given as
\begin{align}
    \intfc{R}_{[ab}{}^n{}_{c]} = \epsilon n_{[a}\jump{T^n{}_{bc]}} \ ,
\end{align}
where we used \eqref{covariantderivativeofdistribution} and \eqref{zerodivergencetorsion}. The projections
\begin{equation}\begin{aligned}
\{ 
\eta^{ab}\delta^c_n, \eta^{ab} u_n u^c, \eta^{ab} e_n e^c, \eta^{ab} e_n u^c, \eta^{ab} u_n e^c, 
\ & \\ 
u^a e^b \delta^c_n, u^a e^b \eta^c{}_n & 
\}
\end{aligned}\end{equation}
lead to, respectively, 
\begin{equation}\label{SingBianchi-I-Eqns}\begin{aligned}
    \intfc{M} = -\epsilon \ntime \jump{\tau}
    \ , \\
    \intfc{\mathbb{E}} - \frac{1}{2}\intfc{M} = \epsilon \ntime \jump{\tau}
    \ , \\
    \intfc{\mathbb{E}} + \frac{1}{2}\intfc{M} = 0
    \ , \\
    \intfc{\mathcal{H}_r} + \intfc{\mathbb{H}_t} = 0
    \ , \\
    \intfc{\mathcal{H}_r} + \intfc{\mathcal{H}_t} = 2\epsilon \nspace \jump{\tau}
    \ , \\
    \intfc{\mathbb{Q}} = 0
    \ , \\
    2\intfc{\mathbb{H}_r} + \intfc{\mathbb{H}_t} + \intfc{\mathcal{H}_t} = 0
    \ .
\end{aligned}\end{equation}
Note that the projection $\eta^{ab} \eta^c{}_n$, only remaining independent projection of Equation \eqref{BianchiIdType1}, leads to a trivial equation.

\subsection{Singular Part of the Weyl Equation}\label{SubSec:DivPartofWeylEqn}

Let us now evaluate the singular part of equation \eqref{oncecontractedBianchiId}. This equation is the most complicated to work with and among the possible projections
\begin{align*}
    \{ u^a g^{bc}, e^a g^{bc}, u^a e^b u^c, u^a e^b e^c, \eta^{ab} e^c, \eta^{ab} u^c, u^a \eta^{bc}, e^a\eta^{bc} \} \ ,
\end{align*}
most cases can be proven to be redundant. Here we limit ourselves to giving just the final independent equations considering separately the timelike and the spacelike normal. We have:
\begin{enumerate}
    \item For timelike normal: From projection $u^a g^{bc}$ and $u^a e^b e^c$, we obtain, respectively
    \begin{equation}\begin{aligned}
    \jump{\mu} + \ifavg{\Theta}\jump{\Sigma-\frac{2}{3}\Theta} + \ifavg{\frac{3}{2}\Sigma - \Theta} \jump{\Sigma} 
    \quad & \\
    - 2\ifavg{\Omega}\jump{\tau} &=0 \label{TimelikeEnergyConsDiv}
    \ , \\
    \jump{\mathcal{E}+\frac{\mu}{6}} + \ifavg{\Sigma + \frac{\Theta}{3}}\jump{\frac{\Sigma}{2} - \frac{\Theta}{3}}
    \quad & \\
    + \ifavg{\Omega}\jump{\tau} + \frac{3}{2} \ifavg{\frac{\Sigma}{2} - \frac{\Theta}{3}}\jump{\Sigma}  &=0 \ .
    \end{aligned}\end{equation}
    \item For spacelike normal: From projections $e^a g^{bc}$ and $u^a e^b u^c$, we obtain, respectively
    \begin{equation}\begin{aligned}
    \jump{p} + \jump{\Pi} - \ifavg{\phi}\jump{\mathcal{A} + \frac{\phi}{2}} 
    \quad & \\
    - \ifavg{\mathcal{A}}\jump{\phi} - 2\ifavg{\tau}\jump{\tau} &= 0
    \ , \\
    \jump{\mathcal{E}- \frac{\mu}{3} - \frac{p}{2}} + \frac{1}{2}\ifavg{\phi}\jump{\mathcal{A}-\frac{\phi}{2}}
    \quad & \\
    +\frac{1}{2}\ifavg{\mathcal{A}}\jump{\phi} + \ifavg{4\Omega-\tau}\jump{\tau} &= 0 \ .
    \end{aligned}\end{equation}
\end{enumerate}
In Appendix \ref{App:DivPartofWeylEqn}, as an example, we present the explicitly evaluation for the singular part of the Weyl equation for the projection $u^a g^{bc}$ and further, report all the projections of singular part of the Weyl equation, including redundant ones.

\subsection{Double Layer Part of the Weyl Equation}\label{SubSec:DoublePartofWeylEqn}
As discussed in Sections \ref{SubSec:NoDoubleLayer}-\ref{SubSec:SingularMatterAtInterface}, no covariant variable possesses a non-vanishing double layer term in its distribution. However, the double layer distribution appears in the Weyl equation \eqref{oncecontractedBianchiId} due to the covariant derivative of Weyl tensor and Ricci tensor, which \textit{a priori} have a non-vanishing singular term. The coefficient of the double layer distribution term in Equation \eqref{oncecontractedBianchiId}, evaluated using Equation \eqref{covariantderivativeofdistribution}, is given as
\begin{align}
    \epsilon n^d \intfc{C}_{abcd} = -\epsilon n_{[a}\intfc{R}_{b]c} - \frac{\epsilon}{6} g_{c[a}n_{b]} \intfc{R} \ .
\end{align}
Upon taking the following independent projections
\begin{align}
\{ u^a g^{bc}, e^a g^{bc}, u^a e^b u^c, u^a e^b e^c, \eta^{ab} e^c, \eta^{ab} u^c, u^a \eta^{bc}, e^a\eta^{bc} \} \ , \label{IndepWeylEqnProj}
\end{align}
we obtain, respectively,
\begin{equation}\label{DoubleWeylEqns}\begin{aligned}
    \ntime\intfc{\mu} + \nspace ( \intfc{\mathbb{Q}} - \intfc{Q} ) = 0
    \ , \\
    \nspace ( \intfc{p} + \intfc{\Pi} ) - \ntime ( \intfc{\mathbb{Q}} + \intfc{Q} ) = 0
    \ , \\
    \ntime ( \intfc{\mathbb{Q}} + \intfc{Q} ) + 2\nspace ( \intfc{\mathcal{E}} - \frac{1}{3} \intfc{\mu} - \frac{1}{2} \intfc{p} ) = 0
    \ , \\
    2\ntime ( \intfc{\mathcal{E}} + \frac{1}{6} \intfc{\mu} + \frac{1}{2} \intfc{\Pi} ) + \nspace ( \intfc{\mathbb{Q}} - \intfc{Q} ) = 0
    \ , \\
    \ntime \intfc{\mathcal{H}_r} = 0
    \ , \\
    \nspace \intfc{\mathcal{H}_r} = 0
    \ , \\
    \ntime ( 2 \intfc{\mathbb{E}} + \intfc{M} ) + \nspace \intfc{\mathbb{H}_t} = 0
    \ , \\
    \nspace ( 2 \intfc{\mathbb{E}} - \intfc{M} ) - \ntime \intfc{\mathcal{H}_t} = 0
    \ .
\end{aligned}\end{equation}

\subsection{Constraint Equations To Jump Relations}\label{SubSec:ConstraintEqnJumps}
Reference \citep{UjjwalSante} provides the constraint equations for both of the bulk manifold $\mathcal{M}^\pm$. By writing these constraint equations for each bulk manifold and then subtracting them at the interface (identified with both bulk manifolds at the junction), we derive following relations between the jumps of quantities:
\begin{equation} \begin{aligned}
    \jump{\mathbb{E}} &= \jump{\tau \left( \Sigma + \frac{\Theta}{3} \right)}
    \ , \\
    \jump{\mathcal{H}_r} &= \jump{3\xi\Sigma - \left( 2\mathcal{A} - \phi \right) \Omega  + 2\mathcal{A}\tau}
    \ , \\
    \jump{\mathbb{H}_t + \mathcal{H}_r} &= \jump{2\mathcal{A}\tau}
    \ , \\
    \jump{2\mathbb{H}_r + \mathbb{H}_t + \mathcal{H}_t} &= 0
    \ .
\end{aligned} \label{ConstraintJumpsEqns} \end{equation}
These relations apply for both timelike and spacelike normals.

\subsection{General Properties of Type-II Covariant Junction Conditions} \label{SubSec:SummaryTypeIIJC}
We conclude this Section by giving a synthetic perspective on the CJC-II and highlighting their main properties.  

The CJC-II we have derived can be combined and further simplified. The final results are given in Tables \ref{TableJumps} and \ref{TablesingularParts}. Table \ref{TableJumps} gives the relations between the jumps of the covariant variables which need to be satisfied in order to form the glued spacetime, while Table \ref{TablesingularParts} provides information on the singular part of the matter variables and of the Weyl tensor in terms of the jumps of the covariant variables. We will use these tables as starting points for our considerations.

As we can see in  Table \ref{TableJumps}, independently of the nature of the normal, the conditions
\begin{align}
    \jump{\Omega-\tau} = \jump{\xi} &= 0 \ , \label{JumpofVortices}
\end{align}
must hold. Combining with \eqref{InnerClassIIConditon}, satisfied by TLRS-II manifold $\mathcal{M}^-$, we obtain
\begin{align}
    (\Omega-\tau)^+\bigg|_\mathcal{I} = 0 = \xi^+\bigg|_\mathcal{I} \ . \label{OuterClassIICond}
\end{align}
From Reference \citep{UjjwalSante}, we know that local values (vanishing or non-vanishing) of the variables ${\{(\Omega-\tau),\xi\}}$ are (usually) enough to determine the class of the TLRS spacetimes. Therefore, the condition \eqref{OuterClassIICond} implies that the manifold $\mathcal{M}^+$ must also belong to TLRS class II, i.e.
\begin{align}
    \xi^+=0=(\Omega-\tau)^+ \ , \label{OuterClassIICondGlobal}
\end{align}
globally on $\mathcal{M}^+$. Hence, we can conclude that a TLRS class II spacetime can only be glued to another TLRS class II spacetime. Additionally, due to \eqref{OuterClassIICond}, the condition \eqref{NormalParametricOrthogonality} is satisfied for the interface considered here.

Among  the rest of the  conditions in  Table \ref{TableJumps}, the following ones
\begin{align}
\jump{\phi}&=0 \quad \mbox{for} \quad   n_a=u_a  \ ,  \label{JumpTimeNormPhi}\\
\jump{\Sigma - \frac{2}{3}\Theta}&=0  \quad \mbox{for} \quad n_a=e_a  \label{JumpSpaceNormChi}
\end{align}
are particularly relevant. In fact, since the quantities involved characterize the covariant derivative of the frame $(u_a,e_a)$ in the bulk manifolds, they can be used to determine whether there exists a compatible frame between the interior and the exterior spacetimes.

As we mentioned, since covariant variables are frame dependent, a major problem in the construction of a glued spacetime is the determination of a global distributional frame on $\mathcal{M}$.  If these conditions are not satisfied, then the two spacetimes $\mathcal{M}^\pm$ cannot be glued. The relations \eqref{JumpTimeNormPhi} or \eqref{JumpSpaceNormChi}, depending on the nature of the normal,  constitute exactly the conditions that ensure the existence of such a frame.  In our specific case, since there is a natural frame  $(u_a^-,e_a^-)$ in $\mathcal{M}^-$, the comoving one, the relations \eqref{JumpTimeNormPhi} or \eqref{JumpSpaceNormChi} can be used to determine the compatible frame $(u_a^+,e_a^+)$ in $\mathcal{M}^+$.

%%%%%%%%%%%%%%%%%%%%%%%%%%%%%%%%%%%%%%%%%%
\begin{table}[]
\centering
\begin{tabular}{|c|c|c|}
\hline
Jump                                  & $n_a=u_a$     & $n_a=e_a$ \\
\hline
$\jump{\Omega-\tau}$                  & $0$                          & $0$
\\
$\jump{\xi}$                          & $0$                          & $0$
\\
$\jump{\phi}$                         & $0$                          & $\cross$
\\
$\jump{\Sigma - \frac{2}{3}\Theta}$   & $\cross$                  & $0$
\\
$\jump{\mu}$                          & $\jump{\tau^2 + \frac{1}{3}\Theta^2 - \frac{3}{4}\Sigma^2}$ & $\cross$
\\
$\jump{p}$                            & $\cross$ & $\jump{\tau^2 + \mathcal{A}\phi +\frac{1}{4}\phi^2}$
\\
$\jump{\mathcal{E} - \frac{\mu}{3}}$  & $\jump{- \tau^2 - \left( \frac{\Sigma}{2} - \frac{\Theta}{3} \right)^2 }$            & $\jump{\frac{1}{4}\phi^2 - \tau^2}$
\\
$\jump{\mathbb{E}}$ & $\jump{\tau \left( \Sigma + \frac{\Theta}{3} \right)}$ & $\jump{\tau \left( \Sigma + \frac{\Theta}{3} \right)}$
\\
$\jump{\mathcal{H}_r}$ & $\jump{\phi \tau}$ & $\jump{\phi\tau}$
\\
$\jump{\mathbb{H}_t}$ & $\jump{(2\mathcal{A} - \phi) \tau}$ & $\jump{(2\mathcal{A} - \phi) \tau}$
\\
$\jump{\mathbb{H}_r}$ & $-\frac{1}{2}\jump{\mathbb{H}_t + \mathcal{H}_t}$ & $-\frac{1}{2}\jump{\mathbb{H}_t + \mathcal{H}_t}$
\vspace{1mm}\\ 
\hline
$\jump{K}$                            & $0$                          & $0$
\\
\hline
\end{tabular}
\caption{Summary of the simplified CJC-II for the jump of the covariant variables and the Gaussian curvature. The cross ($\cross$) indicates that the jump of a quantity is not constrained by the CJC-II.}
\label{TableJumps}
\end{table}

%%%%%%%%%%%%%%%%%%%%%%%%%%%%%%%%%%%%%%%%%%%%%%

\begin{table}[]
\centering
\begin{tabular}{|c|c|c|}
\hline
Quantity                      & $n_a=u_a$     & $n_a=e_a$ \\
\hline
$\intfc{\mathcal{A}},\intfc{\Theta},\intfc{\Sigma},\intfc{\Omega},\intfc{\phi},\intfc{\xi},\intfc{\tau}$ & $0$ & $0$ \\
\hline
$\intfc{\mu}$                          & $0$                           & $-\jump{\phi}$ 
\\
$\intfc{p}$                            & $- \frac{2}{3} \jump{\Theta}$ & $ \frac{2}{3}\jump{\mathcal{A}+\frac{\phi}{2}}$ 
\\
$\intfc{\Pi}$                          & $\jump{\Sigma}$               & $-\frac{2}{3}\jump{\mathcal{A}+\frac{\phi}{2}}$ 
\\
$\intfc{M}$                            & $\jump{\tau}$                 & $0$ 
\\
$\intfc{Q}$                            & $0$                           & $0$ 
\\
$\intfc{\mathbb{Q}}$                 & $0$                           & $0$ 
\\
$\intfc{\mathcal{E}}$                  & $- \frac{1}{2} \jump{\Sigma}$ & $\frac{1}{3}\jump{\mathcal{A}-\frac{\phi}{2}}$ 
\\
$\intfc{\mathbb{E}}$       & $-\frac{1}{2}\jump{\tau}$     & $0$ 
\\
$\intfc{\mathcal{H}_r}$                & $0$                           & $0$ 
\\
$\intfc{\mathcal{H}_t}$                & $0$                           & $2\jump{\tau}$ 
\\
$\intfc{\mathbb{H}_r}$     & $0$                           & $-\jump{\tau}$ 
\\
$\intfc{\mathbb{H}_t}$     & $0$                           & $0$ 
\\
\hline
$\intfc{K}$ & $0$ & $0$
\\
\hline
\end{tabular}
\caption{Summary of the simplified CJC-II for the singular part of the associated distributions for covariant variables and the Gaussian curvature.}
\label{TablesingularParts}
\end{table}
%%%%%%%%%%%%%%%%%%%%%%%%%%%%%%%%%%%%%%%%%%%%%%

Let us now focus on Table \ref{TablesingularParts}. By providing an expression for the singular part of the covariant variables, the information on this table makes it easy to determine the smooth matching conditions, introduced in Section \ref{Sec:JunctionsAndDistributionsFull}. 

The bulk manifolds $\mathcal{M}^\pm$ are smoothly matched, and the thin shell disappears if the following continuity conditions are imposed on the covariant variables:
\begin{enumerate}
    \item Case of Timelike normal ($\ntime=1$, $\nspace=0$, $n_a=u_a$, $\epsilon=-1$):
     \begin{equation}\label{SmoothMatchingConditionTimelike}
        \begin{aligned}
        \jump{\Sigma} &= 0 \ , \\
        \jump{\Theta} & = 0 \ , \\
        \jump{\tau} & = 0 \ . 
        \end{aligned}
     \end{equation}
    \item Case of Spacelike normal ($\ntime=0$, $\nspace=1$, $n_a=e_a$, $\epsilon=1$):
    \begin{equation}\label{SmoothMatchingConditionSpacelike}
    \begin{aligned}
        \jump{\mathcal{A}} &= 0 \ , \\
         \jump{\phi} &= 0 \ , \\
         \jump{\tau} &= 0 \ . 
    \end{aligned}
    \end{equation}
\end{enumerate}
In terms of differentiability, the smooth matching conditions imply that, at the interface, the glued manifold $\mathcal{M}$ is of class $\mathcal{C}^2$ and the torsion tensor is continuous ($\mathcal{C}^0$).  Notice that the conditions above ensure that the singular part of the entire Riemann curvature tensor vanishes. This is a stronger result than the one normally found in the literature. For example, the analysis in \citep{Arkuszewski} and other traditional approaches to study junctions focus only on the field equations and the energy-momentum tensor. As a result, the prescribed smooth matching conditions therein only regularise the Ricci tensor.

Finally, Tables \ref{TableJumps} and \ref{TablesingularParts} allow us to deduce the properties of the Gaussian curvature. Using  \eqref{gaussiancurvatureLRS}, it is not difficult to see that the singular part of the Gaussian curvature $K$ vanishes for either case of the normal:
\begin{align}
    \intfc{K} &= 0 \ .
\end{align}
Moreover, taking the jump of the relation \eqref{gaussiancurvatureLRS} (as we did in Section \ref{SubSec:ConstraintEqnJumps}) and using the results in Table \ref{TableJumps}, for either case of the normal, we obtain:
\begin{align}
\jump{K} = 0 \ . \label{GaussianCurvatureJumpZero}
\end{align}
This result, along with the discussion following Equation \eqref{JumpofVortices}, implies that the glued manifold $\mathcal{M}$ belongs to the TLRS class II, and the $2-$surface $N_{ab}$ forms an integral submanifold. The Gaussian curvature is directly related to the volume form (area-element) of the $2-$surface $N_{ab}$ [see Ref. \citep{ClarksonBetschart_2004} or equation \eqref{GaussianCurvatureMetricComponents}]. Therefore, Eq.~\eqref{GaussianCurvatureJumpZero} ensures that the integral on the $2-$surface $N_{ab}$ is unique everywhere in $\mathcal{M}$, as one would require from a consistent distribution-based description of the glued spacetime (cf. Section \ref{Sec:JunctionsAndDistributionsFull}).

\section{Examples}\label{Sec:Examples}

We now turn to the application of the results derived in the previous sections to a specific example. Before proceeding, to ease understanding of the advantages of the covariant approach to junctions that we just presented, it is worth making some considerations.

In the traditional approach to junction conditions, one uses the metric components to evaluate the singular part of the Einstein Field equations  \citep{RelToolkitPoisson}. This approach requires choosing a coordinate system that is well-defined across the interface. Since there is no algorithmic approach to derive such a coordinate system, one is left with only a trial-and-error approach. The drawback is that, if the junction cannot be formed at the chosen coordinates,  there is no guarantee that two spacetimes cannot be glued altogether, or that there exists another choice of coordinates in which the gluing is possible.

As we have seen, in the covariant approach, one does not rely on a specific choice of coordinates; rather, one defines a frame based on the timelike and the spacelike congruences $(u^a, e^a)$. Like in the standard approach, the covariant junction conditions require the frames chosen in $\mathcal{M}^\pm$ to be compatible, but they also give {\it precise prescriptions} for such compatibility in the form of Eqs. \eqref{JumpTimeNormPhi} and \eqref{JumpSpaceNormChi}. In breaking covariance, i.e., in choosing a specific coordinate system in a given frame, we can translate these prescriptions in terms of the coordinate compatibility condition, thereby solving the problem of the coordinate compatibility. 

Let us characterize the breaking of covariance in more detail.

As discussed in Section \ref{SubSec:TLRSClassII}, the TLRS class II spacetimes are characterized by hypersurface orthogonality of the congruences $(u_a,e_a)$, which allows us to write them as $1-$forms parallel to an exact form:
\begin{align}
    u = fdt
    \ , \quad
    e = gdr
    \ , \label{coordinateframerelation}
\end{align}
where $(t,r)$ describe parameters along the timelike and spacelike congruences that can be interpreted as coordinate time and coordinate radius, respectively, and $\{f,g\}$ are functions which relate the congruences to these coordinates \citep{UjjwalSante}.

In terms of these coordinates, a TLRS-II line interval is given as
\begin{align}
    ds^2 = -f^2 dt^2 + g^2 dr^2 + h ( d\theta^2 + \sin^2{\theta} d\varphi^2 ) \ , \label{MetricForm-TLRSII}
\end{align}
where $(\theta,\varphi)$ are angular coordinates spanning the $2-$surface $N_{ab}$ and $h$ is a function which describes the area of an infinitesimal element on the $2-$surface. Due to local rotational symmetry, we have the dependence $f=f(r,t)$, $g=g(r,t)$ and $h=h(r,t)$. Using Equations \eqref{LeviCivitaDForm}, \eqref{LeviCivita2Form} and the determinant of the metric in the given coordinates being $\mathfrak{g}(r,t)=-f^2g^2h^2\sin^2{\theta}$, the non-vanishing components of $\eta_{ab}$ can be evaluated as
\begin{align}
\eta_{\theta\varphi} = - \eta_{\varphi\theta} &= - h\sin\theta \ .
\end{align}

We can now write the covariant variables in the given coordinate system. The kinematic variables can be related to the metric components by directly evaluating their expression in terms of  covariant derivatives 
\begin{equation}\begin{aligned}
    \Sigma + \frac{\Theta}{3} &= e^a e^b \nabla_a u_b
    \ , &
    \Sigma - \frac{2}{3}\Theta &= - N^{ab} \nabla_a u_b
    \ , \\
    \mathcal{A} &= -u^a u^b \nabla_a e_b
    \ , &  
    \phi &= N^{ab} \nabla_a e_b 
    \ .
\end{aligned}\end{equation}
For the TLRS-II spacetimes  \eqref{MetricForm-TLRSII} filled with Weyssenhoff fluid (with $u_a$ being the fluid $4-$velocity) and the compatible torsion tensor \eqref{SimplifiedTorsionofTheory}, we obtain
\begin{equation}\begin{aligned}
     \Sigma + \frac{\Theta}{3} &= -\frac{g_{,t}}{fg}
    \ , &
   \Sigma-\frac{2}{3}\Theta &= \frac{h_{,t}}{fh}
    \ , \\
    \mathcal{A} &= \frac{f_{,r}}{fg}  
    \ , &
   \phi &= \frac{h_{,r}}{gh}
    \ ,
\end{aligned} \label{CovToMetricComp} 
\end{equation}
where $(\cdot)_{,t}$ and $(\cdot)_{,r}$ denotes derivative with respect to time coordinate $t$ and radial coordinate $r$ respectively. Finally, because of \eqref{SimplifiedTorsionofTheory}, the torsion on $2-$surface $N_{ab}$ vanishes: 
\begin{align}
    \prescript{2}{}{T}_{abc} = N^p_a N^q_b N^r_c T_{pqr} = 0 \ ,
\end{align} 
 and the Gaussian curvature $K$ reads
 \begin{equation}\begin{aligned}
    K = \frac{\prescript{2}{}{R}}{2} = \frac{1}{h} \ .
\end{aligned} \label{GaussianCurvatureMetricComponents} \end{equation}
As discussed, the spacetime  $\mathcal{M}^-$ is  TLRS class II by definition, and the conditions  \eqref{OuterClassIICondGlobal} require that $\mathcal{M}^+$, is of the same class.  Therefore, the metric form \eqref{MetricForm-TLRSII} applies to both spacetimes and we can write 
\begin{equation}
    (ds^2)^- = -f^2 dt^2 + g^2 dr^2 + h \{ (d\theta^-)^2 + \sin^2{\theta^-} (d\varphi^-)^2 \} \ ,
 \label{PairGenericMetrics-}\end{equation}
in the comoving frame for $\mathcal{M}^-$ and
\begin{equation}    (ds^2)^+ = -F^2 dT^2 + G^2 dR^2 + H \{ (d\theta^-)^2 + \sin^2{\theta^+} (d\varphi^+)^2 \} \ ,
\label{PairGenericMetrics+}\end{equation}
for a generic frame in $\mathcal{M}^+$, where $(T,R)$ are coordinates and $(F,G,H)$ are functions of these coordinates, with same interpretation as drawn from analogy to \eqref{MetricForm-TLRSII}.

Using CJC-II \eqref{GaussianCurvatureJumpZero} and the relation \eqref{GaussianCurvatureMetricComponents}, we can conclude that
\begin{align}
    h|_\mathcal{I} = H|_\mathcal{I} \ . \label{jumpofhzero}
\end{align}
The above result, along with CJC-I $\jump{N_{ab}}=0$ \eqref{ZeroJumpCovariantTensor}, shows that the angular coordinates can be trivially chosen to be aligned
\begin{align}
    \theta^-=\theta^+ &\rightarrow \theta
    \ , &
    \varphi^-=\varphi^+ &\rightarrow \varphi
    \ .
\end{align}
In general, CJC-I allows us to evaluate the transformation law between the coordinates of $\mathcal{M}^+$ and $\mathcal{M}^-$ at the interface (the overlapping region for both coordinate systems).

\subsection{Buchdahl Star}

In this section, we shall use the results derived above to study an example of an interface between the (TLRS-II) spacetime associated with a static spherically symmetric matter distribution composed of Weyssenhoff fluid and the Schwarzschild solution for exterior vacuum in ECSK gravity. More specifically, we will consider the solution derived for the first time by Buchdahl in the context of General Relativity \citep{Buchdahl1967} that has been recently generalized to ECSK theory in \citep{CarloniLuz2019}. This kind of solution is interesting because the limit of maximum compactness is higher for such stellar objects composed of spin matter fluid \citep{CarloniLuz2019,Buchdahl1959}.

The interior manifold $\mathcal{M}^-$ is described by the metric  in \eqref{PairGenericMetrics-} with components
\begin{equation}\begin{aligned}
    f &= \sqrt{\frac{\alpha(1+\alpha-\eta)}{1+\alpha+\eta}} = \frac{1}{g} 
    \ , & 
    h &= \frac{r^2 (1+\alpha+\eta)^2}{4\alpha^2}
    \ , \\
    \eta &= \frac{(\alpha-1)\sin{(\beta r)}}{\beta r}
    \ ,
\end{aligned}\label{BuchdahlMetric}\end{equation}
where $\alpha,\beta \in \mathbb{R}$ are real-parameters. The matter content is described as
\begin{equation}\begin{aligned}
    \mu^- &= \frac{\alpha \beta^2 \eta (3\eta - 2\alpha - 2)}{(\gamma-1)(1+\alpha+\eta)^2}
    \ , \\
    p^- &= \frac{\alpha \beta^2 \eta \{2\gamma(2\eta-\alpha-1)-\eta\}}{(\gamma-1)(1+\alpha+\eta)^2}
    \ , \\
    (\tau^-)^2 &= \gamma\mu^-
    \ ,
\end{aligned}\label{BuchdahlMatter}\end{equation}
where $\gamma$ determines the relationship between the energy density and the spin density of the fluid. 

The manifold $\mathcal{M}^+$ is the Schwarzschild spacetime, satisfying vacuum condition \eqref{OuterVacuumCondition}, with metric components in \eqref{PairGenericMetrics+} given as
\begin{align}
    F &= \sqrt{1-\frac{R_s}{R}} = \frac{1}{G} 
    \ , & 
    H &= R^2 \label{SchwarzMetric}
    \ ,
\end{align}
where $R_s$ is the Schwarzschild radius, related to the mass $M$ of the spherical matter source as $R_s=M/4\pi$ in natural units.

The first step is to identify the normal form.  It is convenient to align the normal with spacelike congruence of  $\mathcal{M}^-$:
\begin{align}
n^- &= e^-= gdr  \ .
\end{align}
In this this way, the congruence $u_a^-$ describes a timelike killing field. From \eqref{CovToMetricComp}, \eqref{BuchdahlMetric} and \eqref{SchwarzMetric}, we must have
\begin{align}
    \Sigma^-= 0 = \Theta^- \quad \Rightarrow \quad\Sigma^+= 0 = \Theta^+ \ ,
\end{align}
thus, the normal in  $\mathcal{M}^+$ can be chosen as:
\begin{align}
    n^+  = e^+= GdR \ .
\end{align}
Any other congruence that satisfies the above conditions on $\Theta$ and $\Sigma$ would equally lead to compatible coordinate systems.

At the interface we can calculate the relation between internal coordinates $(t,r)$ and external coordinates $(T,R)$  using CJC-I \eqref{ZeroJumpCovariantTensor} and \eqref{GaussianCurvatureJumpZero} [or rather \eqref{jumpofhzero}]. We have
\begin{equation}\begin{aligned}
    \jump{K}=0 & \implies
    \\ &  R_\mathcal{I} = \frac{r_\mathcal{I}}{2\alpha}\left( 1+\alpha+\eta_\mathcal{I} \right)
    \ , \\
    \jump{u} = 0 & \implies (fdt)|_\mathcal{I}=(FdT)|_\mathcal{I} \implies
    \\ & \left(\frac{\partial T}{\partial t}\right)\bigg|_\mathcal{I} = \left(\frac{f}{F}\right)\bigg|_\mathcal{I} \ , 
    \\ & \left(\frac{\partial T}{\partial r}\right)\bigg|_\mathcal{I} = 0
    \ , \\
    \jump{e} = 0 & \implies (gdr)|_\mathcal{I}=(GdR)|_\mathcal{I} \implies
    \\ & \left(\frac{\partial R}{\partial t}\right)\bigg|_\mathcal{I} = 0 \ , 
    \\ & \left(\frac{\partial R}{\partial r}\right)\bigg|_\mathcal{I} = \left(\frac{g}{G}\right)\bigg|_\mathcal{I} = \left(\frac{F}{f}\right)\bigg|_\mathcal{I}
    \ .
\end{aligned}\label{coordinaterelations}\end{equation}

Let us now derive the condition for smooth matching. As the normal is spacelike, we have to refer to the conditions \eqref{SmoothMatchingConditionSpacelike}.  Considering the third of  such conditions and using \eqref{BuchdahlMatter} and \eqref{OuterVacuumCondition}, we obtain
\begin{equation}\label{muInterfacezero}\begin{aligned}
    \jump{\tau} = 0 
    & \implies \tau^-\bigg|_\mathcal{I} = (\gamma\mu^-)\bigg|_\mathcal{I} = 0
    \\ 
    & \implies \mu^-\bigg|_\mathcal{I} = 0 \ ,
\end{aligned}\end{equation}
and further the relation for $\jump{p}$ in Table \ref{TableJumps} leads to
\begin{align}\label{pInterfacezero}
    p^-\bigg|_\mathcal{I} = 0 \ .
\end{align}
Utilising equation \eqref{BuchdahlMatter} and $\mu^-|_\mathcal{I} = 0 = p^-|_\mathcal{I}$, we can conclude that
\begin{align} \label{etaInterfacezero}
    \eta\bigg|_\mathcal{I} = \frac{(\alpha-1)\sin{(\beta r_\mathcal{I})}}{\beta r_\mathcal{I}} = 0 \ ,
\end{align}
which introduces a constraint upon the radial coordinate at the interface as
\begin{align}
    \sin{(\beta r_\mathcal{I})} = 0 
    \implies 
    r_\mathcal{I} = \frac{m \pi}{\beta} \ , \quad m \in \mathbb{N} \ , \label{RadiusofStar}
\end{align}
where $\mathbb{N}$ is set of natural numbers. Note that we can also achieve $\eta|_\mathcal{I} =0$ by imposing a condition on the parameter
\begin{align}
    \alpha=1 \ ,
\end{align}
which we will discuss later.

Imposing  the second of the conditions \eqref{SmoothMatchingConditionSpacelike}, and using Equations \eqref{CovToMetricComp}, \eqref{BuchdahlMetric}, \eqref{SchwarzMetric} and \eqref{coordinaterelations}, we obtain
\begin{equation}\label{InterJPhiZero}\begin{aligned}
    \jump{\phi} = 0 & \implies
    \\ & F(R_\mathcal{I}) = \frac{f(r_\mathcal{I})}{2\alpha}\left\{ 1+ \alpha + (\alpha-1) \cos{(\beta r_\mathcal{I})} \right\}
    \ .
\end{aligned}\end{equation}
The above equation can be further simplified by using the constraint \eqref{RadiusofStar}, leading to two separate cases:
\begin{enumerate}
    \item Solution for even $m$ in \eqref{RadiusofStar}: Equation \eqref{InterJPhiZero} simplifies to
    \begin{align}
        F(R_\mathcal{I}) = f(r_\mathcal{I})
        \implies
        R_\mathcal{I} = \frac{1}{1-\alpha}R_s
        \ ,
    \end{align}
    where we used Equations \eqref{BuchdahlMetric}, \eqref{SchwarzMetric} and \eqref{etaInterfacezero}. Since $R_\mathcal{I}>R_s$ must be satisfied, we obtain the following constraint
    \begin{align}
        \alpha \in (0,1) \ .
    \end{align}
    Further, using the relation \eqref{RadiusofStar}, the observed radius and mass  of the stellar object can be evaluated to be
    \begin{align}
        R_\mathcal{I} &= \frac{1+\alpha}{2\alpha\beta}\pi m 
        \ , &
        R_s &= \frac{1-\alpha^2}{2\alpha\beta}\pi m
        \ . \label{evenprediction}
    \end{align}
    \item Solution for odd $m$ in \eqref{RadiusofStar}: Similar to the previous case, using Equations \eqref{BuchdahlMetric}, \eqref{SchwarzMetric} and \eqref{etaInterfacezero}, Equation \eqref{InterJPhiZero} simplifies to
    \begin{align}
        F(R_\mathcal{I}) = \frac{1}{\alpha}f(r_\mathcal{I}) \implies R_\mathcal{I} = \frac{\alpha}{\alpha-1}R_s
        \ .
    \end{align}
    Now we obtain the constraint
    \begin{align}
        \alpha \in (1,\infty) \ ,
    \end{align}
    from the condition $R_\mathcal{I}>R_s$. The observed radius and mass of the stellar object are given as, using \eqref{RadiusofStar},
    \begin{align}
        R_\mathcal{I} &= \frac{1+\alpha}{2\alpha\beta}\pi m 
        \ , &
        R_s &= \frac{\alpha^2-1}{2\alpha^2\beta}\pi m
        \ . \label{oddprediction}
    \end{align}
\end{enumerate}
For either case, it is easy to show that the first of conditions \eqref{SmoothMatchingConditionSpacelike},  $\jump{\mathcal{A}}=0$ is satisfied.

For completeness, we can now remark on the case $\alpha=1$. For this value of $\alpha$, the Buchdahl metric \eqref{BuchdahlMetric} and matter content \eqref{BuchdahlMatter} simplifies to
\begin{align}
    \eta&=0 \ , & f&=1=g \ , &  h &= \frac{r^2}{4}
    \ , \\
    \mu^- &= 0 \ , & p^- &=0 \ ,
\end{align}
for the entire manifold $\mathcal{M}^-$. Therefore, $\mathcal{M}^-$ describes a Minkowski spacetime. Further imposing $\jump{\phi}=0$, we obtain
\begin{align}
    F(R_\mathcal{I}) = f(r_\mathcal{I}) = 1 \implies R_\mathcal{I} = \infty = r_\mathcal{I} \ ,
\end{align}
where we used \eqref{SchwarzMetric} and \eqref{coordinaterelations}. Therefore, the smooth junction between an interior Minkowski spacetime and an exterior Schwarzschild spacetime is only a theoretical one, forming at infinity. In fact, the gluing of a Minkowski spacetime and Schwarzschild spacetime was also studied in Reference \citep{RosaCarloni}, where it was found that a non-vanishing thin shell is necessary to glue these spacetimes at a finite radius.

\section{Conclusions}\label{Sec:Conclusion}

In this paper, we have developed a coordinate-independent tensor distribution theory for non-smooth spacetimes with torsion and applied it to the study of glued manifolds formed by two bulk regions. The construction provides a framework in which distributional objects, junction conditions, and singular terms can be treated without committing to a particular coordinate system.

After recalling the necessary background in Sections \ref{Sec:LinearDistTheory} and \ref{sec:IntegrableDensities}, we formulated tensor distributions in Section \ref{Sec:TensorDistTheory} in terms of integrable densities. This formulation preserves coordinate independence, allows the SLS theorem to be extended to tensor distributions, and incorporates torsion in a geometrically consistent manner. In particular, the derivative in the distributional sense is defined through the Lie derivative. This leads to a modification, in the presence of torsion, of the postulates of distribution theory given in Ref. \citep{Senovilla2015}. The use of integrable densities also provides a natural route towards extensions to non-orientable manifolds and to spacetimes with non-vanishing non-metricity.

In Section \ref{Sec:JunctionsAndDistributionsFull}, tensor distribution theory was used to analyze junctions obtained by gluing two bulk spacetimes with torsion. We derived the fundamental junction conditions required for the glued spacetime to be mathematically well-defined in this context. These conditions follow from two basic requirements: (i) the absence of $\diracU^2$ terms, so that all relevant objects on the glued manifold are distributions, and  (ii) the well-posedness of operations such as integration over the interface and differentiation with respect to a continuous vector field.  Requirement (i) might lead to the introduction of additional constraints that depend directly on the gravitational theory under consideration, other than the geometry of the bulk spacetimes. 

In considering spacetimes with torsion, we have extended the fundamental junction conditions by developing analogous geometrical conditions on the torsion tensor. In particular,  we found that torsion must not contain a singular part at the interface.  The shared differentiability property of the Levi-Civita connection and (con)torsion suggests that the condition imposed on torsion may be more naturally understood as an analogous regularity condition on the bundle of connections over a generic manifold\footnote{The Levi-Civita connection is then a special case in which the required differentiability property also follows from the $\mathcal{C}^1$ differentiability of the manifold.}. Also notice that the condition obtained here relaxes the restriction on the torsion tensor imposed in Ref. \citep{Bressange}.

Having established the distributional framework, we introduced, in Section \ref{Sec:TLRS-spacetimes}, a covariant formalism for studying the interface and the glued manifold without reference to a coordinate system. In Section \ref{Sec:JuncStarVacFunda}, we derived the type-I covariant junction conditions, or CJC-I, which express the implications of the fundamental junction conditions for congruences and for the covariant variables characterizing the spacetimes. In Section \ref{Sec:TypeIIJC}, we then evaluated the singular parts of the matter variables and of the Weyl covariant variables, which together encode the Riemann curvature tensor, in terms of discontinuities in the kinematic variables and torsion. Special attention was given to junctions between TLRS class-II spacetimes and vacuum spacetime. The resulting relations were identified as the type-II covariant junction conditions, or CJC-II.

The covariant formulation of junction conditions brings several advantages. Since the full suite of covariant variables characterizing the manifold is available, the CJC-II can be evaluated in a generic frame, independently of coordinates. This makes it possible to derive additional relations that are otherwise difficult to identify. The relations \eqref{JumpofVortices}, for example, show explicitly that the foliation admitted by the bulk manifolds plays a decisive role in determining whether two spacetimes can be glued. In particular, we found that a TLRS class-II spacetime can be joined only to another spacetime belonging to the same class. Relations \eqref{JumpTimeNormPhi} and \eqref{JumpSpaceNormChi} provide a further practical advantage: they can be used to determine, in an algorithmic way, a suitable pair of coordinate systems for describing the junction. Without the covariant formalism, identifying such coordinate systems is typically reduced to a trial-and-error procedure.

The CJC-II also reveal new physical effects associated with torsion. A discontinuity in torsion across the interface can induce antisymmetric terms in the singular part of the energy-momentum tensor, even when the energy-momentum tensor is symmetric in each bulk region. It can also generate singular contributions to the magnetic part of the Weyl tensor, specifically $(\intfc{\mathcal{H}_t} , \ \intfc{\mathbb{H}_r})$. In particular, a singular Weyl tensor may signal the presence of standing gravitational waves at the interface. A detailed analysis of this possible physical phenomenon lies beyond the scope of the present work and will be addressed elsewhere. More generally, because the covariant variables have direct physical interpretations, the CJC-II provide a transparent way to interpret the singular structure of the junction.

The CJC-II were also used to derive the smooth matching conditions \eqref{SmoothMatchingConditionTimelike} and \eqref{SmoothMatchingConditionSpacelike}. These conditions ensure that the full Riemann curvature tensor is a regular distribution, namely that its singular part vanishes. This is a stronger requirement than the regularisation considered in Ref. \citep{Arkuszewski}, where the focus is on the energy-momentum tensor and the field equations, so that only the Ricci tensor is regularised by the corresponding smooth matching conditions. Is such a stronger requirement really necessary? The answer depends on the physical phenomenon one wants to investigate on the glued spacetimes. We feel that phenomenology that involves the magnetic part of the Weyl tensor, and therefore gravitational waves, can only be properly treated in glued spacetimes with the additional conditions associated to the regularity of the Riemann tensor.   For spacetimes with torsion, whose conformal structure is considerably richer \citep{UjjwalSante}, the Weyl tensor must also be treated explicitly in order to regularise the full curvature tensor. The covariant formalism developed here provides an effective method for carrying out precisely this analysis.

Finally, in Section \ref{Sec:Examples}, we addressed the standard problem of matching two spacetimes given by metric solutions in specific coordinate systems. We showed how the covariant junction conditions can be used to solve this problem systematically. As an illustration, we considered the junction between the interior solution of a stellar object composed of Weyssenhoff fluid and an exterior Schwarzschild spacetime. By analyzing the properties of the interface and imposing the smooth matching conditions, we obtained theoretical predictions for the size and mass of the stellar object. In particular, the observed radius and mass of the Buchdahl star smoothly matched to the exterior Schwarzschild spacetime is given by \eqref{evenprediction} if one chooses $\alpha\in (0,1)$, and is given by \eqref{oddprediction} if $\alpha \in (1,\infty)$.

\appendix
\section{General Tensor Distributions}\label{App:TensorDenDis}
In this appendix, we will give, for completeness, some basic information on tensor distributions derived from general tensor densities.

An object in the space of tensor densities of weight-$\mathbf{w}$, represented as $\mathfrak{A} \in \mathbb{D}^\mathbf{w}_\mathcal{M} \otimes \mathcal{T}^r_s(\mathcal{M})$, is described by the transformation law \citep{SchutzBook}
\begin{align}
    \tilde{\mathfrak{A}}^a{}_b (\tilde{x}) = \left|\frac{\partial x}{\partial \tilde{x}}\right|^\mathbf{w} \frac{\partial \tilde{x}^a}{\partial x^m} \frac{\partial x^n}{\partial \tilde{x}^b} \mathfrak{A}^m{}_n(x) \ .
\end{align}
For example, from the transformation of the metric tensor, a rank $(0,2)$ tensor, we obtain that its  determinant  transforms as 
\begin{align}
    \tilde{\mathfrak{g}}(\tilde{x}) = \left| J \right|^2 \mathfrak{g}(x) \ ,
\end{align}
where $J$ is the determinant of transformation matrix $\partial x^i / \partial \tilde{x}^j$ for the coordinate system $x$ and $\tilde{x}$ [cf. \eqref{TransDForm}]. Therefore, the determinant of the metric $\mathfrak{g}$ is a density of weight-$2$, and, consequently, the volume form
\begin{align*}
    \mathfrak{G}=\sqrt{|\mathfrak{g}|(x)}d^dx
\end{align*}
transforms as given in \eqref{TransDensity}, thereby justifying the use of the square root in the definition of the volume form \eqref{LeviCivitaDForm}.

As already done in the main text, we can construct the distributions associated to the tensor densities defined above. For simplicity, we will start with tensor densities of weight  $\mathbf{w}=1$.

An easy way to achieve this goal is to start with the operation of multiplying a distribution by a tensor, an analog of multiplying a distribution by a scalar discussed in \eqref{DensityDistVctrSpace}. Considering a distribution $\disd{a} \in \mathbb{D}'_\mathcal{M}$ and a generic tensor $R\in\mathcal{T}^r_s(\mathcal{M})$ we have
\begin{align}
    \nolfuncal{R\disd{a}}{S} = \dfuncal{a}{R^{a_1...a_r}{}_{b_1 ... b_s} S_{a_1 ... a_r}{}^{b_1...b_s}} \ , \label{MultOfDistributionTensor}
\end{align}
where $S\in\mathcal{T}^s_r(\mathcal{M})$ is a generic tensor on the manifold. Both $R$ and $S$ satisfy the properties of sufficient differentiability and boundary condition already defined for test functions in Linear Distribution Theory. 

Using relation \eqref{MultOfDistributionTensor} and the extension of \eqref{DensityDistFundOp}, we can define a tensor density distribution $\disd{R}$ as
\begin{align}
   \mathbb{D}'_\mathcal{M} \otimes \mathcal{T}^r_s(\mathcal{M}) \ni \disd{R}: S\mapsto  \dfuncal{R}{S} \in \mathbb{R}  \ ,
\end{align}
where
\begin{align}
    \dfuncal{R}{S} &= \lim_{\epsilon \to 0} \int_\mathcal{M} (\mathfrak{R}_\epsilon)^{a_1...a_r}{}_{b_1 ... b_s}(x) S_{a_1 ... a_r}{}^{b_1...b_s}(x)  d^dx \ ,
\end{align}
where $\mathfrak{R}_\epsilon \in \mathbb{D}_\mathcal{M} \otimes \mathcal{T}^r_s(\mathcal{M})$ is a basis representation of $\disd{R}$,  $S\in\mathcal{T}^s_r(\mathcal{M})$ is a generic tensor, and we assume that the space of test objects is formed by objects for which integrability is ensured.

The generalization of the above procedure to densities with weights $\mathbf{w}\neq1$ is straightforward. However, in that case the interpretation is delicate because the issue of integrability becomes even more central.

Among the general tensor density distributions described above, some can be expressed as the product of a density (function) distribution and a tensor:
\begin{equation}
\disd{R} = R \disd{a} \ .
\end{equation}
These tensor distributions are called {\it separable}.

Separable tensor density distributions can be mapped  to  tensor distributions using the relation \eqref{DensityFunctionRelDist}:
\begin{align}
    \disd{R} = R\disd{a}=\mathfrak{G}\dis{a}R\equiv \dis{aR} \in \mathcal{T}'{}^r_s(\mathcal{M})\ .
\end{align}
The tensor distribution $\dis{aR}$ operates on $S\in \mathcal{T}^s_r(\mathcal{M})$ as:
\begin{align}
    \dis{aR}: S \mapsto \funcal{aR}{S} \in \mathbb{R} \ ,
\end{align}
where,
\begin{align}
    \funcal{aR}{S} = \lim_{\epsilon \to 0} \int_\mathcal{M} a_\epsilon R^{a_1...a_r}{}_{b_1 ... b_s} S_{a_1 ... a_r}{}^{b_1...b_s} d\mathcal{V} \ .
\end{align}

\section{Some Details on the Mappings between Integrable Densities and other objects}\label{App:DensityComp}

In this appendix, we give some additional information on the way in which integrable densities can be mapped to other geometrical objects. We will start with the isomorphism between $ d$-forms and integrable densities, and then comment on the Hodge dual.

The mapping \eqref{DensityDformRel} between $ d$-forms and integrable densities is sensitive to the orientation of the coordinate system, specifically to inversions \citep{Nicolaescu}. This is because, as evident by the transformation laws \eqref{TransDForm} and \eqref{TransDensity}, the components of $d-$forms are sensitive to the orientation of the coordinate system, while the same is not true for densities. In fact, due to this reason, densities are better suited to describe the integrals when non-orientable manifolds, for which inversions can be achieved via continuous transformations, are considered. This presents another reason why we have preferred integrable densities rather than $ d$-forms to develop the theory of tensor distributions, thereby making its extensions to broader topologies more direct.

Nonetheless, in this work, the discussion has been explicitly limited to continuous transformations on orientable manifolds (with right-handed orientation) that do not induce inversions and, therefore, satisfy $J>0$, thereby making the mapping between $d$-forms and densities isomorphic. A broader discussion on densities and the above mapping can be found in References \citep{JohnLeeBook,Nicolaescu}.

The Hodge-dual ($\star$) of a generic function ($0-$form) $a$ maps it to a $d-$form given as \citep{SchutzBook}
\begin{align} \label{Firsthodgesetup}
    \alpha = \star a = a \eta \ .
\end{align}
Taking the Hodge dual of $\alpha$, gives us, by properties of the Hodge dual,
\begin{align}
    \star \alpha = \star (\star a) = sign(\mathfrak{g}) a \ ,
\end{align}
where $sign(\mathfrak{g}) = \mathfrak{g}/|\mathfrak{g}|$ is the sign of the determinant $\mathfrak{g}$ of the metric $g_{ab}$. We can also evaluate the above operations in terms of components. We have, by definition,
\begin{equation}
\star \alpha = \frac{1}{d!}g^{i_0 j_0}...g^{i_{(d-1)}j_{(d-1)}} \eta_{i_0...i_{(d-1)}}\alpha_{j_0...j_{(d-1)}}
\end{equation}
 and
\begin{equation}\begin{aligned}
\star \alpha 
    &= \frac{1}{d!}g^{i_0 j_0}...g^{i_{(d-1)}j_{(d-1)}} \mathfrak{G}\varepsilon_{i_0...i_{(d-1)}} \mathfrak{a}\varepsilon_{j_0...j_{(d-1)}}
    \\
    &= \frac{sign(\mathfrak{g})}{\mathfrak{G}} \mathfrak{a} \ ,
\end{aligned}\end{equation}
where we used the map \eqref{DensityDformRelComp} between $d-$forms and densities in component form, and the contraction between the Levi-Civita symbol and the inverse metric $g^{ab}$ is resolved to give its determinant as: 
\begin{align}
    g^{i_0 j_0}...g^{i_{(d-1)}j_{(d-1)}} \varepsilon_{i_0...i_{(d-1)}} \varepsilon_{j_0...j_{(d-1)}} = \frac{d!}{\mathfrak{g}} \ .
\end{align}
Combining the two results above, we obtain the mapping between functions and integrable densities , given as
\begin{align}
    a = \frac{\mathfrak{a}}{\mathfrak{G}} \ .
\end{align}
We can then extend this mapping to density distributions and function distributions as
\begin{align}
    \dis{a} = \frac{\disd{a}}{\mathfrak{G}} \ , \label{DensityFunctionRelDistAppendix}
\end{align}
which coincides with \eqref{DensityFunctionRelDist}.

\section{The Relation Between Lie and Covariant Derivatives} \label{App:LieCovDerivRel}
As we have seen, for a generic $d-$dimensional manifold equipped with a covariant derivative $\nabla$, the Lie derivative and the covariant derivative are related as \citep{NomizuInterscience}
\begin{align}
 \mathcal{L}_X R  - \nabla_X R =   \mathbb{A}_X R \ , \label{LieCovDerivRelation}
\end{align}
where $R$ is a generic tensor (with its indices suppressed), $X$ is a vector field and $\mathbb{A}_X$ is an operator generated by the endomorphism given as
\begin{align}
    (A_X)^m{}_n = -\nabla_n X^m - T^m{}_{kn}X^k \ ,
\end{align}
where $T^a{}_{bc}$ is the torsion tensor as described in Equation \eqref{DefofTorsion}. The operation of derivative $\mathbb{A}_X$ on a generic tensor $R^a{}_b$ of rank $(1,1)$ is given as
\begin{align}
    (\mathbb{A}_X R)^a{}_b = (A_X)^a{}_m R^m{}_b - (A_X)^m{}_b R^a{}_m \ ,
\end{align}
which extends similarly for each index of a generic tensor of rank $(r,s)$. Furthermore, one can evaluate that, the operation of $\mathbb{A}_X$ on a generic $d-$form $\alpha\in\Omega^d_\mathcal{M}$ is simply given as
\begin{align}
    (\mathbb{A}_X \alpha)_{a_0 ... a_{d-1}} = - (A_X)^m{}_m \alpha_{a_0 ... a_{d-1}} \ .
\end{align}

\section{Fundamental Distributions on Glued Spacetimes}\label{App:DistributionsDerivations}

In this appendix, we present a rigorous derivation of the three fundamental distributions mentioned in Section \ref{Sec:JunctionsAndDistributionsFull}. The compact support of the distribution depends on the structure of the glued manifold $\mathcal{M}$ \eqref{StructureManifold}. In the following evaluations, the boundary of the bulk manifolds $\mathcal{M}^+$ is going to be relevant. It is given by $\partial \mathcal{M}^+ = \partial \mathcal{M}^+_\infty \cup \mathcal{I}$, where $\mathcal{I} $ is the interface and $\partial \mathcal{M}^+_\infty$ is the boundary at infinity (if any). The steps followed in this derivation can be generalized to other irregular manifolds by changing the compact support of distributions.

We begin by defining the step distribution $\stepU$ (associated with a step density distribution) for a generic manifold $\mathcal{M}$ as
\begin{align}
   \mathcal{F}'_\mathcal{M} \ni \stepU : \varphi 
   \mapsto
   \funcal{\step}{\varphi} = \int_\mathcal{M} \varphi\stepU d\mathcal{V} = \int_{\mathcal{M}^+} \varphi d\mathcal{V} \ ,
\end{align}
where $\varphi \in \mathcal{F}_\mathcal{M}$. 

Now we shall take the derivative of the step distribution. We first notice that for a function distribution $\dis{f}$, we have $\dis{\mathcal{L}_X f} = \dis{\nabla_X f}$. We then proceed with the covariant derivative, as it will be instrumental in decomposing the covariant gradient $\nabla_a$ of the step distribution. Upon taking the covariant derivative $\nabla_X \stepU$, using equation \eqref{CovDevDist}, we obtain
\begin{equation}\begin{aligned}\label{Der_Theta}
    \funcal{\nabla_X \step}{\varphi} 
        &= - \funcal{\step}{\div(\varphi X)} 
        = -\int_{\mathcal{M}^+}\div(\varphi X) d\mathcal{V}
        \\
        &= \int_\mathcal{I} \epsilon \varphi X^a n_a d\sigma \ ,
\end{aligned}\end{equation}
where we used the divergence theorem \eqref{DivergenceTheorem}. The change in sign in the last step is due to the fact that on the interface $\mathcal{I}$, that is included in the boundary of $\mathcal{M}^+$, the normal $n_a$ points from $\mathcal{M}^-$ to $\mathcal{M}^+$. Since, by definition, $\nabla_X \stepU = X^a \dis{\nabla_a\step}$ and the relation \eqref{MultOfDistributionTensor} holds, we can also write \eqref{Der_Theta} as
\begin{align}\label{DefNabTh}
    \funcal{\nabla_a\step}{Y^a} = \int_\mathcal{I} \epsilon Y^a n_a d\sigma \ ,
\end{align}
where $Y^a=\varphi X^a$.
Decomposing the covector distribution $\dis{\nabla_a\step}$ along the normal and orthogonal to it, we obtain
\begin{align}
    \funcal{\nabla_a\step}{Y^a} 
        &= \nolfuncal{\epsilon n_a n^b\dis{\nabla_b\step} + P_a^b \dis{\nabla_b\step}}{Y^a}
    \nonumber \\
        &= \nolfuncal{n^b\dis{\nabla_b\step}}{\epsilon n_a Y^a} + \nolfuncal{P_a^b \dis{\nabla_b\step}}{Y^a} \label{DecNabTh}\ .
\end{align}
Combining  \eqref{DecNabTh} and \eqref{DefNabTh}, we obtain:
\begin{align}
    \nolfuncal{n^b\dis{\nabla_b\step}}{\epsilon n_a Y^a} + \nolfuncal{P_a^b \dis{\nabla_b\step}}{Y^a} 
    =
    \int_\mathcal{I} \epsilon Y^a n_a d\sigma \ .
\end{align}
We now define the Dirac delta distribution $\diracU$  on  $\mathcal{M}$, taking into account  \eqref{lindiracdefinition}, as
\begin{align}
\mathcal{F}'_\mathcal{M} \ni \diracU : \varphi \mapsto \funcal{\dirac}{\varphi} = \int_\mathcal{I} \varphi d\sigma \ ,
\end{align}
where $\varphi \in \mathcal{F}_\mathcal{M}$. Comparing this definition with  the above result, we can conclude that
\begin{align}
    n^a\dis{\nabla_a\step} &= \diracU &&\text{and}& P_a^b\dis{\nabla_a\step} &= 0 \ .
\end{align}
Therefore, the derivative relation between step distribution and Dirac delta distribution is evaluated to be
\begin{align}
    \dis{\nabla_a\step} = \epsilon n_a \diracU \ .
\end{align}

Furthermore, we can similarly evaluate the derivative of $\diracU$ as
\begin{align}
    \nolfuncal{\nabla_X \diracU}{\varphi} = - \funcal{\dirac}{\div(\varphi X)} = - \int_\mathcal{I} \div(\varphi X) d\sigma \ ,
\end{align}
and perform similar steps to obtain:
\begin{align}
    \funcal{\nabla_a\dirac}{Y^a} = - \int_\mathcal{I} \div Y d\sigma \ , \label{AppDoubleLayerInter1}
\end{align}
where, for torsional spacetimes\footnote{Notice that in our definition we have assumed that the test vector $Y^a$ is sufficiently differentiable, and in fact, it is typically assumed to be smooth. In this way, the covariant derivative we considered in \eqref{AppDoubleLayerInter1} does not involve terms which would make the integration ill-defined.},
\begin{equation}
\div Y = \nabla_a Y^a + T^k{}_{ak} Y^a \ .
\end{equation}
We now need to evaluate the above divergence. Before we start, we should remark that such evaluation is done at the interface, which is identified with both $\mathcal{I}^\pm$ in the bulk manifolds $\mathcal{M}^\pm$. Since the basic steps are similar, we present here a treatment that is agnostic about the bulk spacetime under consideration. In fact, as we will see, this aspect of the derivation hides a minor ambiguity that we will resolve later. 

The first step is the decomposition of the vector field $Y^a$ along the normal and orthogonal to it:
\begin{align}
    Y^a = \epsilon n^a y + y^a \ , \label{DecompOfYa}
\end{align}
where $y=n_a Y^a$ and $y^a= P^a_b Y^b$. Then the term $\nabla_a Y^a$ can also be written as follows
\begin{align}
    \nabla_a Y^a 
        &= (\epsilon n^a n^b + P^{ab})\nabla_a (\epsilon n_b y + y_b)
    \nonumber \\
        &= \epsilon n^a \nabla_a y + \epsilon n^a n^b \nabla_a y_b + \epsilon y P^{ab} \nabla_a n_b 
    \nonumber \\ & \quad
         + P^{ab} \nabla_a y_b \ .
\end{align}
Furthermore, the $3-$torsion on the interface is defined as \citep{EllisBruniHwang,ClarksonBetschart_2004}:
\begin{align}
    {}^3T_{abc} = P^p_a P^q_b P^r_c T_{pqr} \ ,
\end{align}
where we used hypersurface orthogonality condition \eqref{HypersurfaceOrthogonalityCondition}, which is satisfied in the neighborhood of the interface $\mathcal{I}$. Decomposing the semi trace  $T^k{}_{ak} = g_k^b T^k{}_{ab}$ of the torsion tensor along the normal and orthogonal to it, we obtain
\begin{align}
    T^k{}_{ak} 
        &= {}^3T^k{}_{ak} + 2 \epsilon n_{[a}n^pP^q_{r]}T^r{}_{pq} \ .
\end{align}
Combing the above decompositions, we rewrite $\div Y$ as:
\begin{equation}\label{AppDecompDiv}\begin{aligned}
\div Y 
    &= \nabla_a Y^a + T^k{}_{ak}Y^a
\\
    &=  \epsilon n^a \nabla_a y + \epsilon n^a n^b \nabla_a y_b 
\\ & \quad    
    + \epsilon y P^{ab} \nabla_a n_b + P^{ab} \nabla_a y_b 
\\ & \quad 
        + (\epsilon n^a y + y^a) ({}^3T^k{}_{ak} + 2 \epsilon n_{[a}n^pP^q_{r]}T^r{}_{pq}) \ .
\end{aligned}\end{equation}
Hence, for the integral on the $3-$hypersurface $\mathcal{I}$, the integrand is now broken into pieces which are explicitly on the interface or orthogonal to it. However, we can eliminate some terms using the divergence theorem on the $3-$hypersurface. As we know, the $3-$hypersurface has an induced torsion given by ${}^3T_{abc}$, and the term $P^{ab} \nabla_a y_b$ represents the covariant derivative of a $3-$vector projected on the $3-$hypersurface. Therefore, one can write the divergence theorem on the $3-$hypersurface as
\begin{align}\label{3D_Int}
    \int_\mathcal{I} \prescript{3}{}{\div} y d\sigma = \int_{\partial\mathcal{I}} \kappa m_a y^a d\sigma_2 \ ,
\end{align}
where
\begin{align}
    \prescript{3}{}{\div}y = P^{ab} \nabla_a y_b + {}^3T^k{}_{ak} y^a \ , \label{3-divergenceformula}
\end{align}
$\partial \mathcal{I}$ is the boundary of the interface $\mathcal{I}$, $d\sigma_2$ is the surface element of the $2-$surface, and $m_a$ is the normal to $\partial \mathcal{I}$ with normalisation given by a normalized parameter $\kappa$.
Since the interface here is considered to be boundary-less (or one may say the boundary is at infinity), the integral \eqref{3D_Int} vanishes, that is, 
\begin{align}
\int_\mathcal{I} \left( P^{ab} \nabla_a y_b + {}^3T^k{}_{ak} y^a \right) d\sigma = 0 \ . \label{AppZero3Div}
\end{align}

Finally, using \eqref{AppDecompDiv} and \eqref{AppZero3Div} and after few minor manipulations, we can write \eqref{AppDoubleLayerInter1} as:
\begin{equation}\label{AppDoubleLayerInter2}\begin{aligned}
    \funcal{\nabla_a\dirac}{Y^a}
    &= - \int_\mathcal{I} d\sigma \div Y
    \\
        &=  - \int_\mathcal{I} d\sigma (\epsilon n^a \nabla_a y - \epsilon n^a y^b \nabla_a n_b + \epsilon y \nabla_a n^a 
        \\ & \qquad \qquad \qquad
        + \epsilon n^p T^r{}_{pr}y - \epsilon n_r n^p y^q T^r{}_{pq})
    \\
        &= - \int_\mathcal{I} d\sigma \div(\epsilon y n) 
        \\
        & \qquad + \int_\mathcal{I} d\sigma \epsilon n^a y^b (\nabla_a n_b + n^k T_{kab})
    \ ,
\end{aligned}\end{equation}
where we used \eqref{DivVctrFldDef} to write\footnote{Note that, since $\div(\epsilon yn)$ does not represent a divergence term on the $3-$hypersurface, one should not mistakenly apply the divergence theorem.}
\begin{align}
\nabla_a(\epsilon yn^a) + T^k{}_{ak} \epsilon yn^a = \div(\epsilon yn) \ .
\end{align}

Now, we decompose the covariant derivative $\dis{\nabla_a\dirac}$ along the normal and orthogonal to it as
\begin{equation}\begin{aligned}
    \funcal{\nabla_a\dirac}{Y^a} 
    &= \nolfuncal{\epsilon n_a n^b\dis{\nabla_b\dirac} + P_a^b \dis{\nabla_b\dirac}}{Y^a}
    \\
    &= \nolfuncal{n^b\dis{\nabla_b\dirac}}{\epsilon y} + \nolfuncal{P^b_a\dis{\nabla_b \dirac}}{y^a} \ ,
\end{aligned}\end{equation}
where we utilised Equations \eqref{DecompOfYa} and \eqref{MultOfDistributionTensor}. Combining the above relation with \eqref{AppDoubleLayerInter2}, we obtain:
\begin{align*}
    \nolfuncal{n^b\dis{\nabla_b\dirac}}{\epsilon y} + \nolfuncal{P^b_a\dis{\nabla_b \dirac}}{y^a}
    &=
    - \int_\mathcal{I} d\sigma \ \div(\epsilon y n)
    \\ & \quad
    + \int_\mathcal{I} d\sigma \epsilon n^l y^a (\nabla_l n_a + n^k T_{kla})
    \ .
\end{align*}
We now define the so-called \textit{double layer distribution} $\DoublayerU$ as:
\begin{align}
    \mathcal{F}'_\mathcal{M} \ni \DoublayerU : \varphi \mapsto \nolfuncal{\DoublayerU}{\varphi} = - \int_\mathcal{I} d\sigma \div(\varphi n) \ ,
\end{align}
where $\varphi \in \mathcal{F}_\mathcal{M}$.

Combining all of the above results, we can write the derivative of the $\diracU$ distribution as: 
\begin{align}
    n^a \dis{\nabla_a \dirac} &= \DoublayerU 
    \ , &
    P_a^b\dis{\nabla_b\dirac} &= \epsilon n^l \diracU ( \nabla_l n_a + n^k T_{kla})|_\mathcal{I}
    \ , \label{CovDerofDiracSeparatedApp}
\end{align}
or
\begin{align}
    \dis{\nabla_a \dirac} = \epsilon n_a \DoublayerU + \epsilon n^l \diracU ( \nabla_l n_a + n^k T_{kla} )|_\mathcal{I} \ , \label{CovDerofDiracUnfixed}
\end{align}
where $(\cdot)|_\mathcal{I}$ denotes the value of the expression in brackets at the interface $\mathcal{I}$. Here we run into an ambiguity that concerns the second term in the above equation: while we know $n_a$ is continuous across the interface, we  also have
\begin{align}
    \lim_{x^+ \xrightarrow[\mathcal{M^+}]{}\mathcal{I}} \left( \nabla_l n_a + n^k T_{kla} \right) \neq \lim_{x^- \xrightarrow[\mathcal{M^-}]{}\mathcal{I}} \left( \nabla_l n_a + n^k T_{kla} \right) \ .
\end{align}
So, which value of the above quantity should be chosen at the interface? This choice can be made by using the local representation of step distribution $\stepU$ at the interface $\mathcal{I}$ to write
\begin{equation}\begin{aligned}
(\nabla_l n_a + n^k T_{kla})|_\mathcal{I} 
    &= \stepU|_\mathcal{I} (\nabla_l n_a + n^k T_{kla})^+ 
    \\ & \qquad 
    + (1-\stepU|_\mathcal{I}) (\nabla_l n_a + n^k T_{kla})^- \ .
\end{aligned}\end{equation}
In this way, the value of $(\nabla_l n_a + n^k T_{kla})|_\mathcal{I}$ is determined by the choice of the value of the step distribution at the interface $\mathcal{I}$. As we have chosen such value to be $1/2$ in \eqref{StepDistPointwise}, we can write
\begin{equation}\begin{aligned}
(\nabla_l n_a + n^k T_{kla})|_\mathcal{I} 
    &=\ifavg{\nabla_l n_a + n^k T_{kla}}
    \\
    &= \ifavg{\nabla_l n_a} + n^k \ifavg{T_{kla}} \ ,
\end{aligned}\end{equation}
where used notation \eqref{InterfaceAverage}. Consequently, the covariant derivative of the Dirac delta \eqref{CovDerofDiracUnfixed} distribution can be given as
\begin{align}
    \nabla_a \diracU = \epsilon n_a \DoublayerU + \epsilon n^l \left( \ifavg{\nabla_l n_a} + n^k \ifavg{T_{kla}} \right) \diracU \ . \label{CovDerofDiracFixedApp}
\end{align}

Additionally, with the distributions\footnote{Note that singular parts of both torsion and covariant derivative of the normal, $\intfc{T}_{abc}$ and $\intfc{\nabla_a n_b}=\epsilon n_a [n_b]$, vanish due to the fundamental junction conditions. See Section \ref{SubSec:JunctionsFundamentals}.}
\begin{align}
    \dis{\nabla_a n_b} &= (\nabla_a n_b)^+ \stepU + (\nabla_a n_b)^- (1-\stepU) + \cancel{\epsilon n_a \jump{n_b} \diracU} \label{nablanormaldistApp}
    \ , \\
    \dis{T_{abc}} &= (T_{abc})^+ \stepU + (T_{abc})^- (1-\stepU) + \cancel{\intfc{T}_{abc} \diracU}
    \ ,
\end{align}
and using the ad-hoc rule \eqref{stepdiracadhoc}, we can manipulate the above expression to write:
\begin{align}
     \nabla_a \diracU = \epsilon n_a \DoublayerU + \epsilon n^l \left( \dis{\nabla_l n_a} + n^k \dis{T_{kla}} \right) \diracU \ .
\end{align}

Finally, we present the direct comparison between the double layer $\DoublayerU$ defined above and the co-vector double layer distribution defined in Ref. \citep{Senovilla2015} for torsion-free spacetimes and spacelike normal. In \citep{Senovilla2015}, the double layer distribution is defined as
\begin{align}
\mathcal{T}'{}^0_1(\mathcal{M}) \ni  \dis{\Delta_a}: Y^a \mapsto \funcal{\Delta_a}{Y^a} = -\int_\mathcal{I}d\sigma n^a n^b \nabla_a Y_b
   \end{align}
 or, equivalently,
\begin{align}
    \dis{\Delta_a} = \dis{\nabla_b (n_a n^b \dirac)} \ ,
\end{align}
 and it is related to the similarly defined Dirac delta distribution as:
\begin{align}
    \dis{\nabla_a \dirac} = \dis{\Delta_a} - n_a \ifavg{\nabla_b n^b}\diracU \ . \label{covecdoublelayerdiracrelation}
\end{align}
From the above definition, it is easy to evaluate
\begin{equation*}\begin{aligned}
    \dis{\Delta_a} 
    &= n_a n^b \dis{\nabla_b \dirac} + n^b \dis{\nabla_b n_a} \diracU + n_a \dis{\nabla_b n^b} \diracU
    \\
    &= n_a \DoublayerU + n^b \ifavg{\nabla_b n_a} \diracU + n_a \ifavg{\nabla_b n^b} \diracU
    \ ,
\end{aligned}\end{equation*}
where we used \eqref{CovDerofDiracSeparatedApp} and manipulated the product of $\dis{\nabla_b n_a}$ with $\diracU$ using \eqref{nablanormaldistApp}. Using the above result in Equation \eqref{covecdoublelayerdiracrelation}, we obtain:
\begin{align}
    \dis{\nabla_a \dirac} = n_a \DoublayerU + n^b \ifavg{\nabla_b n_a} \diracU \ ,
\end{align}
therefore, reproducing relation \eqref{CovDerofDiracFixedApp} for the case of vanishing torsion and spacelike normal ($\epsilon=1$).

\section{Derivation of Type-II Junction Conditions from Singular Part of Weyl Equation}\label{App:DivPartofWeylEqn}

As mentioned in Section \ref{SubSec:DivPartofWeylEqn}, the singular part of the Weyl equation and its projections are most complicated to evaluate. In the following, we shall explicit evaluate the projection $u^a g^{bc}$ of the singular part of \eqref{oncecontractedBianchiId}, before giving all the decompositions of the Weyl equation. The singular part of Equation \eqref{ConservationEqn} is given as
\begin{equation}\begin{aligned}
\epsilon n_b \jump{S^{ab}} &= 
     - \nabla_b \intfc{S}^{ab}  - \epsilon n^p \intfc{S}^{ab} \ifavg{\nabla_p n_b} 
    \\ & \qquad
    + \frac{\ifavg{T^{n}{}_{n}{}^{a}} }{2} g^{km} \intfc{S}_{km} - \ifavg{T^{nma}} \intfc{S}_{nm} 
    \\ & \qquad
    - \frac{\ifavg{T_{nmk}}}{2} \intfc{R}^{anmk} \ ,
\end{aligned}\end{equation}
where we used Equation \eqref{covariantderivativeofdistribution} and eliminated a term using the torsion \eqref{SimplifiedTorsionofTheory} and \eqref{NormalParametric}.  Taking the projection along $u^a$, we obtain
\begin{equation}\begin{aligned}
    0 =&
    \dot{\intfc{\mu}} - \epsilon s \jump{\mu} - \epsilon t \intfc{\mu} \ifavg{\Sigma + \frac{\Theta}{3}} + \ifavg{\Theta} ( \intfc{\mu} + \intfc{p} + \intfc{\Pi} )
    \\ &
    + \ifavg{\frac{3}{2}\Sigma - \Theta} \intfc{\Pi} - 2 \ifavg{\Omega} \intfc{M} + \epsilon t \jump{Q} - \epsilon t \jump{\mathbb{Q}} 
    \\ &
    + \hat{\intfc{Q}} - \hat{\intfc{\mathbb{Q}}} + ( \epsilon s \ifavg{\mathcal{A}} + \ifavg{\mathcal{A} + \phi}) \left( \intfc{Q} - \intfc{\mathbb{Q}} \right) 
    \\ &
    + \ifavg{\mathcal{A}} \left( \intfc{Q} + \intfc{\mathbb{Q}} \right) \ .
\end{aligned}\label{FullEnergyConsDiv}\end{equation}
As we know, we have $\jump{Q} = 0 = \jump{\mathbb{Q}}$. In fact, from equations \eqref{SingBianchi-I-Eqns} and \eqref{DoubleWeylEqns}, one can also derive that $\intfc{Q} = 0 = \intfc{\mathbb{Q}}$. Therefore, it is evident that the above equation can be substantially simplified by utilizing the results from Sections \ref{SubSec:DivPartofRicciId}, \ref{SubSec:DivPartofBianchiI} and \ref{SubSec:DoublePartofWeylEqn}. Furthermore, from Equation \eqref{DoubleWeylEqns}, we can notice that for timelike normal, $\intfc{\mu}$ vanishes. After the aforementioned simplifications, we obtain that Equation \eqref{FullEnergyConsDiv}, for timelike normal $n_a=u_a$, becomes
\begin{equation}\begin{aligned}
    \jump{\mu} + \ifavg{\Theta}\jump{\Sigma-\frac{2}{3}\Theta} + \ifavg{\frac{3}{2}\Sigma - \Theta} \jump{\Sigma} &
    \\
    - 2\ifavg{\Omega}\jump{\tau} &= 0 \ ,\label{TimelikeEnergyConsDivApp}
\end{aligned}\end{equation}
and, for spacelike normal $n_a=e_a$, becomes
\begin{align}
    \dot{\intfc{\mu}} - \ifavg{\Sigma - \frac{2}{3}\Theta} \intfc{\mu} + \ifavg{\frac{3}{2}\Sigma - \Theta}\intfc{\Pi} = 0 \ . \label{SpacelikeEnergyConsDiv}
\end{align}
Therefore, the physical implication of the equation \eqref{FullEnergyConsDiv} changes considerably for spacelike and timelike normal, and therefore, each case is treated separately in the following.

For spacelike normal, the Equation \eqref{SpacelikeEnergyConsDiv} can be derived directly from the covariant equations given in Reference \citep{UjjwalSante}. From Equations \eqref{singRicci-U-Eqns}, \eqref{singRicci-E-Eqns} and \eqref{DoubleWeylEqns}, for spacelike normal, we obtain
\begin{align}
    \intfc{\mu} &= - \jump{\phi} 
    \ , & 
    \intfc{\Pi} &= -\frac{2}{3}\jump{\mathcal{A} + \frac{\phi}{2}} 
    \ , &
    \jump{\Sigma - \frac{2}{3}\Theta} &= 0
    \ . \label{IntermidiatePropertiesforSpaclikeNormal}
\end{align}
Therefore, Equation \eqref{SpacelikeEnergyConsDiv} becomes
\begin{align}\label{IntDerivJuncEqn}
    \dotjump{\phi} = \ifavg{\Sigma - \frac{2}{3}\Theta}\jump{\frac{\phi}{2} - \mathcal{A}} \ .
\end{align}
For TLRS spacetimes filled with Weyssenhoff fluid described in the comoving frame and governed by ECSK theory, we have \citep{UjjwalSante}
\begin{align}
    \dot{\phi} &= - \left( \frac{\Sigma}{2} - \frac{\Theta}{3} \right) \left( 2\mathcal{A}-\phi \right) + 2\xi\Omega \ . 
\end{align}
Taking the jump of the above equation and using \eqref{IntermidiatePropertiesforSpaclikeNormal}, \eqref{jumpofproductdist} and the condition \eqref{NormalParametricOrthogonality} for spacelike normal (${\xi^\pm|_\mathcal{I}=0 \implies \jump{\xi}=\ifavg{\xi}=0}$), we obtain
\begin{align} \label{IntDerivCovEqn}
    \jump{\dot{\phi}} = \ifavg{\Sigma - \frac{2}{3}\Theta}\jump{\frac{\phi}{2} - \mathcal{A}} \ .
\end{align}
Comparing Equations \eqref{IntDerivJuncEqn} and \eqref{IntDerivCovEqn}, both of the equations can be shown to be identical. For this, we use the relation
\begin{align}
    \jump{\nabla_a \psi} = \epsilon n_a \jump{n^b\nabla_b \psi} + P^b_a \nabla_b \jump{\psi} \ ,
\end{align}
where $\psi \in \mathcal{F}_\mathcal{M}$ is generic scalar \citep{Senovilla2015}. Taking its projection along $u_a$ and utilising $n_a = e_a$ and \eqref{ZeroJumpCovariantTensor}, the LHS and RHS can be resolved as
\begin{equation}\begin{aligned}
    u^a \jump{\nabla_a \psi} &= u^a \left( \epsilon e_a \jump{e^b\nabla_b \psi} + (N_a^b - u_au^b) \nabla_b \jump{\psi} \right)
    \\
    \jump{u^a \nabla_a \psi} &= u^b \nabla_b \jump{\psi}
    \\
    \jump{\dot{\psi}} &= \dotjump{\psi}
    \ .
\end{aligned}\end{equation}
In conclusion, Equation \eqref{SpacelikeEnergyConsDiv} can be obtained from the covariant equations directly and is redundant.

We now give the projections of the singular part of the Weyl equation \eqref{oncecontractedBianchiId} for the case of timelike and spacelike normals separately, after the simplifications allowed by the results of Sections \ref{SubSec:No2quantities}, \ref{SubSec:DivPartofRicciId}-\ref{SubSec:DivPartofBianchiI}, and \ref{SubSec:DoublePartofWeylEqn}. 

The results are:\\
\textit{Timelike Normal: $\ntime=1, \ \nspace=0, \ \epsilon = -1$:}
The projections 
\begin{align}
\{ u^a g^{bc}, e^a g^{bc}, \eta^{ab} u^c, \eta^{ab} e^c, u^a \eta^{bc}, e^a\eta^{bc}, u^a e^b u^c, u^a e^b e^c \}
\end{align}
of \eqref{oncecontractedBianchiId} lead to
\begin{equation} \begin{aligned}
    \jump{\mu} + \ifavg{\Theta}\jump{\Sigma-\frac{2}{3}\Theta} + \ifavg{\frac{3}{2}\Sigma - \Theta} \jump{\Sigma} 
    \quad & \\
    - 2\ifavg{\Omega}\jump{\tau} &= 0
    \ , \\
    \hatjump{\Sigma} - \frac{2}{3}\hatjump{\Theta} + \frac{3}{2}\ifavg{\phi}\jump{\Sigma} + 2\ifavg{\xi}\jump{\tau}  &= 0 
    \ , \\
    \ifavg{\Omega-\tau} \jump{\Sigma + \frac{\Theta}{3}} &= 0
    \ , \\
    \jump{\mathcal{H}_r} - 3\ifavg{\xi}\jump{\Sigma} - \ifavg{\phi}\jump{\tau} &= 0
    \ , \\
    \jump{2\mathbb{E}+M} - 2\ifavg{\Omega} \jump{\Sigma+\frac{\Theta}{3}} - 2\ifavg{\Sigma+\frac{\Theta}{3}}\jump{\tau} &=0
    \ , \\
    2\hatjump{\tau} + \ifavg{\phi}\jump{\tau} - 3\ifavg{\xi}\jump{\Sigma} - \jump{\mathcal{H}_t} &=0
    \ , \\
    \hatjump{\Sigma} - \frac{2}{3}\hatjump{\Theta} + \frac{3}{2} \ifavg{\phi} \jump{\Sigma} + 2 \ifavg{\xi} \jump{\tau} &= 0
    \ , \\
    \jump{\mathcal{E} + \frac{\mu}{6}} + \ifavg{\Sigma + \frac{\Theta}{3}} \jump{\frac{\Sigma}{2} - \frac{\Theta}{3}} + \ifavg{\Omega}\jump{\tau} 
    \quad & \\
    + \frac{3}{2} \ifavg{\frac{\Sigma}{2} - \frac{\Theta}{3}} \jump{\Sigma} &=0
    \ .
\end{aligned} \end{equation}
\textit{Spacelike Normal: $\ntime=0, \ \nspace=1, \ \epsilon = 1$:}
The projections 
\begin{align}
\{ u^a g^{bc}, e^a g^{bc}, \eta^{ab} u^c, \eta^{ab} e^c, u^a \eta^{bc}, e^a\eta^{bc}, u^a e^b u^c, u^a e^b e^c \}
\end{align}
of \eqref{oncecontractedBianchiId} lead to
\begin{equation} \begin{aligned}
    \dotjump{\phi} - \ifavg{\Sigma - \frac{2}{3}\Theta}\jump{\frac{\phi}{2} - \mathcal{A}} &= 0
    \ , \\
    \jump{p} + \jump{\Pi} - 2\ifavg{\tau} \jump{\tau} - \ifavg{\mathcal{A}} \jump{\phi} 
    \quad & \\
    - \ifavg{\phi} \jump{\mathcal{A} + \frac{\phi}{2}} &= 0
    \ , \\
    \jump{\mathcal{H}_r} - \jump{\tau \phi} + \ifavg{\Omega-\tau} \jump{2\mathcal{A} - \phi} &= 0
    \ , \\
    \ifavg{\xi}\jump{\mathcal{A}} &= 0
    \ , \\
    \jump{\mathbb{H}_t} + \ifavg{\phi-2\mathcal{A}}\jump{\tau} + \ifavg{\Omega} \jump{2\mathcal{A}-\phi} &= 0
    \ , \\
    \dotjump{\tau} - \ifavg{\Sigma-\frac{2}{3}\Theta} \jump{\tau} - \ifavg{\xi}\jump{\mathcal{A}} + \jump{\mathbb{E}} &= 0
    \ , \\
    \jump{\mathcal{E}- \frac{\mu}{3} - \frac{p}{2}} + \frac{1}{2}\ifavg{\phi}\jump{\mathcal{A}-\frac{\phi}{2}}
    \quad & \\
    +\frac{1}{2} \ifavg{\mathcal{A}} \jump{\phi} + \ifavg{4\Omega-\tau}\jump{\tau} &= 0
    \ , \\
    \dotjump{\phi} - \ifavg{\Sigma - \frac{2}{3}\Theta} \jump{\frac{\phi}{2} - \mathcal{A}} - 4\ifavg{\xi}\jump{\tau} &= 0
    \ .
\end{aligned} \end{equation}
The projections which lead to independent CJC-II are reported in Section \ref{SubSec:DivPartofWeylEqn}.

\bibliography{biblio}%the standard way to do bibliographies in overleaf is with "natbib" (not biblatex). The corresponding comment to print the bibliography (with .bib file name "biblio") is given here.

\begin{thebibliography}{75}%
\makeatletter
\providecommand \@ifxundefined [1]{%
 \@ifx{#1\undefined}
}%
\providecommand \@ifnum [1]{%
 \ifnum #1\expandafter \@firstoftwo
 \else \expandafter \@secondoftwo
 \fi
}%
\providecommand \@ifx [1]{%
 \ifx #1\expandafter \@firstoftwo
 \else \expandafter \@secondoftwo
 \fi
}%
\providecommand \natexlab [1]{#1}%
\providecommand \enquote  [1]{``#1''}%
\providecommand \bibnamefont  [1]{#1}%
\providecommand \bibfnamefont [1]{#1}%
\providecommand \citenamefont [1]{#1}%
\providecommand \href@noop [0]{\@secondoftwo}%
\providecommand \href [0]{\begingroup \@sanitize@url \@href}%
\providecommand \@href[1]{\@@startlink{#1}\@@href}%
\providecommand \@@href[1]{\endgroup#1\@@endlink}%
\providecommand \@sanitize@url [0]{\catcode `\\12\catcode `\$12\catcode `\&12\catcode `\#12\catcode `\^12\catcode `\_12\catcode `\%12\relax}%
\providecommand \@@startlink[1]{}%
\providecommand \@@endlink[0]{}%
\providecommand \url  [0]{\begingroup\@sanitize@url \@url }%
\providecommand \@url [1]{\endgroup\@href {#1}{\urlprefix }}%
\providecommand \urlprefix  [0]{URL }%
\providecommand \Eprint [0]{\href }%
\providecommand \doibase [0]{http://dx.doi.org/}%
\providecommand \selectlanguage [0]{\@gobble}%
\providecommand \bibinfo  [0]{\@secondoftwo}%
\providecommand \bibfield  [0]{\@secondoftwo}%
\providecommand \translation [1]{[#1]}%
\providecommand \BibitemOpen [0]{}%
\providecommand \bibitemStop [0]{}%
\providecommand \bibitemNoStop [0]{.\EOS\space}%
\providecommand \EOS [0]{\spacefactor3000\relax}%
\providecommand \BibitemShut  [1]{\csname bibitem#1\endcsname}%
\let\auto@bib@innerbib\@empty
%</preamble>
\bibitem [{\citenamefont {Oppenheimer}\ and\ \citenamefont {Snyder}(1939)}]{OppSnyderColl}%
  \BibitemOpen
  \bibfield  {author} {\bibinfo {author} {\bibfnamefont {J.~R.}\ \bibnamefont {Oppenheimer}}\ and\ \bibinfo {author} {\bibfnamefont {H.}~\bibnamefont {Snyder}},\ }\href {\doibase 10.1103/PhysRev.56.455} {\bibfield  {journal} {\bibinfo  {journal} {Phys. Rev.}\ }\textbf {\bibinfo {volume} {56}},\ \bibinfo {pages} {455} (\bibinfo {year} {1939})}\BibitemShut {NoStop}%
\bibitem [{\citenamefont {Steinbauer}\ and\ \citenamefont {Vickers}(2006)}]{Steinbauer2006}%
  \BibitemOpen
  \bibfield  {author} {\bibinfo {author} {\bibfnamefont {R.}~\bibnamefont {Steinbauer}}\ and\ \bibinfo {author} {\bibfnamefont {J.~A.}\ \bibnamefont {Vickers}},\ }\href {\doibase 10.1088/0264-9381/23/10/r01} {\bibfield  {journal} {\bibinfo  {journal} {Classical and Quantum Gravity}\ }\textbf {\bibinfo {volume} {23}},\ \bibinfo {pages} {R91–R114} (\bibinfo {year} {2006})}\BibitemShut {NoStop}%
\bibitem [{\citenamefont {Lichnerowicz}(1971)}]{lichnerowicz}%
  \BibitemOpen
  \bibfield  {author} {\bibinfo {author} {\bibfnamefont {A.}~\bibnamefont {Lichnerowicz}},\ }\href@noop {} {\bibfield  {journal} {\bibinfo  {journal} {C.R. Acad. Sci.}\ }\textbf {\bibinfo {volume} {273}},\ \bibinfo {pages} {528} (\bibinfo {year} {1971})}\BibitemShut {NoStop}%
\bibitem [{\citenamefont {Taub}(1980)}]{Taub}%
  \BibitemOpen
  \bibfield  {author} {\bibinfo {author} {\bibfnamefont {A.~H.}\ \bibnamefont {Taub}},\ }\href {\doibase 10.1063/1.524568} {\bibfield  {journal} {\bibinfo  {journal} {Journal of Mathematical Physics}\ }\textbf {\bibinfo {volume} {21}},\ \bibinfo {pages} {1423} (\bibinfo {year} {1980})}\BibitemShut {NoStop}%
\bibitem [{\citenamefont {Choquet-Bruhat}\ and\ \citenamefont {DeWitt-Morette}(1982)}]{Bruhat}%
  \BibitemOpen
  \bibfield  {author} {\bibinfo {author} {\bibfnamefont {Y.}~\bibnamefont {Choquet-Bruhat}}\ and\ \bibinfo {author} {\bibfnamefont {C.}~\bibnamefont {DeWitt-Morette}},\ }\href {https://shop.elsevier.com/books/analysis-manifolds-and-physics-revised-edition/choquet-bruhat/978-0-444-86017-0} {\emph {\bibinfo {title} {Analysis, Manifolds and Physics (Revised Edition)}}}\ (\bibinfo  {publisher} {Elsevier Science},\ \bibinfo {year} {1982})\BibitemShut {NoStop}%
\bibitem [{\citenamefont {Reina}\ \emph {et~al.}(2016)\citenamefont {Reina}, \citenamefont {Senovilla},\ and\ \citenamefont {Vera}}]{Senovilla2015}%
  \BibitemOpen
  \bibfield  {author} {\bibinfo {author} {\bibfnamefont {B.}~\bibnamefont {Reina}}, \bibinfo {author} {\bibfnamefont {J.~M.~M.}\ \bibnamefont {Senovilla}}, \ and\ \bibinfo {author} {\bibfnamefont {R.}~\bibnamefont {Vera}},\ }\href {\doibase 10.1088/0264-9381/33/10/105008} {\bibfield  {journal} {\bibinfo  {journal} {Classical and Quantum Gravity}\ }\textbf {\bibinfo {volume} {33}},\ \bibinfo {pages} {105008} (\bibinfo {year} {2016})}\BibitemShut {NoStop}%
\bibitem [{\citenamefont {Mars}\ and\ \citenamefont {Senovilla}(1993)}]{SenovillaMars}%
  \BibitemOpen
  \bibfield  {author} {\bibinfo {author} {\bibfnamefont {M.}~\bibnamefont {Mars}}\ and\ \bibinfo {author} {\bibfnamefont {J.~M.~M.}\ \bibnamefont {Senovilla}},\ }\href {\doibase 10.1088/0264-9381/10/9/026} {\bibfield  {journal} {\bibinfo  {journal} {Classical and Quantum Gravity}\ }\textbf {\bibinfo {volume} {10}},\ \bibinfo {pages} {1865} (\bibinfo {year} {1993})}\BibitemShut {NoStop}%
\bibitem [{\citenamefont {{Marsden}}(1968)}]{Marsden}%
  \BibitemOpen
  \bibfield  {author} {\bibinfo {author} {\bibfnamefont {J.~E.}\ \bibnamefont {{Marsden}}},\ }\href {\doibase 10.1007/BF00251661} {\bibfield  {journal} {\bibinfo  {journal} {Archive for Rational Mechanics and Analysis}\ }\textbf {\bibinfo {volume} {28}},\ \bibinfo {pages} {323} (\bibinfo {year} {1968})}\BibitemShut {NoStop}%
\bibitem [{\citenamefont {Hehl}\ \emph {et~al.}(1976)\citenamefont {Hehl}, \citenamefont {von~der Heyde}, \citenamefont {Kerlick},\ and\ \citenamefont {Nester}}]{Hehl}%
  \BibitemOpen
  \bibfield  {author} {\bibinfo {author} {\bibfnamefont {F.~W.}\ \bibnamefont {Hehl}}, \bibinfo {author} {\bibfnamefont {P.}~\bibnamefont {von~der Heyde}}, \bibinfo {author} {\bibfnamefont {G.~D.}\ \bibnamefont {Kerlick}}, \ and\ \bibinfo {author} {\bibfnamefont {J.~M.}\ \bibnamefont {Nester}},\ }\href {\doibase 10.1103/RevModPhys.48.393} {\bibfield  {journal} {\bibinfo  {journal} {Rev. Mod. Phys.}\ }\textbf {\bibinfo {volume} {48}},\ \bibinfo {pages} {393} (\bibinfo {year} {1976})}\BibitemShut {NoStop}%
\bibitem [{\citenamefont {Hehl}\ \emph {et~al.}(1974)\citenamefont {Hehl}, \citenamefont {von~der Heyde},\ and\ \citenamefont {Kerlick}}]{HehlSingularity}%
  \BibitemOpen
  \bibfield  {author} {\bibinfo {author} {\bibfnamefont {F.~W.}\ \bibnamefont {Hehl}}, \bibinfo {author} {\bibfnamefont {P.}~\bibnamefont {von~der Heyde}}, \ and\ \bibinfo {author} {\bibfnamefont {G.~D.}\ \bibnamefont {Kerlick}},\ }\href {\doibase 10.1103/PhysRevD.10.1066} {\bibfield  {journal} {\bibinfo  {journal} {Phys. Rev. D}\ }\textbf {\bibinfo {volume} {10}},\ \bibinfo {pages} {1066} (\bibinfo {year} {1974})}\BibitemShut {NoStop}%
\bibitem [{\citenamefont {Kibble}(1961)}]{Kibble}%
  \BibitemOpen
  \bibfield  {author} {\bibinfo {author} {\bibfnamefont {T.~W.~B.}\ \bibnamefont {Kibble}},\ }\href {\doibase 10.1063/1.1703702} {\bibfield  {journal} {\bibinfo  {journal} {J. Math. Phys.}\ }\textbf {\bibinfo {volume} {2}},\ \bibinfo {pages} {212} (\bibinfo {year} {1961})}\BibitemShut {NoStop}%
\bibitem [{\citenamefont {Sciama}(1962)}]{Sciama}%
  \BibitemOpen
  \bibfield  {author} {\bibinfo {author} {\bibfnamefont {D.~W.}\ \bibnamefont {Sciama}},\ }\href@noop {} {\emph {\bibinfo {title} {Recent Developments in General Relativity}}}\ (\bibinfo  {publisher} {Pergamon, New York},\ \bibinfo {year} {1962})\BibitemShut {NoStop}%
\bibitem [{\citenamefont {Schwartz}(1966)}]{schwartzBook}%
  \BibitemOpen
  \bibfield  {author} {\bibinfo {author} {\bibfnamefont {L.}~\bibnamefont {Schwartz}},\ }\href {https://www.editions-hermann.fr/livre/theorie-des-distributions-laurent-schwartz} {\emph {\bibinfo {title} {Th{\'e}orie des Distributions}}}\ (\bibinfo  {publisher} {Hermann, Paris},\ \bibinfo {year} {1966})\BibitemShut {NoStop}%
\bibitem [{\citenamefont {Vladimirov}(1979)}]{Vladimirov}%
  \BibitemOpen
  \bibfield  {author} {\bibinfo {author} {\bibfnamefont {V.~S.}\ \bibnamefont {Vladimirov}},\ }\href {https://cir.nii.ac.jp/crid/1130282270916627584} {\emph {\bibinfo {title} {Generalized functions in mathematical physics}}}\ (\bibinfo  {publisher} {Mir},\ \bibinfo {year} {1979})\BibitemShut {NoStop}%
\bibitem [{\citenamefont {Strichartz}(2003)}]{RobertDistributions}%
  \BibitemOpen
  \bibfield  {author} {\bibinfo {author} {\bibfnamefont {R.~S.}\ \bibnamefont {Strichartz}},\ }\href {\doibase 10.1142/5314} {\emph {\bibinfo {title} {A Guide to Distribution Theory and Fourier Transforms}}}\ (\bibinfo  {publisher} {World Scientific},\ \bibinfo {year} {2003})\BibitemShut {NoStop}%
\bibitem [{\citenamefont {Darmois}(1927)}]{Darmois}%
  \BibitemOpen
  \bibfield  {author} {\bibinfo {author} {\bibfnamefont {G.}~\bibnamefont {Darmois}},\ }\href {http://eudml.org/doc/192556} {\emph {\bibinfo {title} {Les équations de la gravitation einsteinienne}}}\ (\bibinfo  {publisher} {Gauthier-Villars},\ \bibinfo {year} {1927})\BibitemShut {NoStop}%
\bibitem [{\citenamefont {Israel}(1966)}]{Israel}%
  \BibitemOpen
  \bibfield  {author} {\bibinfo {author} {\bibfnamefont {W.}~\bibnamefont {Israel}},\ }\href {\doibase 10.1007/BF02710419} {\bibfield  {journal} {\bibinfo  {journal} {Il Nuovo Cimento B (1965-1970)}\ }\textbf {\bibinfo {volume} {44}},\ \bibinfo {pages} {1} (\bibinfo {year} {1966})}\BibitemShut {NoStop}%
\bibitem [{\citenamefont {Poisson}(2004)}]{RelToolkitPoisson}%
  \BibitemOpen
  \bibfield  {author} {\bibinfo {author} {\bibfnamefont {E.}~\bibnamefont {Poisson}},\ }\href@noop {} {\emph {\bibinfo {title} {A Relativist’s Toolkit: The Mathematics of Black-Hole Mechanics}}}\ (\bibinfo  {publisher} {Cambridge University Press},\ \bibinfo {year} {2004})\BibitemShut {NoStop}%
\bibitem [{\citenamefont {Mukohyama}(2001)}]{Mukohyama}%
  \BibitemOpen
  \bibfield  {author} {\bibinfo {author} {\bibfnamefont {S.}~\bibnamefont {Mukohyama}},\ }\href {\doibase 10.1103/PhysRevD.65.024028} {\bibfield  {journal} {\bibinfo  {journal} {Phys. Rev. D}\ }\textbf {\bibinfo {volume} {65}},\ \bibinfo {pages} {024028} (\bibinfo {year} {2001})}\BibitemShut {NoStop}%
\bibitem [{\citenamefont {Ellis}\ and\ \citenamefont {van Elst}(2008)}]{EllisElst1998CM}%
  \BibitemOpen
  \bibfield  {author} {\bibinfo {author} {\bibfnamefont {G.~F.~R.}\ \bibnamefont {Ellis}}\ and\ \bibinfo {author} {\bibfnamefont {H.}~\bibnamefont {van Elst}},\ }\href {https://arxiv.org/abs/gr-qc/9812046} {\enquote {\bibinfo {title} {Cosmological models (carg\`{e}se lectures 1998)},}\ } (\bibinfo {year} {2008}),\ \Eprint {http://arxiv.org/abs/gr-qc/9812046} {arXiv:gr-qc/9812046 [gr-qc]} \BibitemShut {NoStop}%
\bibitem [{\citenamefont {Clarkson}\ and\ \citenamefont {Barrett}(2003)}]{Clarkson_2003}%
  \BibitemOpen
  \bibfield  {author} {\bibinfo {author} {\bibfnamefont {C.~A.}\ \bibnamefont {Clarkson}}\ and\ \bibinfo {author} {\bibfnamefont {R.~K.}\ \bibnamefont {Barrett}},\ }\href {\doibase 10.1088/0264-9381/20/18/301} {\bibfield  {journal} {\bibinfo  {journal} {Classical and Quantum Gravity}\ }\textbf {\bibinfo {volume} {20}},\ \bibinfo {pages} {3855–3884} (\bibinfo {year} {2003})}\BibitemShut {NoStop}%
\bibitem [{\citenamefont {Carloni}\ and\ \citenamefont {Vernieri}(2018{\natexlab{a}})}]{CarloniTOVIso1Fluid}%
  \BibitemOpen
  \bibfield  {author} {\bibinfo {author} {\bibfnamefont {S.}~\bibnamefont {Carloni}}\ and\ \bibinfo {author} {\bibfnamefont {D.}~\bibnamefont {Vernieri}},\ }\href {\doibase 10.1103/PhysRevD.97.124056} {\bibfield  {journal} {\bibinfo  {journal} {Phys. Rev. D}\ }\textbf {\bibinfo {volume} {97}},\ \bibinfo {pages} {124056} (\bibinfo {year} {2018}{\natexlab{a}})}\BibitemShut {NoStop}%
\bibitem [{\citenamefont {Carloni}\ and\ \citenamefont {Vernieri}(2018{\natexlab{b}})}]{CarloniTOVAniso1Fluid}%
  \BibitemOpen
  \bibfield  {author} {\bibinfo {author} {\bibfnamefont {S.}~\bibnamefont {Carloni}}\ and\ \bibinfo {author} {\bibfnamefont {D.}~\bibnamefont {Vernieri}},\ }\href {\doibase 10.1103/PhysRevD.97.124057} {\bibfield  {journal} {\bibinfo  {journal} {Phys. Rev. D}\ }\textbf {\bibinfo {volume} {97}},\ \bibinfo {pages} {124057} (\bibinfo {year} {2018}{\natexlab{b}})},\ \Eprint {http://arxiv.org/abs/1709.03996} {arXiv:1709.03996 [gr-qc]} \BibitemShut {NoStop}%
\bibitem [{\citenamefont {Naidu}\ \emph {et~al.}(2022)\citenamefont {Naidu}, \citenamefont {Carloni},\ and\ \citenamefont {Dunsby}}]{CarloniNaidu}%
  \BibitemOpen
  \bibfield  {author} {\bibinfo {author} {\bibfnamefont {N.~F.}\ \bibnamefont {Naidu}}, \bibinfo {author} {\bibfnamefont {S.}~\bibnamefont {Carloni}}, \ and\ \bibinfo {author} {\bibfnamefont {P.}~\bibnamefont {Dunsby}},\ }\href {\doibase 10.1103/PhysRevD.106.124023} {\bibfield  {journal} {\bibinfo  {journal} {Phys. Rev. D}\ }\textbf {\bibinfo {volume} {106}},\ \bibinfo {pages} {124023} (\bibinfo {year} {2022})},\ \Eprint {http://arxiv.org/abs/2210.06867} {arXiv:2210.06867 [gr-qc]} \BibitemShut {NoStop}%
\bibitem [{\citenamefont {Betschart}\ and\ \citenamefont {Clarkson}(2004)}]{ClarksonBetschart_2004}%
  \BibitemOpen
  \bibfield  {author} {\bibinfo {author} {\bibfnamefont {G.}~\bibnamefont {Betschart}}\ and\ \bibinfo {author} {\bibfnamefont {C.~A.}\ \bibnamefont {Clarkson}},\ }\href {\doibase 10.1088/0264-9381/21/23/018} {\bibfield  {journal} {\bibinfo  {journal} {Classical and Quantum Gravity}\ }\textbf {\bibinfo {volume} {21}},\ \bibinfo {pages} {5587–5607} (\bibinfo {year} {2004})}\BibitemShut {NoStop}%
\bibitem [{\citenamefont {Clarkson}(2007)}]{Clarkson_2007}%
  \BibitemOpen
  \bibfield  {author} {\bibinfo {author} {\bibfnamefont {C.}~\bibnamefont {Clarkson}},\ }\href {\doibase 10.1103/physrevd.76.104034} {\bibfield  {journal} {\bibinfo  {journal} {Physical Review D}\ }\textbf {\bibinfo {volume} {76}} (\bibinfo {year} {2007}),\ 10.1103/physrevd.76.104034}\BibitemShut {NoStop}%
\bibitem [{\citenamefont {Luz}\ and\ \citenamefont {Carloni}(2024{\natexlab{a}})}]{LuzCarloniGaugeInvPert}%
  \BibitemOpen
  \bibfield  {author} {\bibinfo {author} {\bibfnamefont {P.}~\bibnamefont {Luz}}\ and\ \bibinfo {author} {\bibfnamefont {S.}~\bibnamefont {Carloni}},\ }\href {\doibase 10.1088/1361-6382/ad8a14} {\bibfield  {journal} {\bibinfo  {journal} {Classical and Quantum Gravity}\ }\textbf {\bibinfo {volume} {41}},\ \bibinfo {pages} {235012} (\bibinfo {year} {2024}{\natexlab{a}})}\BibitemShut {NoStop}%
\bibitem [{\citenamefont {Luz}\ and\ \citenamefont {Carloni}(2024{\natexlab{b}})}]{LuzCarloniComovingPert}%
  \BibitemOpen
  \bibfield  {author} {\bibinfo {author} {\bibfnamefont {P.}~\bibnamefont {Luz}}\ and\ \bibinfo {author} {\bibfnamefont {S.}~\bibnamefont {Carloni}},\ }\href {\doibase 10.1103/PhysRevD.110.084055} {\bibfield  {journal} {\bibinfo  {journal} {Phys. Rev. D}\ }\textbf {\bibinfo {volume} {110}},\ \bibinfo {pages} {084055} (\bibinfo {year} {2024}{\natexlab{b}})},\ \Eprint {http://arxiv.org/abs/2405.10359} {arXiv:2405.10359 [gr-qc]} \BibitemShut {NoStop}%
\bibitem [{\citenamefont {Luz}\ and\ \citenamefont {Carloni}(2024{\natexlab{c}})}]{LuzCarloniAdiabaticPert}%
  \BibitemOpen
  \bibfield  {author} {\bibinfo {author} {\bibfnamefont {P.}~\bibnamefont {Luz}}\ and\ \bibinfo {author} {\bibfnamefont {S.}~\bibnamefont {Carloni}},\ }\href {\doibase 10.1103/PhysRevD.110.084054} {\bibfield  {journal} {\bibinfo  {journal} {Phys. Rev. D}\ }\textbf {\bibinfo {volume} {110}},\ \bibinfo {pages} {084054} (\bibinfo {year} {2024}{\natexlab{c}})}\BibitemShut {NoStop}%
\bibitem [{\citenamefont {Bradley}\ \emph {et~al.}(2012)\citenamefont {Bradley}, \citenamefont {Dunsby}, \citenamefont {Forsberg},\ and\ \citenamefont {Keresztes}}]{Bradley1}%
  \BibitemOpen
  \bibfield  {author} {\bibinfo {author} {\bibfnamefont {M.}~\bibnamefont {Bradley}}, \bibinfo {author} {\bibfnamefont {P.~K.~S.}\ \bibnamefont {Dunsby}}, \bibinfo {author} {\bibfnamefont {M.}~\bibnamefont {Forsberg}}, \ and\ \bibinfo {author} {\bibfnamefont {Z.}~\bibnamefont {Keresztes}},\ }\href {\doibase 10.1088/0264-9381/29/9/095023} {\bibfield  {journal} {\bibinfo  {journal} {Classical and Quantum Gravity}\ }\textbf {\bibinfo {volume} {29}},\ \bibinfo {pages} {095023} (\bibinfo {year} {2012})}\BibitemShut {NoStop}%
\bibitem [{\citenamefont {T\"ornkvist}\ and\ \citenamefont {Bradley}(2019)}]{Bradley2}%
  \BibitemOpen
  \bibfield  {author} {\bibinfo {author} {\bibfnamefont {R.}~\bibnamefont {T\"ornkvist}}\ and\ \bibinfo {author} {\bibfnamefont {M.}~\bibnamefont {Bradley}},\ }\href {\doibase 10.1103/PhysRevD.100.124043} {\bibfield  {journal} {\bibinfo  {journal} {Phys. Rev. D}\ }\textbf {\bibinfo {volume} {100}},\ \bibinfo {pages} {124043} (\bibinfo {year} {2019})}\BibitemShut {NoStop}%
\bibitem [{\citenamefont {Rosa}\ and\ \citenamefont {Carloni}(2024)}]{RosaCarloni}%
  \BibitemOpen
  \bibfield  {author} {\bibinfo {author} {\bibfnamefont {J.~L.}\ \bibnamefont {Rosa}}\ and\ \bibinfo {author} {\bibfnamefont {S.}~\bibnamefont {Carloni}},\ }\href {\doibase 10.1103/physrevd.109.104037} {\bibfield  {journal} {\bibinfo  {journal} {Physical Review D}\ }\textbf {\bibinfo {volume} {109}} (\bibinfo {year} {2024}),\ 10.1103/physrevd.109.104037}\BibitemShut {NoStop}%
\bibitem [{\citenamefont {Khambule}\ \emph {et~al.}(2021)\citenamefont {Khambule}, \citenamefont {Goswami},\ and\ \citenamefont {Maharaj}}]{Khambule}%
  \BibitemOpen
  \bibfield  {author} {\bibinfo {author} {\bibfnamefont {P.~N.}\ \bibnamefont {Khambule}}, \bibinfo {author} {\bibfnamefont {R.}~\bibnamefont {Goswami}}, \ and\ \bibinfo {author} {\bibfnamefont {S.~D.}\ \bibnamefont {Maharaj}},\ }\href {\doibase 10.1088/1361-6382/abe2dd} {\bibfield  {journal} {\bibinfo  {journal} {Classical and Quantum Gravity}\ }\textbf {\bibinfo {volume} {38}},\ \bibinfo {pages} {075006} (\bibinfo {year} {2021})}\BibitemShut {NoStop}%
\bibitem [{\citenamefont {Arkuszewski}\ \emph {et~al.}(1975)\citenamefont {Arkuszewski}, \citenamefont {Kopczyński},\ and\ \citenamefont {Ponomariev}}]{Arkuszewski}%
  \BibitemOpen
  \bibfield  {author} {\bibinfo {author} {\bibfnamefont {W.}~\bibnamefont {Arkuszewski}}, \bibinfo {author} {\bibfnamefont {W.}~\bibnamefont {Kopczyński}}, \ and\ \bibinfo {author} {\bibfnamefont {V.~N.}\ \bibnamefont {Ponomariev}},\ }\href {\doibase 10.1007/BF01629248} {\bibfield  {journal} {\bibinfo  {journal} {Communications in Mathematical Physics}\ }\textbf {\bibinfo {volume} {45}},\ \bibinfo {pages} {183} (\bibinfo {year} {1975})}\BibitemShut {NoStop}%
\bibitem [{\citenamefont {Bressange}(2000)}]{Bressange}%
  \BibitemOpen
  \bibfield  {author} {\bibinfo {author} {\bibfnamefont {G.~F.}\ \bibnamefont {Bressange}},\ }\href {\doibase 10.1088/0264-9381/17/13/304} {\bibfield  {journal} {\bibinfo  {journal} {Classical and Quantum Gravity}\ }\textbf {\bibinfo {volume} {17}},\ \bibinfo {pages} {2509} (\bibinfo {year} {2000})}\BibitemShut {NoStop}%
\bibitem [{\citenamefont {Potekhin}(2010)}]{PotekhinPhysicsofNS}%
  \BibitemOpen
  \bibfield  {author} {\bibinfo {author} {\bibfnamefont {A.~Y.}\ \bibnamefont {Potekhin}},\ }\href {https://api.semanticscholar.org/CorpusID:119231427} {\bibfield  {journal} {\bibinfo  {journal} {Physics-Uspekhi}\ }\textbf {\bibinfo {volume} {53}},\ \bibinfo {pages} {1235 } (\bibinfo {year} {2010})}\BibitemShut {NoStop}%
\bibitem [{\citenamefont {Cai}\ \emph {et~al.}(2016)\citenamefont {Cai}, \citenamefont {Capozziello}, \citenamefont {De~Laurentis},\ and\ \citenamefont {Saridakis}}]{CaiCapozziello-f(T)gravity}%
  \BibitemOpen
  \bibfield  {author} {\bibinfo {author} {\bibfnamefont {Y.-F.}\ \bibnamefont {Cai}}, \bibinfo {author} {\bibfnamefont {S.}~\bibnamefont {Capozziello}}, \bibinfo {author} {\bibfnamefont {M.}~\bibnamefont {De~Laurentis}}, \ and\ \bibinfo {author} {\bibfnamefont {E.~N.}\ \bibnamefont {Saridakis}},\ }\href {\doibase 10.1088/0034-4885/79/10/106901} {\bibfield  {journal} {\bibinfo  {journal} {Rept. Prog. Phys.}\ }\textbf {\bibinfo {volume} {79}},\ \bibinfo {pages} {106901} (\bibinfo {year} {2016})},\ \Eprint {http://arxiv.org/abs/1511.07586} {arXiv:1511.07586 [gr-qc]} \BibitemShut {NoStop}%
\bibitem [{\citenamefont {Kopczyński}(1973)}]{Kopczynski-Singularity}%
  \BibitemOpen
  \bibfield  {author} {\bibinfo {author} {\bibfnamefont {W.}~\bibnamefont {Kopczyński}},\ }\href {\doibase https://doi.org/10.1016/0375-9601(73)90546-X} {\bibfield  {journal} {\bibinfo  {journal} {Physics Letters A}\ }\textbf {\bibinfo {volume} {43}},\ \bibinfo {pages} {63} (\bibinfo {year} {1973})}\BibitemShut {NoStop}%
\bibitem [{\citenamefont {Weyssenhoff}\ and\ \citenamefont {Raabe}(1947)}]{WeyssenhoffRaabe}%
  \BibitemOpen
  \bibfield  {author} {\bibinfo {author} {\bibfnamefont {J.}~\bibnamefont {Weyssenhoff}}\ and\ \bibinfo {author} {\bibfnamefont {A.}~\bibnamefont {Raabe}},\ }\href {https://www.actaphys.uj.edu.pl/fulltext?series=T&vol=9&no=1&page=7} {\bibfield  {journal} {\bibinfo  {journal} {Acta Phys. Polon.}\ }\textbf {\bibinfo {volume} {9}},\ \bibinfo {pages} {7} (\bibinfo {year} {1947})}\BibitemShut {NoStop}%
\bibitem [{\citenamefont {Obukhov}\ and\ \citenamefont {Korotky}(1987)}]{Korotky}%
  \BibitemOpen
  \bibfield  {author} {\bibinfo {author} {\bibfnamefont {Y.~N.}\ \bibnamefont {Obukhov}}\ and\ \bibinfo {author} {\bibfnamefont {V.~A.}\ \bibnamefont {Korotky}},\ }\href {\doibase 10.1088/0264-9381/4/6/021} {\bibfield  {journal} {\bibinfo  {journal} {Classical and Quantum Gravity}\ }\textbf {\bibinfo {volume} {4}},\ \bibinfo {pages} {1633} (\bibinfo {year} {1987})}\BibitemShut {NoStop}%
\bibitem [{\citenamefont {Halbwachs}(1960)}]{Halbwachs}%
  \BibitemOpen
  \bibfield  {author} {\bibinfo {author} {\bibfnamefont {F.}~\bibnamefont {Halbwachs}},\ }\href {\doibase 10.1016/0029-5582(60)90446-6} {\bibfield  {journal} {\bibinfo  {journal} {Nuclear Physics}\ }\textbf {\bibinfo {volume} {18}},\ \bibinfo {pages} {716} (\bibinfo {year} {1960})}\BibitemShut {NoStop}%
\bibitem [{\citenamefont {Ellis}(1967)}]{Ellis1967}%
  \BibitemOpen
  \bibfield  {author} {\bibinfo {author} {\bibfnamefont {G.~F.~R.}\ \bibnamefont {Ellis}},\ }\href {\doibase 10.1063/1.1705331} {\bibfield  {journal} {\bibinfo  {journal} {Journal of Mathematical Physics}\ }\textbf {\bibinfo {volume} {8}},\ \bibinfo {pages} {1171} (\bibinfo {year} {1967})}\BibitemShut {NoStop}%
\bibitem [{\citenamefont {Stewart}\ and\ \citenamefont {Ellis}(1968)}]{EllisStewart1968}%
  \BibitemOpen
  \bibfield  {author} {\bibinfo {author} {\bibfnamefont {J.~M.}\ \bibnamefont {Stewart}}\ and\ \bibinfo {author} {\bibfnamefont {G.~F.~R.}\ \bibnamefont {Ellis}},\ }\href {\doibase 10.1063/1.1664679} {\bibfield  {journal} {\bibinfo  {journal} {Journal of Mathematical Physics}\ }\textbf {\bibinfo {volume} {9}},\ \bibinfo {pages} {1072} (\bibinfo {year} {1968})}\BibitemShut {NoStop}%
\bibitem [{\citenamefont {Elst}\ and\ \citenamefont {Ellis}(1996)}]{EllisElst}%
  \BibitemOpen
  \bibfield  {author} {\bibinfo {author} {\bibfnamefont {H.~v.}\ \bibnamefont {Elst}}\ and\ \bibinfo {author} {\bibfnamefont {G.~F.~R.}\ \bibnamefont {Ellis}},\ }\href {\doibase 10.1088/0264-9381/13/5/023} {\bibfield  {journal} {\bibinfo  {journal} {Classical and Quantum Gravity}\ }\textbf {\bibinfo {volume} {13}},\ \bibinfo {pages} {1099–1127} (\bibinfo {year} {1996})}\BibitemShut {NoStop}%
\bibitem [{\citenamefont {Agarwal}\ and\ \citenamefont {Carloni}(2025)}]{UjjwalSante}%
  \BibitemOpen
  \bibfield  {author} {\bibinfo {author} {\bibfnamefont {U.}~\bibnamefont {Agarwal}}\ and\ \bibinfo {author} {\bibfnamefont {S.}~\bibnamefont {Carloni}},\ }\href {\doibase 10.1103/9m6d-rgv3} {\bibfield  {journal} {\bibinfo  {journal} {Phys. Rev. D}\ }\textbf {\bibinfo {volume} {112}},\ \bibinfo {pages} {064075} (\bibinfo {year} {2025})}\BibitemShut {NoStop}%
\bibitem [{\citenamefont {Luz}\ and\ \citenamefont {Carloni}(2019)}]{CarloniLuz2019}%
  \BibitemOpen
  \bibfield  {author} {\bibinfo {author} {\bibfnamefont {P.}~\bibnamefont {Luz}}\ and\ \bibinfo {author} {\bibfnamefont {S.}~\bibnamefont {Carloni}},\ }\href {\doibase 10.1103/physrevd.100.084037} {\bibfield  {journal} {\bibinfo  {journal} {Physical Review D}\ }\textbf {\bibinfo {volume} {100}} (\bibinfo {year} {2019}),\ 10.1103/physrevd.100.084037}\BibitemShut {NoStop}%
\bibitem [{\citenamefont {Clarke}\ and\ \citenamefont {Dray}(1987)}]{Clarke_1987}%
  \BibitemOpen
  \bibfield  {author} {\bibinfo {author} {\bibfnamefont {C.~J.~S.}\ \bibnamefont {Clarke}}\ and\ \bibinfo {author} {\bibfnamefont {T.}~\bibnamefont {Dray}},\ }\href {\doibase 10.1088/0264-9381/4/2/010} {\bibfield  {journal} {\bibinfo  {journal} {Classical and Quantum Gravity}\ }\textbf {\bibinfo {volume} {4}},\ \bibinfo {pages} {265} (\bibinfo {year} {1987})}\BibitemShut {NoStop}%
\bibitem [{\citenamefont {Colombeau}(1992)}]{Colombeau}%
  \BibitemOpen
  \bibfield  {author} {\bibinfo {author} {\bibfnamefont {J.~F.}\ \bibnamefont {Colombeau}},\ }\href@noop {} {\emph {\bibinfo {title} {Multiplication of Distributions - A tool in Mathematics, Numerical Engineering and Theoretical Physics, Lect. Notes in Math. 1532}}}\ (\bibinfo  {publisher} {Springer-Verlag, Berlin},\ \bibinfo {year} {1992})\BibitemShut {NoStop}%
\bibitem [{\citenamefont {Gsponer}(2008)}]{Gsponer}%
  \BibitemOpen
  \bibfield  {author} {\bibinfo {author} {\bibfnamefont {A.}~\bibnamefont {Gsponer}},\ }\href {\doibase 10.1088/0143-0807/30/1/011} {\bibfield  {journal} {\bibinfo  {journal} {European Journal of Physics}\ }\textbf {\bibinfo {volume} {30}},\ \bibinfo {pages} {109} (\bibinfo {year} {2008})}\BibitemShut {NoStop}%
\bibitem [{\citenamefont {Kunzinger}\ and\ \citenamefont {Steinbauer}(2001)}]{Steinbauer2001}%
  \BibitemOpen
  \bibfield  {author} {\bibinfo {author} {\bibfnamefont {M.}~\bibnamefont {Kunzinger}}\ and\ \bibinfo {author} {\bibfnamefont {R.}~\bibnamefont {Steinbauer}},\ }\href@noop {} {\  (\bibinfo {year} {2001})},\ \Eprint {http://arxiv.org/abs/math/0102019} {arXiv:math/0102019} \BibitemShut {NoStop}%
\bibitem [{\citenamefont {Penrose}(1972)}]{Penrose1972}%
  \BibitemOpen
  \bibfield  {author} {\bibinfo {author} {\bibfnamefont {R.}~\bibnamefont {Penrose}},\ }\enquote {\bibinfo {title} {{The geometry of impulsive gravitational waves}},}\ in\ \href@noop {} {\emph {\bibinfo {booktitle} {{General relativity}: {Papers in honour of J.L. Synge}}}},\ \bibinfo {editor} {edited by\ \bibinfo {editor} {\bibfnamefont {L.}~\bibnamefont {O'Raifeartaigh}}}\ (\bibinfo {year} {1972})\ pp.\ \bibinfo {pages} {101--115}\BibitemShut {NoStop}%
\bibitem [{\citenamefont {Griffiths}\ and\ \citenamefont {Podolský}(2009)}]{PodolskyBook}%
  \BibitemOpen
  \bibfield  {author} {\bibinfo {author} {\bibfnamefont {J.~B.}\ \bibnamefont {Griffiths}}\ and\ \bibinfo {author} {\bibfnamefont {J.}~\bibnamefont {Podolský}},\ }\href@noop {} {\emph {\bibinfo {title} {Exact Space-Times in Einstein’s General Relativity}}},\ Cambridge Monographs on Mathematical Physics\ (\bibinfo  {publisher} {Cambridge University Press},\ \bibinfo {year} {2009})\BibitemShut {NoStop}%
\bibitem [{\citenamefont {Podolský}\ and\ \citenamefont {Veselý}(1998)}]{PodolskyPaper1}%
  \BibitemOpen
  \bibfield  {author} {\bibinfo {author} {\bibfnamefont {J.}~\bibnamefont {Podolský}}\ and\ \bibinfo {author} {\bibfnamefont {K.}~\bibnamefont {Veselý}},\ }\href {\doibase https://doi.org/10.1016/S0375-9601(98)00162-5} {\bibfield  {journal} {\bibinfo  {journal} {Physics Letters A}\ }\textbf {\bibinfo {volume} {241}},\ \bibinfo {pages} {145} (\bibinfo {year} {1998})}\BibitemShut {NoStop}%
\bibitem [{\citenamefont {Podolsky}\ and\ \citenamefont {Griffiths}(1999)}]{PodolskyPaper2}%
  \BibitemOpen
  \bibfield  {author} {\bibinfo {author} {\bibfnamefont {J.}~\bibnamefont {Podolsky}}\ and\ \bibinfo {author} {\bibfnamefont {J.~B.}\ \bibnamefont {Griffiths}},\ }\href {\doibase 10.1088/0264-9381/16/9/311} {\bibfield  {journal} {\bibinfo  {journal} {Class. Quant. Grav.}\ }\textbf {\bibinfo {volume} {16}},\ \bibinfo {pages} {2937} (\bibinfo {year} {1999})},\ \Eprint {http://arxiv.org/abs/gr-qc/9907022} {arXiv:gr-qc/9907022} \BibitemShut {NoStop}%
\bibitem [{\citenamefont {Podolsk\'y}\ \emph {et~al.}(2017)\citenamefont {Podolsk\'y}, \citenamefont {\ifmmode~\check{S}\else \v{S}\fi{}varc}, \citenamefont {Steinbauer},\ and\ \citenamefont {S\"amann}}]{PodolskyPaper3}%
  \BibitemOpen
  \bibfield  {author} {\bibinfo {author} {\bibfnamefont {J.}~\bibnamefont {Podolsk\'y}}, \bibinfo {author} {\bibfnamefont {R.}~\bibnamefont {\ifmmode~\check{S}\else \v{S}\fi{}varc}}, \bibinfo {author} {\bibfnamefont {R.}~\bibnamefont {Steinbauer}}, \ and\ \bibinfo {author} {\bibfnamefont {C.}~\bibnamefont {S\"amann}},\ }\href {\doibase 10.1103/PhysRevD.96.064043} {\bibfield  {journal} {\bibinfo  {journal} {Phys. Rev. D}\ }\textbf {\bibinfo {volume} {96}},\ \bibinfo {pages} {064043} (\bibinfo {year} {2017})}\BibitemShut {NoStop}%
\bibitem [{\citenamefont {Geroch}\ and\ \citenamefont {Traschen}(1987)}]{Geroch}%
  \BibitemOpen
  \bibfield  {author} {\bibinfo {author} {\bibfnamefont {R.}~\bibnamefont {Geroch}}\ and\ \bibinfo {author} {\bibfnamefont {J.}~\bibnamefont {Traschen}},\ }\href {\doibase 10.1103/PhysRevD.36.1017} {\bibfield  {journal} {\bibinfo  {journal} {Phys. Rev. D}\ }\textbf {\bibinfo {volume} {36}},\ \bibinfo {pages} {1017} (\bibinfo {year} {1987})}\BibitemShut {NoStop}%
\bibitem [{\citenamefont {Steinbauer}(2008)}]{SteinbauerComparison}%
  \BibitemOpen
  \bibfield  {author} {\bibinfo {author} {\bibfnamefont {R.}~\bibnamefont {Steinbauer}},\ }\href@noop {} {\bibfield  {journal} {\bibinfo  {journal} {Novi Sad J. Math}\ }\textbf {\bibinfo {volume} {38}},\ \bibinfo {pages} {189} (\bibinfo {year} {2008})},\ \Eprint {http://arxiv.org/abs/0812.0173} {arXiv:0812.0173 [gr-qc]} \BibitemShut {NoStop}%
\bibitem [{\citenamefont {Ramirez}\ and\ \citenamefont {Mart\'{\i}nez}(2025)}]{Ramirez}%
  \BibitemOpen
  \bibfield  {author} {\bibinfo {author} {\bibfnamefont {M.~A.}\ \bibnamefont {Ramirez}}\ and\ \bibinfo {author} {\bibfnamefont {C.}~\bibnamefont {Mart\'{\i}nez}},\ }\href {\doibase 10.1103/m5lt-l2jg} {\bibfield  {journal} {\bibinfo  {journal} {Phys. Rev. D}\ }\textbf {\bibinfo {volume} {112}},\ \bibinfo {pages} {024007} (\bibinfo {year} {2025})}\BibitemShut {NoStop}%
\bibitem [{\citenamefont {Huber}(2020)}]{Huber}%
  \BibitemOpen
  \bibfield  {author} {\bibinfo {author} {\bibfnamefont {A.}~\bibnamefont {Huber}},\ }\href {\doibase https://doi.org/10.1140/epjc/s10052-020-08714-0} {\enquote {\bibinfo {title} {Junction conditions and local spacetimes in general relativity},}\ } (\bibinfo {year} {2020}),\ \Eprint {http://arxiv.org/abs/1908.08735} {arXiv:1908.08735 [gr-qc]} \BibitemShut {NoStop}%
\bibitem [{\citenamefont {Wald}(1984)}]{WaldBook}%
  \BibitemOpen
  \bibfield  {author} {\bibinfo {author} {\bibfnamefont {R.~M.}\ \bibnamefont {Wald}},\ }\href@noop {} {\emph {\bibinfo {title} {{General Relativity}}}}\ (\bibinfo  {publisher} {The University od Chicago Press},\ \bibinfo {year} {1984})\BibitemShut {NoStop}%
\bibitem [{\citenamefont {Schwartz}(1954)}]{schwartz1954}%
  \BibitemOpen
  \bibfield  {author} {\bibinfo {author} {\bibfnamefont {L.}~\bibnamefont {Schwartz}},\ }\href@noop {} {\bibfield  {journal} {\bibinfo  {journal} {Sur l'impossibilit{\'e} de la multiplicaciones des distribuciones, Comptes Rendus Acad. Sci. Par{\'\i}s}\ }\textbf {\bibinfo {volume} {239}},\ \bibinfo {pages} {847} (\bibinfo {year} {1954})}\BibitemShut {NoStop}%
\bibitem [{\citenamefont {Kobayashi}\ and\ \citenamefont {Nomizu}(1963)}]{NomizuInterscience}%
  \BibitemOpen
  \bibfield  {author} {\bibinfo {author} {\bibfnamefont {S.}~\bibnamefont {Kobayashi}}\ and\ \bibinfo {author} {\bibfnamefont {K.}~\bibnamefont {Nomizu}},\ }\href@noop {} {{\selectlanguage {English}\emph {\bibinfo {title} {Foundations of differential geometry. {I}}}}},\ \bibinfo {series} {Intersci. Tracts Pure Appl. Math.}, Vol.~\bibinfo {volume} {15}\ (\bibinfo  {publisher} {Interscience Publishers, New York, NY},\ \bibinfo {year} {1963})\BibitemShut {NoStop}%
\bibitem [{\citenamefont {Lee}(2003)}]{JohnLeeBook}%
  \BibitemOpen
  \bibfield  {author} {\bibinfo {author} {\bibfnamefont {J.~M.}\ \bibnamefont {Lee}},\ }\href {https://link.springer.com/book/10.1007/978-1-4419-9982-5} {\emph {\bibinfo {title} {Introduction to Smooth Manifolds}}}\ (\bibinfo  {publisher} {Springer New York, NY},\ \bibinfo {year} {2003})\BibitemShut {NoStop}%
\bibitem [{\citenamefont {Nicolaescu}(2007)}]{Nicolaescu}%
  \BibitemOpen
  \bibfield  {author} {\bibinfo {author} {\bibfnamefont {L.~I.}\ \bibnamefont {Nicolaescu}},\ }\href {\doibase 10.1142/6528} {\emph {\bibinfo {title} {Lectures on the Geometry of Manifolds}}},\ \bibinfo {edition} {2nd}\ ed.\ (\bibinfo  {publisher} {World Scientific},\ \bibinfo {year} {2007})\BibitemShut {NoStop}%
\bibitem [{\citenamefont {Schutz}(1980)}]{SchutzBook}%
  \BibitemOpen
  \bibfield  {author} {\bibinfo {author} {\bibfnamefont {B.~F.}\ \bibnamefont {Schutz}},\ }\href@noop {} {\emph {\bibinfo {title} {Geometrical Methods of Mathematical Physics}}}\ (\bibinfo  {publisher} {Cambridge University Press},\ \bibinfo {year} {1980})\BibitemShut {NoStop}%
\bibitem [{\citenamefont {Luz}\ and\ \citenamefont {Mena}(2020)}]{LuzHypersurface}%
  \BibitemOpen
  \bibfield  {author} {\bibinfo {author} {\bibfnamefont {P.}~\bibnamefont {Luz}}\ and\ \bibinfo {author} {\bibfnamefont {F.~C.}\ \bibnamefont {Mena}},\ }\href {\doibase 10.1063/1.5126220} {\bibfield  {journal} {\bibinfo  {journal} {J. Math. Phys.}\ }\textbf {\bibinfo {volume} {61}},\ \bibinfo {pages} {012502} (\bibinfo {year} {2020})},\ \Eprint {http://arxiv.org/abs/1909.00018} {arXiv:1909.00018 [gr-qc]} \BibitemShut {NoStop}%
\bibitem [{\citenamefont {Hawking}\ and\ \citenamefont {Ellis}(1973)}]{HawkingEllisBook}%
  \BibitemOpen
  \bibfield  {author} {\bibinfo {author} {\bibfnamefont {S.~W.}\ \bibnamefont {Hawking}}\ and\ \bibinfo {author} {\bibfnamefont {G.~F.~R.}\ \bibnamefont {Ellis}},\ }\href@noop {} {\emph {\bibinfo {title} {The Large Scale Structure of Space-Time}}},\ Cambridge Monographs on Mathematical Physics\ (\bibinfo  {publisher} {Cambridge University Press},\ \bibinfo {year} {1973})\BibitemShut {NoStop}%
\bibitem [{\citenamefont {Kundt}\ and\ \citenamefont {Tr{\"u}mper}(2016)}]{TrumperKundt}%
  \BibitemOpen
  \bibfield  {author} {\bibinfo {author} {\bibfnamefont {W.}~\bibnamefont {Kundt}}\ and\ \bibinfo {author} {\bibfnamefont {M.}~\bibnamefont {Tr{\"u}mper}},\ }\href {\doibase 10.1007/s10714-015-2009-y} {\bibfield  {journal} {\bibinfo  {journal} {Gen. Rel. Grav.}\ }\textbf {\bibinfo {volume} {48}},\ \bibinfo {pages} {44} (\bibinfo {year} {2016})}\BibitemShut {NoStop}%
\bibitem [{\citenamefont {Capozziello}\ \emph {et~al.}(2001)\citenamefont {Capozziello}, \citenamefont {Lambiase},\ and\ \citenamefont {StornaioloI}}]{Capozziello}%
  \BibitemOpen
  \bibfield  {author} {\bibinfo {author} {\bibfnamefont {S.}~\bibnamefont {Capozziello}}, \bibinfo {author} {\bibfnamefont {G.}~\bibnamefont {Lambiase}}, \ and\ \bibinfo {author} {\bibfnamefont {C.}~\bibnamefont {StornaioloI}},\ }\href {\doibase 10.1002/andp.20015130803} {\bibfield  {journal} {\bibinfo  {journal} {Annalen der Physik}\ }\textbf {\bibinfo {volume} {513}},\ \bibinfo {pages} {713–727} (\bibinfo {year} {2001})}\BibitemShut {NoStop}%
\bibitem [{\citenamefont {Ellis}\ \emph {et~al.}(1990)\citenamefont {Ellis}, \citenamefont {Bruni},\ and\ \citenamefont {Hwang}}]{EllisBruniHwang}%
  \BibitemOpen
  \bibfield  {author} {\bibinfo {author} {\bibfnamefont {G.~F.~R.}\ \bibnamefont {Ellis}}, \bibinfo {author} {\bibfnamefont {M.}~\bibnamefont {Bruni}}, \ and\ \bibinfo {author} {\bibfnamefont {J.}~\bibnamefont {Hwang}},\ }\href {\doibase 10.1103/PhysRevD.42.1035} {\bibfield  {journal} {\bibinfo  {journal} {Phys. Rev. D}\ }\textbf {\bibinfo {volume} {42}},\ \bibinfo {pages} {1035} (\bibinfo {year} {1990})}\BibitemShut {NoStop}%
\bibitem [{\citenamefont {Mathisson}(1937)}]{Mathisson1937}%
  \BibitemOpen
  \bibfield  {author} {\bibinfo {author} {\bibfnamefont {M.}~\bibnamefont {Mathisson}},\ }\href@noop {} {\bibfield  {journal} {\bibinfo  {journal} {Acta Phys. Polon.}\ }\textbf {\bibinfo {volume} {6}},\ \bibinfo {pages} {163} (\bibinfo {year} {1937})}\BibitemShut {NoStop}%
\bibitem [{\citenamefont {{Frenkel}}(1926)}]{Frenkel}%
  \BibitemOpen
  \bibfield  {author} {\bibinfo {author} {\bibfnamefont {J.}~\bibnamefont {{Frenkel}}},\ }\href {\doibase 10.1007/BF01397099} {\bibfield  {journal} {\bibinfo  {journal} {Zeitschrift fur Physik}\ }\textbf {\bibinfo {volume} {37}},\ \bibinfo {pages} {243} (\bibinfo {year} {1926})}\BibitemShut {NoStop}%
\bibitem [{\citenamefont {Papapetrou}\ and\ \citenamefont {Peierls}(1951)}]{Paparetrou}%
  \BibitemOpen
  \bibfield  {author} {\bibinfo {author} {\bibfnamefont {A.}~\bibnamefont {Papapetrou}}\ and\ \bibinfo {author} {\bibfnamefont {R.~E.}\ \bibnamefont {Peierls}},\ }\href {\doibase 10.1098/rspa.1951.0200} {\bibfield  {journal} {\bibinfo  {journal} {Proceedings of the Royal Society of London. Series A. Mathematical and Physical Sciences}\ }\textbf {\bibinfo {volume} {209}},\ \bibinfo {pages} {248} (\bibinfo {year} {1951})}\BibitemShut {NoStop}%
\bibitem [{\citenamefont {{Buchdahl}}(1967)}]{Buchdahl1967}%
  \BibitemOpen
  \bibfield  {author} {\bibinfo {author} {\bibfnamefont {H.~A.}\ \bibnamefont {{Buchdahl}}},\ }\href {\doibase 10.1086/149001} {\bibfield  {journal} {\bibinfo  {journal} {\apj}\ }\textbf {\bibinfo {volume} {147}},\ \bibinfo {pages} {310} (\bibinfo {year} {1967})}\BibitemShut {NoStop}%
\bibitem [{\citenamefont {Buchdahl}(1959)}]{Buchdahl1959}%
  \BibitemOpen
  \bibfield  {author} {\bibinfo {author} {\bibfnamefont {H.~A.}\ \bibnamefont {Buchdahl}},\ }\href {\doibase 10.1103/PhysRev.116.1027} {\bibfield  {journal} {\bibinfo  {journal} {Phys. Rev.}\ }\textbf {\bibinfo {volume} {116}},\ \bibinfo {pages} {1027} (\bibinfo {year} {1959})}\BibitemShut {NoStop}%
\end{thebibliography}%
%%%%%%%%%%%%%%%%%%%%%%%%%%%%%%%%%%%%%%%%%%%%%%%%%%%%%%%%%%%%%%%%%%%%%%%%%%%%%%%%%%%%%%%%%%%%%%%%%%%%%%%%%%%%
%%%%%%%%%%%%%%%%%%%%%%%%%%%%%%%%%%%%%%%%%%%%%%%%%%%%%%%%%%%%%%%%%%%%%%%%%%%%%%%%%%%%%%%%%%%%%%%%%%%%%%%%%%%%
%%%%%%%%%%%%%%%%%%%%%%%%%%%%%%%%%%%%%%%%%%%%%%%%%%%%%%%%%%%%%%%%%%%%%%%%%%%%%%%%%%%%%%%%%%%%%%%%%%%%%%%%%%%%
\end{document}